\documentclass[a4paper,fleqn,usenatbib]{mnras}

\usepackage{mathptmx}

\usepackage[T1]{fontenc}
\usepackage{ae,aecompl}

\usepackage{graphicx}   
\usepackage{amsmath}    
\usepackage{amssymb}    
\usepackage{longtable}
\usepackage[hyphenbreaks]{breakurl} 

\newcommand{\nova}{YZ\,Ret}

\newcommand{\opticalpeaklum}{$8.1 \times 10^{38}$\,erg\,s$^{-1}$}
\newcommand{\opticalnustareplum}{$2.7 \times 10^{38}$\,erg\,s$^{-1}$}
\newcommand{\latpeaklum}{$3.3 \times 10^{35}$\,erg\,s$^{-1}$}
\newcommand{\latnustareplum}{$1.4 \times 10^{35}$\,erg\,s$^{-1}$}
\newcommand{\nustarlum}{$1 \times 10^{33}$\,erg\,s$^{-1}$}
\newcommand{\nustarlumextrapolated}{$2 \times 10^{33}$\,erg\,s$^{-1}$}

\newcommand{\topicalpeak}{$t_0 + 3.6$\,d}
\newcommand{\tnustarep}{$t_0+10$\,d}

\DeclareSymbolFont{mysymbols}     {OMS}{cmsy}{m}{n}
\SetSymbolFont{mysymbols}  {bold}{OMS}{cmsy}{b}{n}
\DeclareSymbolFont{myoperators}   {OT1}{cmr} {m}{n}
\SetSymbolFont{myoperators}{bold}{OT1}{cmr} {bx}{n}
\DeclareMathSymbol{\forall}{\mathord}{mysymbols}{"38}
\DeclareMathSymbol{\exists}{\mathord}{mysymbols}{"39}
\DeclareMathSymbol{\pm}{\mathbin}{mysymbols}{"06}
\DeclareMathSymbol{+}{\mathbin}{myoperators}{"2B}
\DeclareMathSymbol{-}{\mathbin}{mysymbols}{"00}
\DeclareMathSymbol{=}{\mathrel}{myoperators}{"3D}
\DeclareMathSymbol{\times}{\mathbin}{mysymbols}{"02}

\title[X-ray observations of \nova{}]{The first nova eruption in a novalike variable: \nova{} as seen in X-rays and $\gamma$-rays}

\author[K.~V.~Sokolovsky et al.]{
\parbox{\textwidth}{Kirill~V.~Sokolovsky$^{1,2}$\thanks{E-mail: kirx@kirx.net (KVS)},
Kwan-Lok~Li$^{3}$,
Raimundo~Lopes de Oliveira$^{4,5,6}$,
Jan-Uwe~Ness$^{7}$,
Koji~Mukai$^{8}$,
Laura~Chomiuk$^{1}$,
Elias~Aydi$^{1}$,
Elad~Steinberg$^{9}$,
Indrek~Vurm$^{10}$,
Brian~D.~Metzger$^{11,17}$,
Aliya-Nur~Babul$^{11}$,
Adam~Kawash$^{1}$,
Justin~D.~Linford$^{12}$,
Thomas~Nelson$^{13}$,
Kim~L.~Page$^{14}$,
Michael~P.~Rupen$^{15}$,
Jennifer~L.~Sokoloski$^{11}$,
Jay~Strader$^{1}$,
David~Kilkenny$^{16}$}
\vspace{0.4cm}\\
\parbox{\textwidth}{
$^{1}$Center for Data Intensive and Time Domain Astronomy, Department of Physics and Astronomy, Michigan State University, 567 Wilson Rd, East Lansing, MI 48824, USA\\
$^{2}$Sternberg Astronomical Institute, Moscow State University, Universitetskii~pr.~13, 119992~Moscow, Russia\\
$^{3}$Department of Physics, National Cheng Kung University, 70101 Tainan, Taiwan\\
$^{4}$Departamento de Astronomia, Instituto de Astronomia, Geof\'isica e Ci\^encias Atmosf\'ericas, Universidade de S\~ao Paulo, R. do Mat\~ao 1226, Cidade Universit\'aria, 05508-090, S\~ao Paulo, SP, Brazil\\
$^{5}$Departamento de F\'isica, Universidade Federal de Sergipe, Av. Marechal Rondon, S/N, 49100-000, S\~ao Crist\'ov\~ao, SE, Brazil\\
$^{6}$Observat\'orio Nacional, Rua Gal. Jos\'e Cristino 77, 20921-400, Rio~de~Janeiro, RJ, Brazil\\
$^{7}$European Space Astronomy Centre, Camino Bajo del Castillo s/n, Urb. Villafranca del Castillo, E-28692 Villanueva de la Ca\~nada, Madrid, Spain\\
$^{8}$CRESST and X-ray Astrophysics Laboratory, NASA/GSFC, Greenbelt, MD 20771, USA\\
$^{9}$Racah Institute of Physics, The Hebrew University, 9190401 Jerusalem, Israel\\
$^{10}$Tartu Observatory, University of Tartu, T\~oravere, 61602 Tartumaa, Estonia\\
$^{11}$Department of Physics and Columbia Astrophysics Laboratory, Columbia University, New York, NY 10027, USA\\
$^{12}$National Radio Astronomy Observatory, Domenici Science Operations Center, 1003 Lopezville Road, Socorro, NM 87801, USA\\
$^{13}$Department of Physics and Astronomy, University of Pittsburgh, Pittsburgh, PA 15260, USA\\
$^{14}$School of Physics and Astronomy, University of Leicester, University Road, Leicester, LE1 7RH, UK \\
$^{15}$National Research Council, Herzberg Astronomy and Astrophysics, 717 White Lake Rd, PO Box 248, Penticton, BC V2A 6J9, Canada \\
$^{16}$Department of Physics \& Astronomy, University of the Western Cape, Private Bag X17, Bellville 7535, South Africa\\
$^{17}$Center for Computational Astrophysics, Flatiron Institute, 162 5th Ave, New York, NY 10010, USA} 
}

\date{Accepted 2022 May 14. Received 2022 April 22; in original form 2021 August 6}

\pubyear{2022}

\begin{document}
\label{firstpage}
\pagerange{\pageref{firstpage}--\pageref{lastpage}}
\maketitle

\begin{abstract}
%
Peaking at 3.7\,mag on 2020 July 11, \nova{} was the second-brightest nova of the decade.
The nova's moderate proximity (2.7\,kpc, from \emph{Gaia}) provided an 
opportunity to explore its multi-wavelength properties in great detail.
Here we report on \nova{} as part of a long-term project to identify the physical mechanisms responsible for high-energy
emission in classical novae.
We use simultaneous {\em Fermi}/LAT and {\em NuSTAR} observations
complemented by {\em XMM-Newton} X-ray grating spectroscopy to probe 
the physical parameters of the shocked 
ejecta and 
the nova-hosting white dwarf.
The {\em XMM-Newton} observations revealed a 
super-soft X-ray emission which is dominated by
emission lines of \ion{C}{V}, \ion{C}{VI}, \ion{N}{VI}, \ion{N}{VII}, and \ion{O}{VIII}  
rather than a blackbody-like continuum, suggesting CO-composition of the white dwarf 
in a high-inclination binary system.
{\em Fermi}/LAT detected \nova{} for 15 days with the $\gamma$-ray spectrum best described by a power law with an exponential cut-off at $1.9 \pm 0.6$\,GeV. 
In stark contrast with theoretical predictions and in keeping with previous {\em NuSTAR} observations of 
\emph{Fermi}-detected classical novae (V5855\,Sgr and V906\,Car), 
the 3.5-78\,keV X-ray emission is found to be two orders of magnitude fainter than the GeV emission. 
The X-ray emission observed by {\em NuSTAR} is consistent with a single-temperature thermal plasma  
model. We do not detect a non-thermal tail of the GeV emission expected to extend down to the {\em NuSTAR} band.
{\em NuSTAR} observations continue to challenge theories of high-energy emission from shocks in novae.
\end{abstract}




\section{Introduction}

\subsection{Classical and dwarf novae}
\label{sec:cvintro}

Accreting white dwarf binaries are called cataclysmic variables when the donor is at or near the main sequence, or symbiotic for a giant donor. 
Many of them display two distinct types of violent phenomena that dramatically increase their
brightness \citep{2001cvs..book.....H,2003cvs..book.....W,2011ApJS..194...28K}: classical nova eruptions (powered by nuclear burning on the white dwarf surface) and dwarf nova outbursts (occurring in the accretion disc). 
Nova eruptions may strongly affect the evolutionary path of those binaries
\citep{1998MNRAS.297..633S,2016ApJ...817...69N,2021MNRAS.507..475G}.

The nova eruption results from a thermonuclear
runaway at the bottom of a hydrogen-rich shell of material accreted 
on to the white dwarf
\citep{2008clno.book.....B,2016PASP..128e1001S}. Novae reach optical peak 
absolute magnitudes in the range $-10$ to $-4$\,mag \citep{2017ApJ...834..196S} and are observed across 
the electromagnetic spectrum from GeV $\gamma$-rays to cm-band radio 
\citep[see the recent reviews by][]{2018arXiv180311529P,2020A&ARv..28....3D,2020arXiv201108751C}. 
The less dramatic (peak absolute magnitudes $\sim4.6$; \citealt{2011MNRAS.411.2695P}), but much
more frequent, phenomenon is the dwarf nova outburst. A dwarf nova occurs when 
the accretion disc surrounding a white dwarf switches from a low-viscosity, 
low-accretion-rate state to a high-viscosity, high-accretion-rate state
(\citealt{2005PJAB...81..291O,2020AdSpR..66.1004H}; 
see also \S~2.2.1 of \citealt{2007A&ARv..15....1D}).
Dwarf novae are prominent X-ray sources \citep{2010MNRAS.408.2298B} and faint radio emitters \citep{2016MNRAS.463.2229C}.

The link between classical and dwarf novae has long been 
established by the similarities of the white dwarf hosting binaries
where these phenomena occur. It is believed that all cataclysmic
variables accreting below the rate needed to sustain stable hydrogen
burning on the white dwarf \citep{2010AN....331..140K,2013ApJ...777..136W} 
periodically display nova eruptions
\citep[e.g.][]{1989PASP..101....5S,2013MNRAS.434.1902P,2020NatAs...4..886H}.
It is expected that most observed novae erupt in systems with a high mass transfer
rate. Such systems tend to have long periods above the 2--3\,h period gap where 
the mass transfer is presumably driven by the magnetic braking mechanism
\citep{1984MNRAS.209..227V,2001ApJ...550..897H}.
The magnetic braking may be more efficient than the gravitational wave radiation 
driving the evolution of white dwarf binaries below the period gap.
The typical high mass accretion rate allows white dwarfs in long-period systems to 
quickly accumulate mass needed for the next nova eruption
\citep{2005ApJ...623..398Y}, however with the exception of 10 known recurrent
novae \citep{2010ApJS..187..275S}, all other novae in the Milky Way recur on
time-scales $\gg100$\,years.

Old nova shells are found around some dwarf novae 
(\citealt{2007Natur.446..159S,2012ApJ...758..121S,2016MNRAS.456..633M,2018PASP..130i4201B,2020ATel13825....1B,2020ATel13829....1D}; 
but not others -- \citealt{2015MNRAS.449.2215S}). 
Some systems show dwarf nova outbursts after a classical nova eruption: 
Nova~Per~1901 \citep[GK\,Per; e.g.][]{2009MNRAS.399.1167E}, 
Nova~Ser~1903 \citep[X\,Ser;][]{2018A&A...614A.141S},
Nova~Sgr~1919 \citep[V1017\,Sgr;][]{2017MNRAS.469.4116V}, 
Nova~Cen~2005 \citep[V1047\,Cen;][]{2019ApJ...886L..14G,2021arXiv210807868A}, 
Nova~Oph~1954 \citep[V908\,Oph, OGLE-BLG-DN-0023;][]{2016MNRAS.462.1371T,2015AcA....65..313M},
Nova~Her~1960 \citep[V446\,Her;][]{2011AJ....141..121H} and the historical 
Nova~Sco~1437 \citep{2017Natur.548..558S} and Nova~Lyn~101 \citep[BK\,Lyn;][]{2013MNRAS.434.1902P}. 
The first four systems show long-lasting outbursts that notably differ from those of
ordinary dwarf novae. 
It is debated if some of these outbursts may be related to symbiotic outbursts,
called ZAND-type according to the General Catalogue of Variable Stars \citep[GCVS;][]{2017ARep...61...80S} classification scheme\footnote{\url{http://www.sai.msu.su/gcvs/gcvs/iii/vartype.txt}}. 
ZAND-type outbursts are probably partly powered by nuclear burning \citep{2006ApJ...636.1002S}.
Some old novae show low-amplitude `stunted' outbursts, 
but it is unclear if they are driven by the same disc instability 
mechanism as dwarf novae \citep{1998AJ....115.2527H,2018MNRAS.478.5427V}.
The archival data revealed that V1017\,Sgr and V1213\,Cen \citep{2016Natur.537..649M} 
were showing dwarf nova outbursts prior to the nova eruption when 
the variability of these objects was discovered (a few other objects displayed 
brightness variations prior to nova eruption, but the nature of these 
variations is uncertain; \citealt{2009AJ....138.1846C}). 

\nova{}, the subject of this paper, is only the third classical (rather than recurrent) nova eruption 
observed in a previously known white dwarf hosting 
binary.
The previous cases were the symbiotic (giant donor) system V407\,Cyg \citep{2011MNRAS.410L..52M} and 
V392\,Per \citep{2018RNAAS...2...24D,2020arXiv200713337C}. 
Both V407\,Cyg and V392\,Per were detected as prominent 
GeV \citep{2010Sci...329..817A,2018ATel11590....1L},
X-ray \citep{2012ApJ...748...43N,2018ATel11905....1D} and 
radio sources \citep{2012ApJ...761..173C,2020A&A...638A.130G,2018ATel11647....1L}. Recently, V1405\,Cas became the fourth previously known variable showing a nova eruption \citep{2021ATel14472....1T}.

\subsection{\nova{} as Nova~Reticuli~2020}
\label{sec:discoveryinfo}

The first low-resolution spectra of \nova{} (under the name
EC\,03572$-$5455) were obtained on 1992-12-19 and 1994-01-15. 
The South African Astronomical Observatory 1.9-m telescope was used together with the Reticon spectrograph 
by \cite{2015MNRAS.453.1879K} in the framework of the Edinburgh-Cape Blue Object Survey. 
The spectra covering 3400--5400\,\AA\ were described as `broad Balmer; He I ?' and at the time, 
the object was not recognized as a cataclysmic variable\footnote{To the best of our knowledge, 
this is only the second example of a pre-eruption spectrum of a classical (non-symbiotic,
non-recurrent) nova, the other being V392\,Per \citep{2000ApJS..128..387L}.}.
They are dominated by a blue continuum with superimposed broad
high-order Balmer absorption lines and weak H~$\beta$ absorption (probably filled
with emission).
Such absorption line-dominated spectra are seen in some novalike variables
and dwarf novae in outburst, for example RW\,Sex \citep{1972ApJ...176L..27C}. 

The variability of \nova{} was first noted in August 2019 by Gabriel Murawski, 
who investigated archival photometry from the Siding Spring Survey 
\citep[the southern counterpart of the Catalina Sky Survey;][]{2009ApJ...696..870D}
and the All-Sky Automated Survey for Supernovae \citep[ASAS-SN;][]{2014ApJ...788...48S,2017PASP..129j4502K} and
reported this object to the International Variable Star Index maintained by 
the American Association of Variable Star Observers
(AAVSO VSX\footnote{\url{https://www.aavso.org/vsx/}})
under the name MGAB-V207.
The object displayed fast irregular variations in the range 15.8--16.9\,mag with two
noticeable fadings down to 17.2\,mag and 18.0\,mag (unfiltered magnitudes
with V zero-point). These fadings suggested the object is an `antidwarf
nova' -- a VY\,Scl type cataclysmic variable
\citep{1999MNRAS.305..225L,2002A&A...394..231H,2004AJ....128.1279H}. 
Unlike the ordinary dwarf
novae that spend most of their time around minimum light (low accretion rate --
`cold accretion disc' state), VY\,Scl type systems spend most of their
time near maximum (higher accretion rate -- `hot accretion disc') dropping
to the minimum light only occasionally. Together with the similar non-magnetic
cataclysmic variables that always maintain a hot accretion disc 
(UX\,UMa and SW\,Sex stars), VY\,Scl systems are referred to as novalike variables
\citep{1996ASSL..208....3D}. An explanation of the VY\,Scl fading phenomenon 
solely in terms of variable mass transfer from the donor (without
relying on disc instability) is discussed in the literature 
\citep{1998MNRAS.295L..50H,2004AJ....128.1279H}.

\cite{2020CBET.4811....1M} noticed a 5\,mag object coinciding with \nova{} on
digital single-lens reflex camera images obtained on 2020-07-15.590\,UT (\S~\ref{sec:optical})
and reported the nova candidate to the Central Bureau for Astronomical Telegrams\footnote{\url{http://www.cbat.eps.harvard.edu/index.html}}. 
Pre-discovery all-sky images by M.~A.~Phillips 
show the nova peaking on 2020-07-11.76 (\topicalpeak{}, $t_0$ is defined below) at 3.7\,mag \citep{2020CBET.4812....1K}, 
while pre-discovery ASAS-SN images indicate the eruption started on
2020-07-08.171
($t_0 = {\rm JD(UTC)}2459038.671$). The nova was also detected by {\em Gaia}
Photometric Science Alerts on $t_0 + 42$\,d as Gaia20elz\footnote{\url{http://gsaweb.ast.cam.ac.uk/alerts/alert/Gaia20elz/}}. 
The pre-eruption {\em Gaia} lightcurve spanning 
$t_0 - 2006$\,d to $t_0 - 30$\,d 
showed irregular variations in the range $G=16.0$--$16.9$.

\begin{figure*}
        \includegraphics[width=1.0\linewidth,clip=true,trim=0cm 0.5cm 0.5cm 0cm,angle=0]{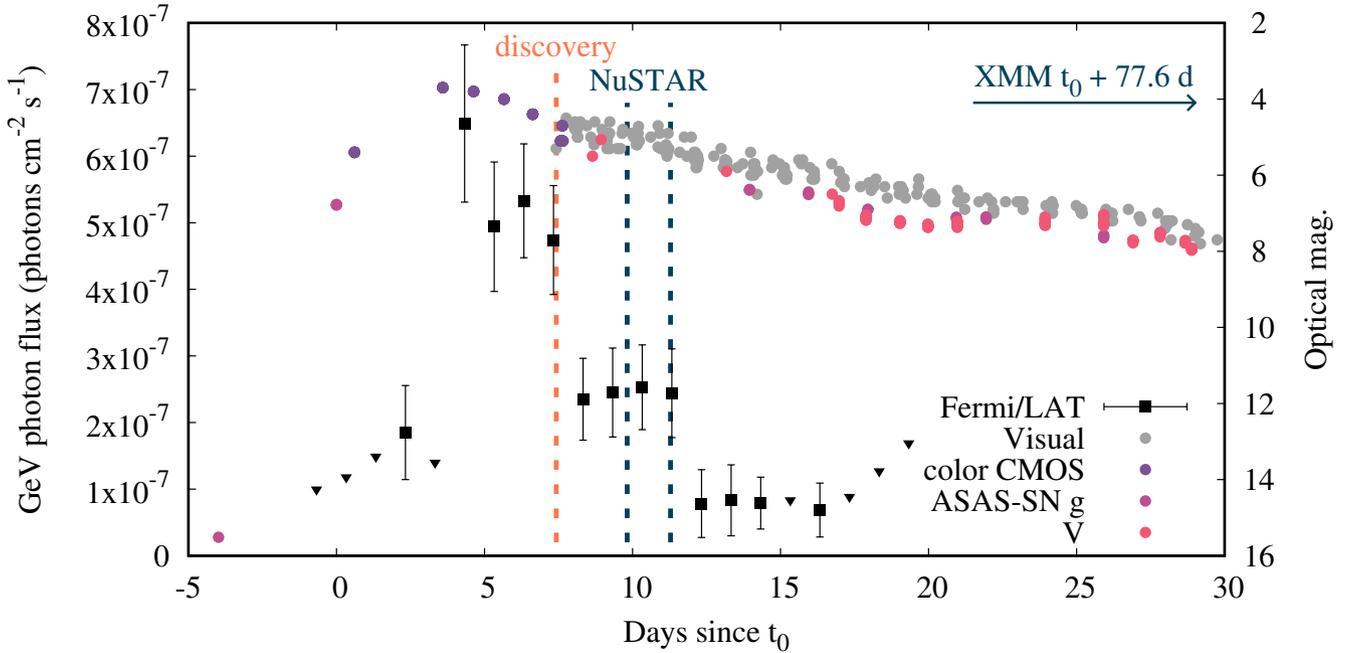}
\caption{{\em Fermi}/LAT $\gamma$-ray and optical lightcurve of \nova{}. 
The time is expressed in days since the first optical detection of the eruption by ASAS-SN on $t_0$ 2020-07-08.171~UT (\S~\ref{sec:discoveryinfo}). 
The black squares represent the {\em Fermi}/LAT detections, while the black triangles are $2\sigma$ upper
limits (\S~\ref{sec:latobs}). The optical observations, including data
collected with CCDs in $g$ and $V$ filters, colour CMOS chips and visual magnitude estimates, are described in \S~\ref{sec:optical}.
The nova discovery time, the duration of the {\em NuSTAR} observation and
the {\em XMM-Newton} observation time are indicated. 
The use of the usual units of $\gamma$-ray and optical flux
in this plot results in the $\gamma$-ray flux being plotted on a linear
scale, while the optical flux is on a logarithmic scale. The optical plot
covers a larger dynamic range than the $\gamma$-ray flux plot 
(over-emphasizing $\gamma$-ray variations) in order to display the latest
pre-eruption optical measurement. This plot aims to present the sequence of
events (eruption, peak, onset of GeV emission, {\em NuSTAR} observation etc.), 
rather than illustrate the relative magnitude of optical and GeV variations.} 
    \label{fig:latoptlc}
\end{figure*}

The fact that the naked-eye transient went unnoticed by the astronomical
community for about a week is alarming in light of our preparedness for
observing the next Galactic supernova \citep{2013ApJ...778..164A}. Existing
surveys relying on image subtraction for transient detection should
implement special procedures for handling new saturated sources. Regular
wide-field imaging of the sky (by both professional and amateur astronomers) 
aimed at detecting rare bright transients should be encouraged. 
To the best of our knowledge, only two Galactic novae have first been discovered 
at wavelengths other then optical or infrared \citep{2021arXiv210104045D}: V959\,Mon first found in
$\gamma$-rays by {\em Fermi}/LAT \citep{2012ATel.4310....1C} and V598\,Pup discovered as an X-ray
transient by {\em XMM-Newton} \citep{2008A&A...482L...1R}. 
\nova{} itself was a $\gamma$-ray transient with a daily flux of $\sim0.5\times10^{-6}$\,photons\,cm$^{-2}$\,s$^{-1}$
(Fig.~\ref{fig:latoptlc}) for three days prior to its optical discovery, but wasn't noticed.

Spectroscopic observations by
\cite{2020CBET.4812....1K}, \cite{2020ATel13867....1A},
\cite{2020ATel13874....1C}, \cite{2020ATel14048....1I},
\cite{2020ATel14149....1G}, \cite{2020ATel14205....1S}
confirmed the optical transient to be a classical nova past the optical peak. 
\nova{} was assigned its permanent GCVS designation following the nova eruption \citep{2020CBET.4826....1K}.
While \cite{2020CBET.4812....1K} describe the spectrum obtained on $t_0 + 8.4$\,d as that of a Fe\,II-type nova
\citep[according to the classification scheme of][]{1992AJ....104..725W},
\cite{2020ATel13874....1C} report He/N-type based on a series spectra obtained on $t_0 + 8.6$\,d with 
the Australian National University 2.3-m telescope. 
From an over-abundance of oxygen and the presence of
[\ion{Ne}{III}]~3342\,\AA{} and [\ion{Ne}{V}]~3426\,\AA{} lines in 
the Very Large Telescope/Ultraviolet and Visual Echelle Spectrograph spectrum obtained on $t_0 + 72$\,d, \cite{2020ATel14048....1I} 
conclude that the nova erupted on an ONe white dwarf (cf. \S~\ref{sec:ejectaabund}).
\cite{2021MNRAS.503..704M} describe their exceptionally dense monitoring of
the line profile evolution in \nova{}, 
while \cite{2021RNAAS...5...48R} report late-time infrared spectroscopy.

\nova{} was detected on $t_0 + 2$\,d in the GeV band by the Large Area Telescope (LAT) on the 
{\em Fermi Gamma-Ray Space Telescope} (\citealt{2020ATel13868....1L}; \S~\ref{sec:latobs}) 
and on $t_0 + 10$\,d at hard X-rays by {\em NuSTAR} (\citealt{2020ATel13900....1S}; \S~\ref{sec:nustarspec}).
By 2020-08-04 ($t_0 + 27$\,d) the emission at the softer 0.3--10\,keV X-ray band
was detected by {\em Swift}/XRT. 
On $t_0 + 59$\,d the soft counts at the XRT 
band started rising dramatically signifying the appearance of 
the super-soft-source \citep[SSS;][]{2020ATel14043....1S}. The super-soft emission
was also observed on $t_0 + 82$\,d 
with the {\em NICER} instrument (0.24--10\,keV) aboard the International Space
Station by \cite{2020ATel14067....1P} who noted aperiodic variations in the
X-ray flux with the amplitude of about 8~per~cent on a time-scale of kiloseconds.
X-ray grating spectroscopy of \nova{} was obtained with {\em Chandra} by \cite{2020ATel14214....1D} on $t_0 + 115$\,d. 
\nova{} was also detected as a faint cm-band radio source at $t_0 + 578$\,d \citep{2022ATel15264....1G}
\cite{2021RNAAS...5..150S} report the pre-eruption orbital period of $0.1324539 \pm0.0000098$\,d
for \nova{} based on {\em TESS} optical photometry.

\subsection{\nova{} position, distance and Galactic extinction}
\label{sec:extinction}

The {\em Gaia}~DR2 \citep{2018A&A...616A...1G} lists the position of \nova{} measured at the mean epoch of 2015.5:
\begin{verbatim}
03:58:29.56 -54:46:41.2 J2000
\end{verbatim}
with the proper motion of  
$7.244 \pm 0.089$ and $2.984 \pm 0.096$\,mas\,yr$^{-1}$ in R.A. and Dec. directions,
respectively. The {\em Gaia}~DR2 parallax of $0.3161 \pm 0.0464$\,mas corresponds
to the distance of $2703_{-293}^{+366}$\,pc according to \cite{2018AJ....156...58B}.
The distance may be underestimated without correction of the apparent motion around the common centre of mass of the binary.

We are lucky to have the trigonometric parallax for \nova{}, 
as progenitors of many other novae lack {\em Gaia} parallaxes due to their faintness. 
\cite{2018MNRAS.481.3033S} estimates that {\em Gaia} provides reliable parallaxes for less than 20~per~cent of the known novae. 
In fact, \cite{2018MNRAS.481.3033S} reports {\em Gaia} parallaxes of 41
novae, 9~per~cent of the 464 novae\footnote{\url{https://github.com/Bill-Gray/galnovae}}
known at the time when that paper was submitted.
For \nova{}, both its relative
proximity and intrinsic brightness (the pre-nova was in the hot accretion
disc state; \S~\ref{sec:discoveryinfo}) helped secure the parallax measurement.
However, the distance uncertainty remains the main contributor to the uncertainty in luminosity of \nova{}. 
\nova{} is located 1.9\,kpc above the Galactic plane at Galactic
coordinates $l=265.39744$, $b=-46.39540$, so it is likely associated with 
the Milky Way's thick disc.

The interstellar reddening towards the nova can be estimated from multicolour 
photometry, assuming a typical intrinsic colour of $(B-V)_0=-0.02$ when the nova is two magnitudes 
below its peak \citep[the dispersion of $(B-V)_0$ is $0.12$\,mag;][]{1987A&AS...70..125V}.
According to photometry reported by A.~Valvasori to AAVSO, on 2020-07-16.817 (JD\,2459047.317)
\nova{} had $V=5.50 \pm 0.05$ and $(B-V) = 0.01 \pm 0.06$. Therefore, the colour excess is $E(B-V)=0.03$, 
which for the standard value of the ratio 
$R=\frac{A_V}{E(B-V)}=3.1$ corresponds to $A_V = 0.08$\,mag.
This is consistent with $E(B-V)<0.1$ derived from the infrared spectroscopy 
by \cite{2021RNAAS...5...48R}. Given the uncertainty in photometry and 
the scatter of nova intrinsic colours, the foreground reddening/absorption 
are consistent with zero.

We can estimate the expected Galactic X-ray absorbing column to \nova{}
using the relation of \cite{2009MNRAS.400.2050G}: 
\begin{equation}
N_\mathrm{H} = 2.21 \times 10^{21}\,{\rm cm}^{-2} \times A_V = 1.86 \times 10^{20}\,{\rm cm}^{-2}
\end{equation}
A small positive value of $E(B-V)$ (and hence $N_\mathrm{H}$)
seems like a better estimate than the hard limit of zero. 
We adopt the above $N_\mathrm{H}$ value for the X-ray spectral analysis
(\S~\ref{sec:nustarspec}).
The adopted $N_\mathrm{H}$ value is close to the total Galactic \ion{H}{I} column in that direction
estimated from radio observations of the 21\,cm hydrogen line: 
$N_\mathrm{HI} = 1.18 \times 10^{20}$\,cm$^{-2}$ \citep{2005A&A...440..775K,2005A&A...440..767B}.
The 21\,cm-derived column density does not account for ionized and molecular
hydrogen, while the abundances of X-ray absorbing atoms are normalized to
the total number of hydrogen atoms. However, these contributions are small
and $N_\mathrm{HI}$ values are often taken as estimates of the total $N_\mathrm{H}$
for the purpose of calculating the X-ray absorbing column.
\cite{2020ATel14048....1I} used the \cite{1985ApJ...298..838F} relation between 
the column density of \ion{Na}{I} (derived from high-resolution optical
spectroscopy) and $N_\mathrm{H} = N_\mathrm{HI} + 2 N_\mathrm{H_2}$ to find 
$N_\mathrm{H} = 10^{19}\,{\rm cm}^{-2}$ for \nova{}, an order of magnitude lower than what we adopt.

\subsection{Novae in $\gamma$-rays and X-rays}
\label{sec:xraynovaintro}

High energy emission of novae may be produced by various mechanisms.
It has long been predicted that decay of radioactive nuclei produced 
in nova nucleosynthesis should emit lines in the MeV band \citep{2014ASPC..490..319H,2016stex.book.....J}.
The 511\,keV electron-positron annihilation line should also be present. 
Comptonization will produce continuum emission at energies below each of
these lines.
The MeV emission from novae has never been observed as the coded aperture mask telescopes
currently operating in the $\sim1$\,MeV band (SPI and IBIS aboard {\em INTEGRAL}) are
probably sufficiently sensitive to detect only a very nearby nova at
a distance $<1$\,kpc, as hinted by the ongoing searches \citep{2002AIPC..637..435H,2018A&A...615A.107S}.  

Another predicted 
phenomenon, that remained unobserved until very recently, is the thermal 
emission from the fireball produced by the thermonuclear runaway. 
Within a few seconds of the onset of the thermonuclear runaway at the bottom
of the accreted envelope, the convection turns on which transports the heat
and decaying radioactive nuclei to the white dwarf surface
\citep{2008ASPC..401..139K,2016PASP..128e1001S}. 
The result is the extreme heating and expansion of the outer layers of the white dwarf.
As the fireball expands, its emission peak shifts from soft X-rays to UV and
then to the optical band \citep{2001MNRAS.320..103S,2002AIPC..637..345K,2007ApJ...663..505N}.
Despite the ongoing searches 
\citep{2016PASJ...68S..11M,2016ApJ...830...40K}, no unambiguous detection 
of the fireball has been reported until now \citep{2013ApJ...779..118M,2012ApJ...761...99L}.
While this manuscript was in review, \citep{Konig2022} presented early {\em Spektr-RG/eROSITA} 
observations of \nova{} that signified the first clear detection of the nova fireball.

Optically thick thermal emission from the heated atmosphere of 
the hydrogen-burning white dwarf becomes visible again 
when the nova ejecta become transparent enough to soft X-rays 
\citep[SSS phase;][]{1994RvMA....7..129H,1997ARA&A..35...69K, 2011ApJS..197...31S}.
According to the modelling by \cite{2013ApJ...777..136W}, the post-nova white
dwarf atmosphere temperature is k$T<0.2$\,keV, while observationally
emission at $<0.5$\,keV is usually considered super-soft.

Shock waves are invoked to explain GeV and hard X-ray emission of 
novae, as well as synchrotron radio emission and high excitation lines in
optical spectra.
Shocks compress and heat plasma to X-ray temperatures \citep[e.g.][]{1967pswh.book.....Z}. 
The shock-heated plasma gives rise to the optically thin thermal emission at
energies $\gtrsim$1\,keV observed in many novae
\citep{2014MNRAS.442..713M,2014ASPC..490..327M,2017PASP..129f2001M,2021ApJ...910..134G}.
Shocks can also amplify any pre-existing magnetic field and use it to
accelerate charged particles to high energies \citep[][]{1978ApJ...221L..29B,2012SSRv..173..491S}. The relativistic particles may emit synchrotron radio as well as high-energy radiation. 
Depending on the balance between the acceleration efficiency and energy
losses, electrons or protons may be the primary particles producing $\gamma$-rays via 
leptonic or hadronic mechanisms \citep{2015MNRAS.450.2739M,2018A&A...612A..38M}. 
In the leptonic scenario electrons are the primary accelerated particles 
that produce $\gamma$-rays via bremsstrahlung and inverse Compton scattering of ambient as well as their own synchrotron photons.
In the hadronic scenario, most of the $\gamma$-ray flux arises 
from the decay of pions produced in interactions of high-energy protons with the surrounding ions and photons. 
The secondary electron/positron pairs from charged pion decay also contribute to the $\gamma$-ray emission via inverse Compton scattering and bremsstrahlung \citep{2018ApJ...852...62V}.
The same mechanisms are believed to be responsible for the high-energy emission of
blazars\footnote{Blazars are active galactic nuclei with relativistic jets pointing
close to the line of sight. The majority of extragalactic GeV
sources are blazars.}, except interactions with matter (bremsstrahlung,
proton-proton collisions) are expected to be less important in blazar jets
than interactions of high-energy particles with photons and the external 
magnetic field \citep{2013ApJ...768...54B,2020Galax...8...72C}.

As of August 2021, GeV emission has been detected from 18 novae: 
the list of \cite{2021ApJ...910..134G}, plus 
V3890\,Sgr \citep{2019ATel13114....1B}, 
V1707\,Sco \citep{2019ATel13116....1L}, \nova{} (\S~\ref{sec:latobs}), 
V1405\,Cas \citep{2021ATel14658....1B},  
V1674\,Her \citep{2021ATel14705....1L}\footnote{\url{https://asd.gsfc.nasa.gov/Koji.Mukai/novae/latnovae.html}}.
\cite{2018A&A...609A.120F} list V679\,Car and V1535\,Sco as low-significance detections.
The $\gamma$-ray properties of novae were investigated by 
\cite{2014Sci...345..554A}, \cite{2016ApJ...826..142C},
\cite{2017NatAs...1..697L}, \cite{2020NatAs...4..776A},
\cite{2020ApJ...905..114L}, \cite{2020arXiv201108751C}.

\subsection{Scope of this work}
\label{sec:thispaper}

We analyse simultaneous GeV $\gamma$-ray 
(0.1--300\,GeV from {\em Fermi}/LAT; \S~\ref{sec:latobs}) and hard X-ray 
(3--79\,keV from {\em NuSTAR}; \S~\ref{sec:nustarspec}) 
observations of the 2020 classical nova eruption of \nova{}, complemented by X-ray grating spectroscopy with 
{\em XMM-Newton} at a later epoch when the nova became bright in the 0.2--10\,keV band (\S~\ref{sec:xmm}). 
We measure the $\gamma$-ray to X-ray flux ratio and
use it to constrain the $\gamma$-ray emission mechanism (\S~\ref{sec:emissionmech}).
We conclude that the hard X-ray emission observed by {\em NuSTAR} is
thermal, based on its spectral shape and speculate about the possible
locations of shocks responsible for the high-energy emission (\S~\ref{sec:location}).
The trigonometric parallax from {\em Gaia}~DR2 (\S~\ref{sec:extinction}) 
allows us to accurately determine the $\gamma$-ray, X-ray and optical luminosity of the nova. 
The paper at hand is a continuation of work by \cite{2019ApJ...872...86N} and \cite{2020MNRAS.497.2569S} 
building a sample of novae simultaneously detected by {\em NuSTAR} and {\em Fermi}/LAT with the aim to characterize shocks in novae.

Throughout this paper we report uncertainties at the $1 \sigma$ level.
For power law emission, we use the positively-defined spectral index $\alpha$ (commonly used in radio astronomy): 
$F_\nu \propto \nu^\alpha$ 
where $F_\nu$ is the spectral flux density and $\nu$ is the frequency; 
the corresponding index in the distribution of the number of photons 
as a function of energy (used in high-energy astronomy) is ${\rm d}N(E)/dE \propto E^{-\Gamma}$, 
where $\Gamma$ is the photon index and $\Gamma = 1 - \alpha$.
The same power law expressed in spectral energy distribution units 
(SED, commonly used in multiwavelength studies and in theoretical studies; \citealt{1997NCimB.112...11G}) is 
$\nu F_\nu \propto \nu^{\alpha + 1} \propto \nu^{-\Gamma + 2}$.
Throughout the text we use the terms `GeV novae' and `$\gamma$-ray novae' 
interchangeably implying the novae detected in the {\em Fermi}/LAT band (0.1--300\,GeV). 
All novae may produce GeV $\gamma$-rays, so `$\gamma$-ray novae' 
are unlikely to be a distinct class and are only the nearest and/or most luminous novae that we can detect.

\section{Observations and analysis}
\label{sec:obs}

\subsection{{\em Fermi}/LAT observations}
\label{sec:latobs}

{\em Fermi}/LAT is a pair-conversion telescope sensitive to $\gamma$-rays in the range 
20\,MeV--300\,GeV with a field of view of 2.4\,sr \citep{2009ApJ...697.1071A,2009APh....32..193A,2012ApJS..203....4A}.
Earlier in the mission, {\em Fermi}/LAT performed a nearly-uniform all-sky survey every day.
The pointing pattern had to be modified after the solar panel drive failure on 2018-03-16, 
resulting in a non-uniform exposure over the sky. 

We downloaded the {\em Fermi}/LAT photon data centred on \nova{} 
(search radius: 20\,degrees; 
energy range: 50\,MeV--300\,GeV; data version: P8R3\_SOURCE\_V2 \citealt{2018arXiv181011394B}) from 
the LAT Data Server at the Fermi Science Support Center\footnote{\url{https://fermi.gsfc.nasa.gov/ssc/data/}}. 
\textsc{Fermitools 1.2.23} with \textsc{fermitools-data 0.18} was used to reduce and analyse the
$\gamma$-ray data. 
We performed the binned analysis with 
a $\gamma$-ray emission model file of the field based on 
the {\em Fermi} Large Area Telescope fourth source catalogue
(4FGL; \citealt{2020ApJS..247...33A}; \texttt{gll\_psc\_v22.fit}). 
The model file includes all the 4FGL sources found within 30 degrees from the target. 
For nearby sources that are within 10 degrees from the nova, we freed the normalization parameters to minimize possible contamination. 
In addition to the catalogued sources, two background emission components, the Galactic (\texttt{gll\_iem\_v07}) 
and isotropic (\texttt{iso\_P8R3\_SOURCE\_V2\_v1}) diffuse emission, were adopted. 

First, we performed a preliminary analysis in the 100\,MeV--300\,GeV energy
range to determine the $\gamma$-ray active period of \nova{} (the normalization parameters of all the 4FGL sources in the model file 
were temporarily fixed to save computational time). Assuming a simple power law $\gamma$-ray
spectrum for \nova{}, we performed analysis with one-day binning in time 
from 2020-06-30 00:00 to 2020-08-04 00:00~UT (MJD 59030.0--59065.0; $t_0-8.2$ to $t_0+26.8$\,d) 
to obtain the $\gamma$-ray lightcurve (Fig.~\ref{fig:latoptlc}). 
With a threshold set at the test statistic \citep{1996ApJ...461..396M} ${\rm TS}>4$ (detection significance $>2\sigma$), 
the analysis gives a detection interval from 2020-07-10 to 2020-07-25 (MJD 59040.0--59055.0; $t_0 + 1.8$ to $t_0 16.8$\,d). 
Using the LAT data taken in this interval, 
we tried to fit the photon data with two spectral models for \nova{}: 
a simple power law 
and a power law with an exponential cut-off, Eq.~(\ref{eq:latpowerlawexp}). 
Both models result in significant detection with ${\rm TS}=676$ (power law) and ${\rm TS}=695$ (exponential cut-off power law). 
A likelihood-ratio test suggests that the exponential cut-off power law is preferred
with a significance of $4.4\sigma$. 
The $\gamma$-ray lightcurve was then updated based on the new exponentially cut-off power law spectral model 
(except for the normalization parameters of \nova{} and the background components, 
all spectral parameters were fixed).

Figure~\ref{fig:latoptlc} presents the daily {\em Fermi}/LAT lightcurve of \nova{}
constructed with the simple power law model. 
If the source was detected with ${\rm TS}<2$ in
a daily integration, its derived photon flux was treated as an upper limit. 
The $\gamma$-ray emission is first detected (${\rm TS}>2$) on 2020-07-10
($t_0+1.8$\,d), peaks two days later ($t_0+4$\,d) at 
$(6.5 \pm 1.2) \times 10^{-7}$\,photons\,cm$^{-2}$\,s$^{-1}$, 
equivalent to the 0.1--300\,GeV peak energy flux of 
$(4.3 \pm 0.8) \times 10^{-10}$\,erg\,cm$^{-2}$\,s$^{-1}$, 
then gradually fades, being last detected on 2020-07-24 ($t_0+16$\,d).

We then analysed {\em Fermi}/LAT data collected simultaneously with the {\em NuSTAR}
observation: 2020-07-17 23:36 -- 2020-07-19 10:46 UT (MJD 59047.98--59049.45;
$t_0+9.81$ -- $t_0+11.28$\,d). 
\nova{} is clearly detected in this time interval 
with ${\rm TS}=104$ and 100\,MeV--300\,GeV photon flux of 
$(2.8 \pm 0.5) \times 10^{-7}$\,photons\,cm$^{-2}$\,s$^{-1}$,
equivalent to the energy flux of 
$(1.9 \pm 0.4) \times 10^{-10}$\,erg\,cm$^{-2}$\,s$^{-1}$.
We adopted the power law with an exponential cut-off model for the $\gamma$-ray spectrum.
The monochromatic flux at 100\,MeV derived from this model using Eq.~(\ref{eq:latpowerlawexpsed}) is 
$\nu F_\nu = (3.6 \pm 0.7) \times 10^{-11}$\,erg\,cm$^{-2}$\,s$^{-1}$. 
The accuracy of {\em Fermi}/LAT absolute calibration at 100\,MeV is
about 5~per~cent \citep{2012ApJS..203....4A}.

\begin{figure}
        \includegraphics[width=1.0\linewidth]{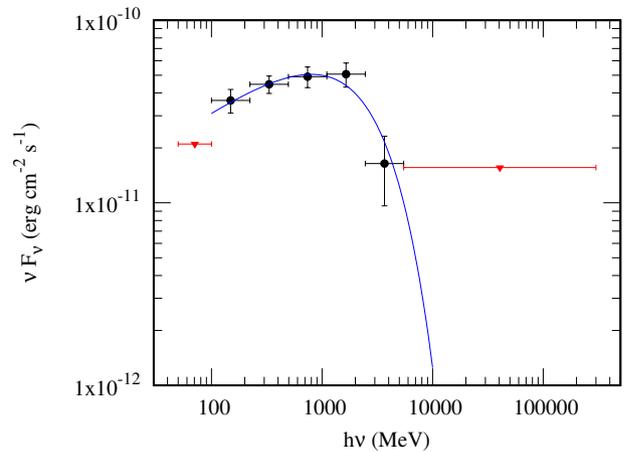}
    \caption{{\em Fermi}/LAT SED of \nova{}. The black points represent the
binned LAT data while the red triangles mark $2\sigma$ upper limits. The blue
curve is the power law with an exponential cut-off,
Eq.~(\ref{eq:latpowerlawexpsed}), model fitted to
0.1--300\,GeV photon data using the maximum likelihood technique.}
    \label{fig:latsed}
\end{figure}

Fig.~\ref{fig:latsed} presents the {\em Fermi}/LAT SED integrated over 
the whole two-week $\gamma$-ray activity phase (all days with ${\rm TS}>2$).
The 0.1--300\,GeV spectrum (number of photons per unit energy) 
is approximated with the power law with an exponential cut-off at high energy:
\begin{equation}
\label{eq:latpowerlawexp}
\frac{{\rm d}N(E)}{{\rm d}E} = N_0 \left(\frac{E}{E_0}\right)^{-\Gamma} e^{-\frac{E}{E_c}},
\end{equation}
where $N_0 = (6.74 \pm 0.72) \times 10^{-10}$\,photons\,MeV$^{-1}$\,cm$^{-2}$\,s$^{-1}$
(fitted) at $E_0 = 200$\,MeV (fixed), 
$\Gamma = 1.59 \pm 0.16$ (fitted)
and the cut-off energy $E_c = 1943 \pm 657$\,MeV (fitted). In monochromatic flux
(SED) units (\S~\ref{sec:thispaper})
the same relation translates to
\begin{equation}
\label{eq:latpowerlawexpsed}
\nu F_\nu = C_{\rm erg/MeV} E_{\rm MeV}^2 N_0 \left(\frac{E_{\rm MeV}}{E_0}\right)^{-\Gamma} e^{-\frac{E_{\rm MeV}}{E_c}},
\end{equation}
where $C_{\rm erg/MeV} = 1.60218 \times 10^{-6}$ is the conversion factor from MeV to
erg. This relation is useful if the monochromatic flux is expressed in ergs 
(the energy and photon fluxes are measured per unit area) while the photon
energy $E$ and the corresponding constants, $E_0$, $E_c$ are expressed in MeV 
($N_0$ is in photons\,MeV$^{-1}$\,cm$^{-2}$\,s$^{-1}$) 
according to the \textsc{Fermitools} convention.

As \nova{} is far from the Galactic plane (\S~\ref{sec:extinction}) 
where contamination at $<100$\,MeV from the Galactic diffuse emission is limited, 
the low-energy LAT data of 50--100\,MeV (which is usually unusable due to the huge Galactic background)
were also analysed. Despite the low noise level, the nova was undetected in this low energy band (${\rm TS}=0$). 
We computed a 95~per~cent upper limit in this band, which is around $2.1 \times 10^{-11}$\,erg\,cm$^{-2}$\,s$^{-1}$. 
However, because of the low collecting area of {\em Fermi}/LAT in this energy range, the limit should be used with caution. 
We used a `flat' power-law ($\Gamma=2$) to derive the photon flux or
its upper limit in each energy bin when reconstructing the {\em Fermi}/LAT
SED of \nova{}.

The model is fit to the 100\,MeV--300\,GeV photon data using the maximum likelihood technique
\citep{1996ApJ...461..396M}. The fitting result is compared to the
{\em Fermi}/LAT SED in Fig.~\ref{fig:latsed}. 
The ${\rm TS}<4$ upper limit on the 50--100\,MeV
photon flux suggests that the spectrum might be departing from the power law
below 100\,MeV (Fig.~\ref{fig:latsed}).
The 0.1--300\,GeV photon flux integrated over the whole $\gamma$-ray activity phase
is $(2.7 \pm 0.2) \times 10^{-7}$\,photons\,cm$^{-2}$\,s$^{-1}$ equivalent 
to the energy flux of $(2.5 \pm 0.3) \times 10^{-10}$\,erg\,cm$^{-2}$\,s$^{-1}$.

\begin{table*}
        \centering
        \caption{X-ray observations log}
        \label{tab:xrayobslog}
        \begin{tabular}{cccccc} 
                \hline
Mission          & ObsID       & PI         & Exposure  & Date       & $t_0$ \\
                \hline
\multicolumn{6}{c}{Pointed observations} \\
{\em NuSTAR}     & 80601317002 & Sokolovsky & 66\,ks    & 2020-07-17 &  $+9.8$\,d  \\
{\em XMM-Newton} & 0871010101  & Sokolovsky & 28\,ks    & 2020-09-23 & $+77.6$\,d  \\
\multicolumn{6}{c}{Slew exposures} \\
{\em ROSAT}      & Survey      &            & 0.5\,ks   & 1990-07-11 & $-10955$\,d \\
{\em XMM-Newton} & 9042100004  &            & 0.005\,ks & 2002-03-28 & $-6677$\,d \\
{\em XMM-Newton} & 9099800003  &            & 0.007\,ks & 2005-05-22 & $-5525$\,d \\
{\em XMM-Newton} & 9175600004  &            & 0.004\,ks & 2009-07-12 & $-4014$\,d \\
{\em XMM-Newton} & 9272700003  &            & 0.002\,ks & 2014-10-30 & $-2078$\,d \\
{\em XMM-Newton} & 9350700002  &            & 0.010\,ks & 2019-02-01 & $-522$\,d \\
{\em XMM-Newton} & 9384600002  &            & 0.005\,ks & 2020-12-08 & $+153.6$\,d \\
{\em XMM-Newton} & 9389300003  &            & 0.009\,ks & 2021-03-13 & $+247.9$\,d \\
                \hline
        \end{tabular}
\end{table*}

\subsection{{\em NuSTAR} observations}
\label{sec:nustarobs}

{\em NuSTAR} \citep{2013ApJ...770..103H} is equipped with a pair of 
identical focusing X-ray telescopes sensitive to hard X-ray photons with 
energies 3--79\,keV \citep{2015ApJS..220....8M}. It is in a low-Earth orbit,
so the observations are periodically interrupted by Earth occultations and
the South Atlantic Anomaly \citep[e.g.][]{2002JASTP..64.1701H} passages.

{\em NuSTAR} observed \nova{} between 2020-07-17 23:36 and 2020-07-19 10:46~UT 
(\tnustarep{}; ObsID~80601317002; PI:~Sokolovsky) for a total exposure of 66\,ks 
(see Table~\ref{tab:xrayobslog} for a summary of X-ray observations). 
The preliminary analysis of this observation was
reported by \cite{2020ATel13900....1S}.
For the analysis we used \textsc{nupipeline} and \textsc{nuproducts} commands from
\textsc{HEASoft\,6.27.2} \citep{2014ascl.soft08004N} to extract source and background spectra and
lightcurves from the focal plane modules A (FPMA) and B (FPMB). 
The nova is clearly detected with signal-to-noise of $\sim11$ by both focal plane modules.
We followed the analysis procedure described by \cite{2020MNRAS.497.2569S}. 
Specifically, we utilized a circular extraction region with 
the radius of 30\,$\arcsec$ centred on the X-ray image of the nova (using \textsc{ds9}; \citealt{2003ASPC..295..489J})
independently for FPMA and FPMB. The background was extracted from five circular regions 
of the same radius placed on the same CZT \citep{2011hxra.book.....A} chip as the nova image.
For an overview of X-ray spectroscopy and timing analysis techniques, see
\cite{2011hxra.book.....A} and \cite{2020tgxg.book.....B}. 

\subsubsection{{\em NuSTAR} spectroscopy}
\label{sec:nustarspec}

The {\em NuSTAR} spectra of previously observed novae where found 
consistent with being emitted by single-temperature optically thin plasma 
\citep{2015MNRAS.448L..35O,2019ApJ...872...86N,2020MNRAS.497.2569S}. 
The plasma is likely heated by a shock \citep{1967pswh.book.....Z}
associated with the nova eruption \citep{2015MNRAS.450.2739M}.
The shock also accelerates high-energy particles responsible for the
$\gamma$-ray emission that may extend down to the {\em NuSTAR} band
\citep{2018ApJ...852...62V}. Based on the previous nova observations and
theoretical expectations, we try two classes of models to describe {\em NuSTAR}
observations of \nova{}: single-temperature optically thin thermal plasma
emission model and a simple power law, as well as a combination of the two models. 
It is also known from optical spectroscopy that nova ejecta are overabundant in CNO elements 
\citep{1985ESOC...21..225W,1998PASP..110....3G,2001MNRAS.320..103S}. 
The composition affects both the spectrum of thermally emitting hot plasma and 
the cold absorber altering the intrinsic thermal and/or non-thermal spectrum. 
In this section we present a detailed description of the spectral fitting 
and explain the adoption of the thermal emission model from a plasma with 
non-solar abundances.

The 3.5--78\,keV emission observed by {\em NuSTAR} is essentially featureless and can be described equally well by
a power law, thermal emission from pure bremsstrahlung \citep{1975ApJ...199..299K}, and thermal emission from bremsstrahlung continuum plus line emission 
\citep[\texttt{vapec;}][]{2005AIPC..774..405B} with non-solar abundances. 
The \texttt{vapec} model with solar abundances results in an
unacceptable fit with $\chi^2 = 43.93$ for 22 degrees of freedom
(Table~\ref{tab:nustarspecmodels}).
In order to suppress the line emission expected for solar-abundance plasma and, specifically, 
the Fe~K~$\alpha$ emission at 6.7\,keV,
the plasma should either be Fe-deficient, or overabundant in nitrogen and oxygen.
While absent in \nova{} and V906\,Car \citep{2020MNRAS.497.2569S}, the 6.7\,keV emission is clearly visible in the {\em NuSTAR} spectrum of
the recurrent nova V745\,Sco, where the shock propagates through the dense wind
of the red giant companion that presumably has nearly-solar abundances
\citep{2015MNRAS.448L..35O}.
The {\em NuSTAR} spectrum of V5855\,Sgr had too few counts to constrain the abundances \citep{2019ApJ...872...86N}.
A combination of both Fe-deficiency and NO overabundance is also possible
and was found in nova V906\,Car by \cite{2020MNRAS.497.2569S}. 
Also in the nova V382\,Vel, a post-outburst X-ray grating spectrum contained no Fe lines 
but strong emission lines of C, N, O, Ne, Mg, and Si \citep{2005MNRAS.364.1015N}. 
A power law provides an adequate fit to the spectrum of \nova{} given the non-solar abundances of the absorber. 
Physically, the power law model may represent non-thermal emission or thermal emission with a very high temperature.
The monochromatic flux at 20\,keV derived from the power law fit using Eq.~(\ref{eq:powerlawsed}) is 
$\nu F_\nu = 2.5 \times 10^{-13}$\,erg\,cm$^{-2}$\,s$^{-1}$. 

For nova V906\,Car the thermal model could be clearly favoured over the power
law thanks to the good statistics. We cannot distinguish between the power law and thermal models for \nova{} 
as both provide a statistically acceptable fit. However, we prefer the thermal model for the X-ray emission of \nova{} 
as we expect the same emission mechanisms at work in nova systems. In addition, the observed soft power law is at odds 
with the theoretical expectations as discussed in \S~\ref{sec:emissionmech}.

The source and background spectra, together with the associated redistribution matrix
and auxiliary response files, were {analysed with} \textsc{XSPEC\,12.11.0}
\citep{1996ASPC..101...17A}.
We restrict the energy range to 3.5--78.0\,keV to avoid calibration uncertainties near 3.0\,keV.
These uncertainties are mostly related to the rip in the protective polyimide film 
\citep{2020arXiv200500569M} 
that covers both front and back sides of {\em NuSTAR} mirror assembly
\citep{2011SPIE.8147E..0HC}.
The 3.5--78.0\,keV spectrum was fit by heavily absorbed, optically thin 
thermal plasma emission \citep[\texttt{vapec;}][]{2005AIPC..774..405B}, with $N_\mathrm{H} \approx 10^{23}$--$10^{24}$\,cm$^{-2}$ 
(depending on the choice of abundances) and ${\rm k}T = 6.5 \pm1.5$\,keV. The unabsorbed 3.5--78\,keV flux is 
$1.1 \times 10^{-12}$\,erg\,cm$^{-2}$\,s$^{-1}$ (or $1.4\times 10^{-12}$\,erg\,cm$^{-2}$\,s$^{-1}$, 
again depending on abundances). 

To obtain a good fit to the \emph{NuSTAR} spectrum, we had to allow for non-solar abundances of N, O and/or Fe for 
both the absorber and emitter. These elements have prominent absorption and emission features in the {\em NuSTAR} band. 
The lower $N_\mathrm{H}$ value in NO overabundance models reduces the Fe~K edge
resulting in the same broadband absorption as the solar abundance model with higher $N_\mathrm{H}$.
Novae are known to show overabundance of CNO elements 
\citep[][and \S\,\ref{sec:ejectaabund}]{1994ApJ...425..797L,1998PASP..110....3G,2001MNRAS.320..103S}.
The shape of the {\em NuSTAR} spectrum is virtually insensitive to the abundance of C (unlike N and O). 
Optical spectra reveal the presence of Fe in the ejecta of \nova{}  \citep{2020ATel13867....1A,2020ATel14048....1I}, but it may be under-abundant with respect to solar values. 

\begin{figure}
\begin{center}
\includegraphics[width=0.48\textwidth,clip=true,trim=0cm 0cm 0cm 0cm,angle=0]{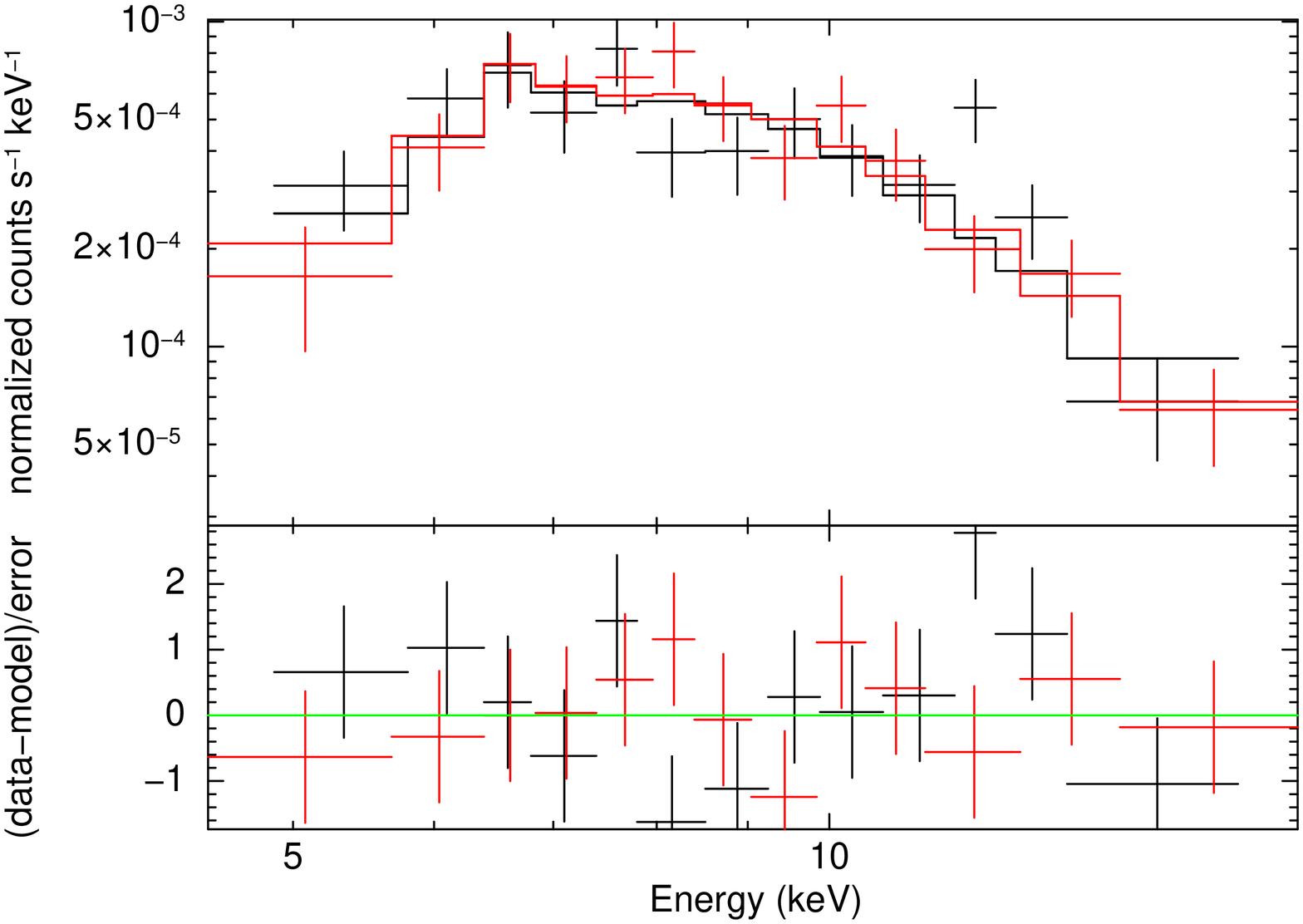}
\end{center}
\caption{Observed {\em NuSTAR} spectra compared with the preferred model in Table~\ref{tab:nustarspecmodels}.
Black and red curves represent spectra obtained with two {\em NuSTAR} telescopes (FPMA and
FPMB, respectively). The top panel shows the spectrum and the model, while
the bottom panel shows the difference between the spectrum and the model in
the units of uncertainty associated with each data bin.
The four models in Table~\ref{tab:nustarspecmodels} that provide an acceptable fit
look very similar when plotted against the data.}
\label{fig:nuspec}
\end{figure}

We simultaneously fit the spectra from the two focal plane modules
using the \textsc{XSPEC} model \texttt{constant*phabs*vphabs*vapec}, where
\texttt{constant} is needed to compensate for the variable 
cross-calibration factor 
between FPMA and FPMB (the average {\em NuSTAR} calibration accuracy is
at a few per~cent level; \citealt{2015ApJS..220....8M}),
\texttt{phabs} represents the foreground Galactic absorber (with solar
abundances and the absorbing column fixed to the value estimated from optical reddening in
\S~\ref{sec:extinction}), 
\texttt{vphabs} represents the intrinsic absorption and is allowed to vary, while \texttt{vapec} is
the plasma emission model. 
We consider two types of models:
\begin{enumerate}
\item the abundances of Fe, Co, Ni are tied together and left free to vary,
while abundances of all other elements are fixed to the solar values of
\citet{2009ARA&A..47..481A}; 
\item the abundances of N and O are tied together and left free to vary, while abundances of all other elements are fixed to the solar values. 
\end{enumerate}
The abundances of the absorber (\texttt{vphabs}) and the emitter (\texttt{vapec}) are tied together in our models. 
The choice of the abundances dramatically affects the intrinsic absorbing column 
(that is expressed in terms of the equivalent, pure hydrogen column).
The same situation was described for nova V906\,Car by \cite{2020MNRAS.497.2569S}. 
Figure~\ref{fig:nuspec} presents the {\em NuSTAR} spectra compared to 
our preferred model described in Table~\ref{tab:nustarspecmodels}.

Following \cite{2019ApJ...872...86N} and \cite{2020MNRAS.497.2569S}, 
we also fit a combination of the thermal plasma and power law emission 
to constrain the non-thermal contribution 
on top of the thermal emission (Table~\ref{tab:nustarspecmodels}). 
We fix the photon index to the theoretically predicted value of $\Gamma = 1.2$ 
(\S~\ref{sec:emissionmech}), manually vary the power law normalization 
and fit for other model parameters. This way we find the
brightest power law emission that, together with the thermal emission
component still provide an acceptable fit (Null hypothesis probability $>0.05$).
The monochromatic flux at 20\,keV for the brightest acceptable power law component
computed with Eq.~(\ref{eq:powerlawsed}) is 
$\nu F_\nu = 1.4 \times 10^{-13}$\,erg\,cm$^{-2}$\,s$^{-1}$. 
If instead of manually setting the power law normalization, we let it free 
to vary, the fit always converges to zero contribution of the power law as 
the observations can be fully explained by thermal emission. 

Table~\ref{tab:nustarspecmodels} summarizes the spectral fitting results.
For each model we list the assumed and/or derived Fe, N and O abundances 
(by number, relative to the solar values of \citealt{2009ARA&A..47..481A}). 
One can see that while the particular choice of abundances fixed to the solar values results in a bad fit, 
a very wide range of Fe, N and O abundances provides acceptable fits, 
to the point that the abundances of these elements are essentially unconstrained. 
The temperature of the thermal model as well as the observed flux 
do not depend strongly on the abundances, the unabsorbed (intrinsic) flux is somewhat dependent 
while the intrinsic absorbing column, $N_\mathrm{H}$, is extremely sensitive to 
the choice of the abundances as detailed in Table~\ref{tab:nustarspecmodels}.

\begin{table*}
        \centering
        \caption{{\em NuSTAR} spectral modelling}
        \label{tab:nustarspecmodels}
        \begin{tabular}{ccc cc cc c@{~~}c@{~~}c} 
                \hline
\texttt{vphabs} $N_\mathrm{H}$  & k$T$  & $\Gamma$ & ${\rm Fe}/{\rm Fe}_{\odot}$ & ${\rm N}/{\rm N}_{\odot}$ & 3.5--78.0\,keV Flux                    & unabs. 3.5--78.0\,keV Flux             & $p$ & $\chi^2$ & d.o.f.\\
($10^{22}$\,cm$^{-2}$)          & (keV) &          &                             & ${\rm O}/{\rm O}_{\odot}$                                          & $\log_{10}$(erg\,cm$^{-2}$\,s$^{-1}$)  & $\log_{10}$(erg\,cm$^{-2}$\,s$^{-1}$)  &     &          &       \\ 
                \hline

\multicolumn{10}{c}{\texttt{constant*phabs*vphabs*powerlaw}} \\
$4.7 \pm 33.2 $ & $  $ & $3.3 \pm 0.7 $ & 1\** & $250 \pm 4300$ & $-12.22 \pm 0.07 $ & $-11.73 \pm 0.11 $ & 0.20 & 26.20 & 21 \\

\multicolumn{10}{c}{\texttt{constant*phabs*vphabs(vapec+powerlaw)}} \\
$6.1 \pm 5.7 $ & $4.5 \pm 0.9 $  & $ $         & 1\** & $72 \pm 66 $ & $-12.10 \pm 0.06 $ & $-11.96 \pm 0.09$ \texttt{vapec}  & 0.05 & 42.19 & 29 \\
               &                 & 1.2\**      &      &              &                    & $<-12.47$\**   \texttt{powerlaw}  &      &       &    \\

\multicolumn{10}{c}{ {\it bad model} \texttt{constant*phabs*vphabs*vapec}} \\
$71.7 \pm 14.0 $ & $11.4 \pm 2.1 $ & $ $ & 1\** & 1\** & $-12.18 \pm 0.05 $ & $-11.91 \pm 0.05 $ & 0.00 & 43.93 & 22 \\

\multicolumn{10}{c}{\texttt{constant*phabs*vphabs*vapec}} \\
$131.3 \pm 25.8 $ & $5.6 \pm 1.2 $ & $ $ & $0.2 \pm 0.1$ & 1\** & $-12.32 \pm 0.04 $ & $-11.84 \pm 0.09 $ & 0.31 & 23.58 & 21 \\

\multicolumn{10}{c}{ {\it preferred model} \texttt{constant*phabs*vphabs*vapec}} \\
$7.3 \pm 7.3 $ & $6.5 \pm 1.5 $ & $ $ & 1\** & $52 \pm 53$ & $-12.30 \pm 0.05 $ & $-11.96 \pm 0.06 $ & 0.29 & 24.12 & 21 \\

                \hline
        \end{tabular}
\begin{flushleft}
The parameters that were kept fixed for the model fit are marked with the \** symbol. 
{\bf Column designation:}
Col.~1~-- intrinsic absorbing column (in excess of the total Galactic value);
Col.~2~-- temperature of the thermal component;
Col.~3~-- photon index of the power law component;
Col.~4~-- Fe abundance by number relative to the solar value;
Col.~5~-- N and O abundances (tied together) by number relative to the solar values;
Col.~6~-- the logarithm of the integrated 3.5--78.0\, keV flux under the model;
Col.~7~-- logarithm of the unabsorbed 3.5--78.0\, keV flux;
Col.~8~-- chance occurrence (null hypothesis) probability;
Col.~9~-- $\chi^2$ value;
Col.~10~-- number of degrees of freedom.

\end{flushleft}
\end{table*}

\subsubsection{{\em NuSTAR} lightcurve}
\label{sec:nustarvar}

\begin{figure}
        \includegraphics[width=1.0\linewidth]{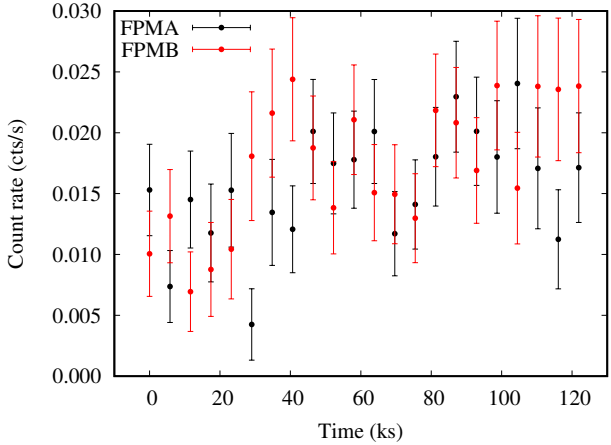}
    \caption{The background-subtracted 3.5--78\,keV {\em NuSTAR} lightcurve of \nova{}.}
    \label{fig:nustarlc}
\end{figure}

Fig.~\ref{fig:nustarlc} presents the 3.5--78\,keV lightcurves of \nova{}
obtained during the {\em NuSTAR} observation described in \S~\ref{sec:nustarspec}. 
The lightcurves were background-subtracted and binned to 5805\,s (corresponding to the {\em NuSTAR}
orbital period at the time of the observations) 
resulting in one count rate measurement per orbit. Comparing the scatter of
the count rate measurements to their error bars using the $\chi^2$ test 
\citep[testing the observations against the null hypothesis that the mean count rate is constant; e.g.][]{2010AJ....139.1269D}
we get about 0.03 chance occurrence probability for each of the lightcurves.
Combining the FPMA and FPMB lightcurves the chance occurrence probability
drops below 0.005, indicating significant variability. 
The visual inspection of Fig.~\ref{fig:nustarlc}
reveals that both FPMA and FPMB lightcurves show an increase in brightness
over the duration of the observations. The $\chi^2$ test 
does not take into account the time and order of the photon flux
measurements, only the measured values and their error bars, 
so the variability significance derived from the $\chi^2$ test may be
considered a lower limit: the probability of chance occurrence of a smoothly
varying lightcurve is smaller than that reported by this simple test 
\citep[see the discussion in][]{2006MNRAS.367.1521T,2013A&A...556A..20F,2017MNRAS.464..274S}.
There is no obvious energy dependence of the variability amplitude, implying 
that the variations are intrinsic rather than related to changing absorption
(that would have mostly affected the lower energies). 

The need to collect enough photons for an accurate count rate measurement
requires long time bins, which in turn limit the time resolution of the lightcurve.
To test for the presence of a periodic signal on time-scales shorter than the {\em NuSTAR}
orbital period we analysed photon arrival times (an unbinned lightcurve). 
The idea is that if the lightcurve is periodic, one can smooth (bin) it in phase rather than in time. 
We used the photon arrival times extracted from an event file 
to compute the power (defined as the squared modulus of the discrete Fourier 
transform) as a function of the variability time-scale \citep[`power spectrum';][]{1975Ap&SS..36..137D,2014MNRAS.445..437M}. 
We also computed the $H_m$-periodogram that for each trial period sums power
over multiple harmonics enhancing sensitivity to variations that do not
look like a sine-wave \citep{1989A&A...221..180D,2010A&A...517L...9D,2011ApJ...732...38K}.
The periodicity search was performed with the  \textsc{patpc}
code\footnote{\url{https://github.com/kirxkirx/patpc}}. 
We found no significant periodicity in the range 0.5--1000\,s that was present in both FPMA and FPMB 
lightcurves and could not be attributed to harmonics of the {\em NuSTAR} orbital period.

\subsection{{\em XMM-Newton} observations}
\label{sec:xmm}

{\em XMM-Newton} is equipped with five X-ray instruments:
the two EPIC-MOS\footnote{European Photon Imaging Camera - Metal Oxide Semiconductor \citep{2001A&A...365L..27T}}
and the EPIC-pn\footnote{European Photon Imaging Camera with the pn-type detector \citep{2001A&A...365L..18S}}
cameras for imaging and low-resolution spectroscopy in the 0.2--10\,keV band
and two Reflection Grating Spectrometers (RGS; \citealt{2001A&A...365L...7D}) 
covering the range 0.33--2.1\,keV (6--38\,\AA{}) with high spectral resolution. 
The X-ray telescopes are supplemented by the Optical Monitor \citep{2001A&A...365L..36M}.
All the instruments are capable of operating simultaneously, with the X-ray photons
not dispersed by the RGS gratings being recorded by the EPIC-MOS cameras.
The 2-day orbital period of {\em XMM-Newton} allows for long uninterrupted observations.

Co-adding data collected prior to eruption, the Upper Limit Server\footnote{\url{http://xmmuls.esac.esa.int/hiligt/}} 
\citep{2011ASPC..442..567S} reports the typical $2\sigma$ EPIC-pn upper limits of $<1$\,cts/s
corresponding to the energy flux limit around 
$2\times10^{-12}$\,erg\,cm$^{-2}$\,s$^{-1}$ on the 0.2--12\,keV flux (for the six {\em XMM-Newton}
slews over the nova position in 2002--2019, ObsIDs 9042100004, 9099800003, 9175600004, 9219500004, 9272700003, 9350700002). 
A ROSAT/PSPC survey observation from 1990 yields an upper limit of
$<0.0131$\,cts/s corresponding to $<10^{-13}$\,erg\,cm$^{-2}$\,s$^{-1}$ of the 0.2--2\,keV flux \citep{2016A&A...588A.103B}.
Two {\em XMM-Newton} slews were performed over the position of \nova{} after 
the eruption resulting in detection of soft (photon energy $<2$\,keV)
emission on 2020-12-08 19:48:42 ($t_0 + 153.6$\,d; $2.0 \pm 0.8$\,cts/s; 
$(4 \pm 2)\times10^{-12}$\,erg\,cm$^{-2}$\,s$^{-1}$; ObsID 9384600002) and 
2021-03-13 02:42:17 ($t_0 + 247.9$\,d; $1.0 \pm 0.4$\,cts/s; 
$(2.3 \pm 0.9)\times10^{-12}$\,erg\,cm$^{-2}$\,s$^{-1}$; the fluxes and
count rates are 0.2--12\,keV; ObsID 9389300003).
The energy fluxes and limits are computed following \cite{1991ApJ...374..344K}, assuming
power law emission with $\Gamma=2$ and the `standard' {\em XMM} Slew Survey
\citep{2008A&A...480..611S} absorbing column of $3 \times 10^{20}\,{\rm cm}^{-2}$
for the count rate to flux conversion.

The dedicated pointed {\em XMM-Newton} observation of \nova{} was performed between
2020-09-23 13:36 and 2020-09-23 21:22~UT ($t_0+77.6$\,days;
ObsID~0871010101; PI:~Sokolovsky) for the total exposure time of 28\,ks.
We did not use the Optical Monitor as the target was too bright, with a visual magnitude $\sim$8.8.
The EPIC was operating with the following configuration: 
pn -- Small Window with Thick Filter,
MOS1 -- Small Window with Thick Filter,
MOS2 -- Timing with Medium Filter.

\subsubsection{{\em XMM-Newton} spectroscopy}
\label{sec:xmmspec}

When choosing the {\em XMM-Newton} instrument setup, we were concerned about the possible 
optical loading (\S~\ref{sec:optical}) and possible low-energy calibration 
issues of the Timing mode (so we choose two different configurations
for the MOS cameras). However, the real problem turned out to be pile-up by
the soft X-ray photons from the SSS component. Essentially, the SSS component
turned out to be much brighter than we anticipated from {\em Swift}/XRT observations \citep{2020ATel14043....1S}.
Pile-up happens when multiple photons arriving almost at the same time are mistaken by the detector for a single event with the sum of their energies. 
This distorts the energy spectrum and results in an underestimate of the count rate (two or more events are counted as one).
Pile-up is so severe in our observations of \nova{} that it makes quantitative analysis of the EPIC spectra
impossible, even when the (most affected) central region of the source image is excluded.

We thus focus on the dispersive RGS where the photons are spread over a much larger area on the chip, making pile-up generally less likely to happen. However, for extremely bright and soft sources such as ours, pile-up can still occur, but can be dealt with following the approach described 
by \cite{2007ApJ...665.1334N}. 
The RGS was operated in standard spectroscopy mode. We extracted the RGS~1 and 2 spectra and co-added them with the {\tt rgsproc} pipeline of the \textsc{SAS}. 
The RGS spectrum was found to be distorted by pile-up and a special procedure has to be applied to correct for it.

The intrinsic energy resolution of the CCD detector that records the dispersed photons is sufficiently high to identify higher dispersion orders from the photon energies. 
The pipeline that extracts second-order spectra does not, however, distinguish between pile-up and second-order dispersion. 
The result is the apparent leakage of counts from the first- to the second-order spectrum. 
Normally, in the case of second order dispersion, a photon of a certain energy $E_\lambda$ is recorded at a position 
that corresponds to half the wavelength.
i.e. $0.5 {\rm h} {\rm c}/{\rm E}_\lambda$ (where ${\rm h}$ is the Planck constant and ${\rm c}$ is the speed of light in vacuum). 
The software recognizes the higher energy of the photon (thanks to the inherent energy resolution of the CCD detector) 
and corrects the corresponding wavelength accordingly. 
Meanwhile, in the case of pile-up, two photons of energy $E_\lambda$ are registered at the chip position that corresponds
to the wavelength $0.5 {\rm h} {\rm c}/{\rm E}_\lambda$, but with the sum of their energies, thus $2E_\lambda$. 
The software then assigns to half the true wavelength resulting in 
the discrepancies between first and second order spectra which is thus owed to pile-up. 

Since there is no first-order emission in the 15--20\,\AA{} range where the piled-up photons are recorded, 
it is easy to correct for pile-up following the approach described by \cite{2007ApJ...665.1334N} by manipulation of the events file. 
We use the columns of wavelength (derived from the photon positions in dispersion direction) and the Pulse Invariant channel number (\texttt{PI}; encoding the photon energy recorded by the CCD). 
For each photon recorded within the wavelength range 12--38\,\AA{} but twice the corresponding photon energy, 
two photons are added with double the wavelength value. 
That way, we re-generated the spectrum with {\tt rgsproc} starting with the manipulated events file. 

\begin{figure*}
\begin{center}
\includegraphics[width=1.0\textwidth,clip=true,trim=1.5cm 1.5cm 5cm 2cm,angle=0]{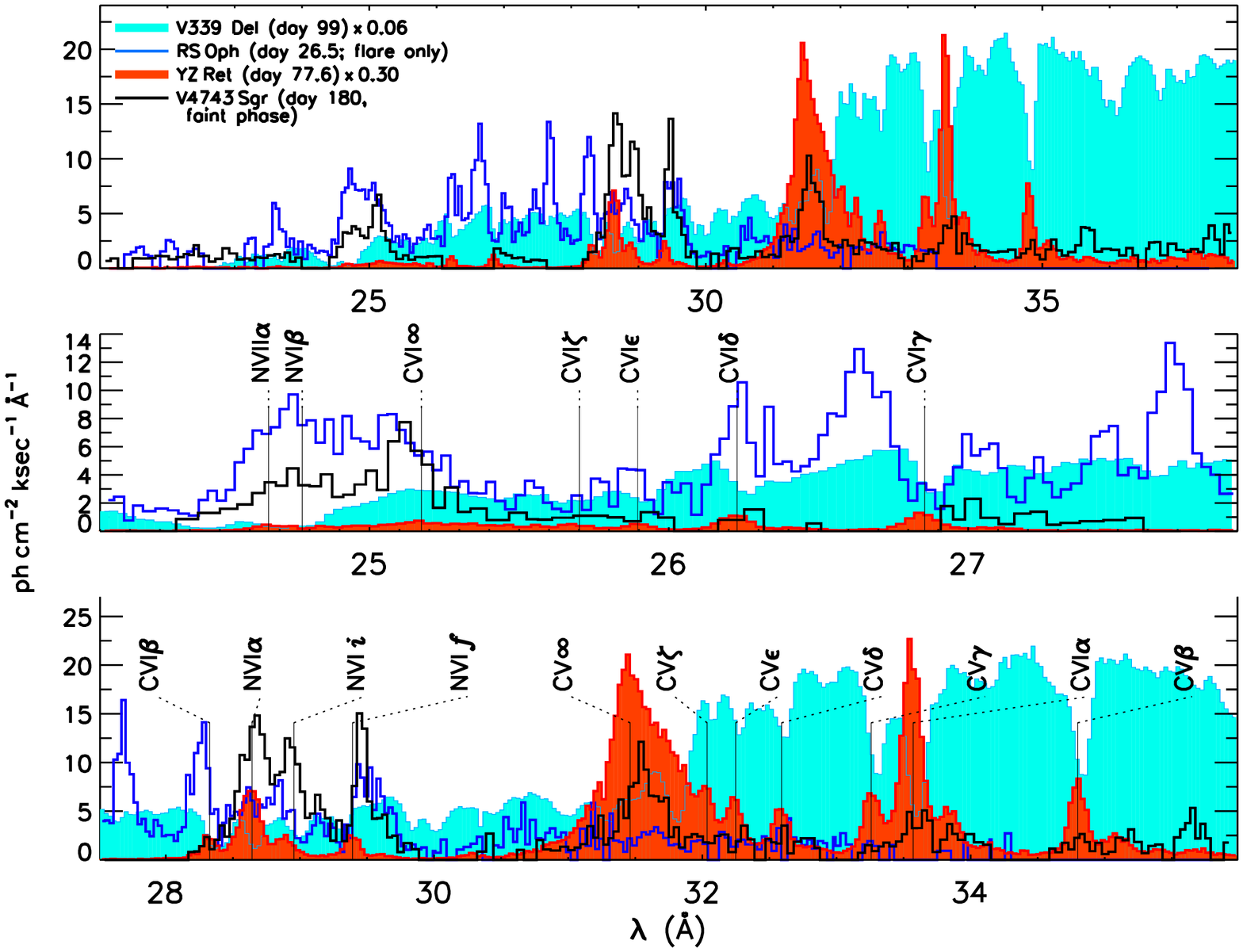}
\end{center}
\caption{\emph{XMM-Newton} RGS1+2 spectra of \nova{} observed on $t_0+77.6$\,days (red) 
compared with three other novae: an RGS spectrum of V339\,Del (cyan shading;
2013-11-21, ObsID~0728200201, PI:~Schwarz), 
RGS spectrum of RS\,Oph (blue line; 2006-03-10, ObsID~0410180201, PI:~Schartel, \citealt{2009AJ....137.3414N}), and 
a {\em Chandra}/LETGS (Low Energy Transmission Grating Spectrometer; \citealt{2000ApJ...530L.111B}) 
spectrum of V4743\,Sgr (black line; 2003-03-19, ObsID~3775, PI:~Starrfield, \citealt{2003ApJ...594L.127N}).
The top panel shows the full spectral range, while the panels below zoom 
in the ranges 24-28\,\AA\, and 28-36\,\AA. The line labels are blue-shifted by 1500km\,s$^{-1}$.
The labels indicate somewhat unusual states of RS\,Oph and V4743\,Sgr when the displayed SSS spectra 
with emission lines were observed: 
on day 26.5, RS\,Oph experienced a small soft flare \citep[see fig.~6 in][]{2012BASI...40..353N};
on day 180.4, V4743\,Sgr experienced a steep decline from very bright to extremely faint emission 
(\citealt{2003ApJ...594L.127N}, see also fig.~5 in \citealt{2012BASI...40..353N}).}
\label{fig:rgsspeccomp}
\end{figure*}

Even after taking into account the leakage of counts from the first to the
second order caused by pile-up, the RGS spectrum (Fig.~\ref{fig:rgsspeccomp}) looks somewhat unusual.
Instead of a soft blackbody-like emission usually found in SSS (and that can be expected from the EPIC spectrum), the spectrum is dominated by emission lines. 
Comparison with previously investigated novae helps to interpret this spectrum.
Fig.~\ref{fig:rgsspeccomp} compares the RGS spectrum of \nova{} to
previously observed novae in the SSS phase: V339\,Del, RS\,Oph and V4743\,Sgr.
The archival RGS spectrum of V339\,Del was extracted by us with the standard \textsc{SAS} tasks,
while the grating spectra of RS\,Oph and V4743\,Sgr were discussed earlier
by \cite{2009AJ....137.3414N} and \cite{2003ApJ...594L.127N}, respectively.
V339\,Del shows a typical SSS spectrum dominated by continuum emission modified by absorption lines while the other novae display prominent emission lines.  
Comparing \nova{} with V339\,Del, one can see some of the \nova{} emission lines have
corresponding absorption lines in V339\,Del, while the huge, broad 
emission line at $\sim$31.5\,\AA{} is also seen in V4743\,Sgr. 

We conclude that the RGS spectrum (Fig.~\ref{fig:rgsspeccomp}) 
is dominated by emission lines of 
H-like Carbon (\ion{C}{VI}) and He-like Carbon (\ion{C}{V}).
The \ion{C}{V} 1s-2p (K~$\alpha$ or Lyman~$\alpha$) line is outside the range of the RGS, 
but all other lines of these ions are seen. With increasing principal quantum number, 
the separation between the lines shrinks, and when the principal quantum number approaches infinity 
(corresponding to the ionization energy \ion{C}{V} to \ion{C}{VI}), 
the lines blend with each other, which explains the shape of the
31.5\,\AA{} feature (labelled \ion{C}{V}$\infty$) where we can still resolve the \ion{C}{V}$\zeta$ (1s-7p) transition in the red wing.
In other words, \ion{C}{V}$\infty$ is equivalent to the Lyman jump in emission,
it is known as `radiative recombination continuum' feature and is observed
in grating X-ray spectra of some active galactic nuclei 
\citep{2007MNRAS.374.1290G,2015A&A...581A..79W}.
For \ion{C}{VI}, the lines are weaker in the spectrum of \nova{}, but we can clearly see all the lines 
and at 25.3\,\AA{}, a small peak can be seen that corresponds to the ionization energy of
\ion{C}{VI}. 

We also see a weak emission line feature corresponding to the \ion{N}{VI}$\alpha$ 1s-2p transition at
28.8\,\AA{} with the resonance, intercombination, and forbidden (1s-2s)
lines as well as the \ion{N}{VI} 1s-3p (24.9\,\AA{}) and \ion{N}{VII} 1s-2p line at 24.8\,\AA. 
Also present are the \ion{O}{VIII} 1s-2p and 1s-3p lines and probably also \ion{O}{VII} 1s-2p at 21.6\,\AA.

Identification of all the emission lines discussed above requires a blue-shift of 1500\,km\,s$^{-1}$. 
Blue-shifted emission lines were previously observed in X-ray grating spectra of 
RS\,Oph \citep{2008ApJ...673.1067N,2009A&A...493.1049O}, 
V959\,Mon \citep{2016ApJ...829....2P,2021MNRAS.500.2798N},  
V906\,Car \citep{2020MNRAS.497.2569S} and V3890\,Sgr \citep{2020ApJ...895...80O,2021MNRAS.501...36S,2022A&A...658A.169N}.
Blueshifts are also observed for absorption lines on top of the continuum SSS emission of novae 
\citep{2007ApJ...665.1334N,2011ApJ...733...70N,2012BASI...40..353N,2013MNRAS.429.1342O,2018ApJ...862..164O,2020AdSpR..66.1193O,2021MNRAS.tmp.1420O,2022A&A...658A.169N}.

The line-dominated emission observed by {\em XMM-Newton}/RGS on $t_0+77.6$ is
characteristic of photoionized or recombining plasma rather than 
collisionally-ionized plasma in thermal equilibrium (the \texttt{vapec} 
model we used to interpret the {\em NuSTAR} spectrum obtained on
\tnustarep{}; \S~\ref{sec:nustarspec}). This is in stark contrast to V906\,Car 
that showed no SSS and allowed modelling its {\em XMM-Newton} spectrum with 
\texttt{vapec} to derive the abundances \citep{2020MNRAS.497.2569S}. 
Quantitative modelling of the line-dominated SSS emission of \nova{} is beyond the scope of
this paper but we emphasise that it does not need a model to see that carbon is unusually 
abundant in \nova{} and in V339 Del. The ejected material originates from the CNO burning layers, 
and C/N is thus expected to be small as carbon is depleted and N is enhanced. 
In most novae, C lines are much weaker while N lines dominate. 
The strong C lines in the {\em XMM-Newton}/RGS spectrum of \nova{} thus indicate 
that the underlying white dwarf is overabundant in carbon, which is typical of CO white dwarfs.

\subsubsection{{\em XMM-Newton} periodicity search}
\label{sec:xmmtiming}

We use the pointed {\em XMM-Newton} observation carried out on 2020-09-23 described
in \S~\ref{sec:xmm} to search for any periodic variation in the X-ray flux of \nova{}. 
We apply the \textsc{patpc} code (\S~\ref{sec:nustarvar}) to search for a
periodicity in the arrival times of photons registered by the MOS2
instrument operating in the timing mode.
We use the full 0.2--10\,keV band, however we note that the counts are
dominated by the super-soft line emission (\S~\ref{sec:linesss}).
No significant periodicity could be identified in the period range 0.5--300\,s
-- (quasi-)periodic variations on these time-scales were reported in other novae, 
during the SSS phase \citep{2015A&A...578A..39N,2018ApJ...855..127W,2020MNRAS.499.2007V,2020MNRAS.499.4814P}.
This is in accordance with the {\em NICER} results reported by \cite{2020ATel14067....1P}.
There is significant power distributed across multiple peaks at longer
periods which can be attributed to variability on a time-scale of a few ks,
either intrinsic to the source or caused by the background variations.

\subsection{Optical photometry of \nova{}}
\label{sec:optical}

In order to track the overall optical brightness evolution of \nova{} (Fig.~\ref{fig:latoptlc}), 
we combined the post-discovery visual (by eye) and $V$-band CCD measurements contributed by the AAVSO 
observers \citep{AAVSODATA} 
with $g$-band CCD photometry from the ASAS-SN survey \citep{2014ApJ...788...48S,2017PASP..129j4502K}
and early observations reported via the Central Bureau Electronic Telegrams 
\citep[CBET;][]{2020CBET.4811....1M,2020CBET.4812....1K,2020CBET.4826....1K}.
The CBET-reported observations were performed using colour (chip with a Bayer filter) CMOS cameras. 
The magnitude zero-point offsets between observations obtained with  
these methods are expected to be small compared to the nova amplitude.
The CMOS and CCD images were measured with aperture photometry techniques
utilizing various sets of comparison stars, while visual magnitude
estimates were made following the AAVSO Visual Observing Manual\footnote{\url{https://www.aavso.org/visual-star-observing-manual}} \citep[see also] []{1984vest.book.....H}.


The latest 
detection in pre-discovery quiescence ($g = 15.51$ on $t_0 - 6.0$\,d) is followed by 
the ASAS-SN detection of the eruption at $t_0$ (2020-07-08.171~UT; $g = 6.77$). 
Subsequently, the lightcurve continued to rise, peaking at 3.7\,mag probably just before \topicalpeak{} (Fig.~\ref{fig:latoptlc}). 
The peak is followed by a nearly linear decline in magnitude (exponential decline in flux). 
At $t_0 + 30$\,d, when the optical decline rate dramatically slows down 
coinciding with the appearance of super-soft X-ray emission (\citealt{2020ATel14043....1S}, see e.g. 
fig.~1 of \citealt{2021MNRAS.503..704M}).

By fitting a straight line to the visual, $V$ band, and colour-CMOS magnitude
estimates obtained between $t_0+7.7$\,d (when the dense observational coverage started) and $t_0+29$\,d 
(just before the lightcurve kink) we estimate the time to decline by 2\,mag 
(3\,mag) to be $t_2=16.0$\,d ($t_3=24.1$\,d). The uncertainties of the $t_2$ 
and $t_3$ values are about a day, dominated by the exact choice of the outlier 
measurements to reject, fitting time interval, the relative weighting of visual 
and CCD measurements, and the choice of the fitting algorithm. The values
above were obtained with the robust linear regression 
\citep[implemented in the \textsc{GNU Scientific Library}][]{gough2009gnu}
effectively assigning equal weights to visual and CCD/CMOS measurements. 
While the CCD measurements are inherently more precise than visual estimates, 
the CCD observations are sparse and have zero-point difference 
with visual and between the different CCD observers. 
(Note the excursion toward the lower fluxes in $V$ band around $t_0+18$\,d
that doesn't seem to have a counterpart in visual data. We attribute this 
discrepancy to a colour change.) \cite{2021RNAAS...5...48R} report a $t_2$ value 
shorter by four days, also citing the AAVSO data.

\section{Discussion}
\label{sec:discussion}

\subsection{Relation between optical and $\gamma$-ray emission}
\label{sec:gammaopt}

The $\gamma$-ray lightcurve peaks in the daily bin centred at $t_0+4.3$\,d,
which is $0.7$\,d past the optical peak (Fig.~\ref{fig:latoptlc}; \S~\ref{sec:optical}). 
The optical peak time is not well constrained (no observations in three days between 
the latest pre-maximum and maximum lightcurve points), so the optical 
to $\gamma$-ray peak delay value should be treated with caution.
Delayed onset of $\gamma$-ray emission with respect to the optical peak has
been observed in other novae \citep[e.g.][]{2016ApJ...826..142C}.
Two possibilities may explain this delay. The $\gamma$-rays may be created 
simultaneously with the optical emission, but initially get absorbed
\citep[e.g.][]{2020ApJ...904....4F}. This scenario is similar to the one 
explaining the delayed onset of shock-powered X-ray emission -- we know
that the X-rays are present early in eruption thanks to {\em NuSTAR}
penetrating through dense absorbing ejecta \citep{2019ApJ...872...86N,2020MNRAS.497.2569S}.
The other possibility is that the shock accelerating the $\gamma$-ray emitting
particles needs time to form. \cite{2017MNRAS.469.4341M} suggest there
may be two peaks in optical lightcurves of $\gamma$-ray novae: the first one
from the freely expanding nova fireball (common to all novae) and the second
peak powered by shocks (specific to the $\gamma$-ray novae). 
According to \cite{2020ApJ...905...62A}, the $\gamma$-ray emitting shock
forms when a fast radiation-driven wind from the white dwarf catches up with
the slowly expanding shell ejected early in the eruption (perhaps through 
common envelope interaction).
Correlated $\gamma$-ray and optical variations 
\citep{2017NatAs...1..697L,2020NatAs...4..776A} 
suggest that shocks within the nova ejecta can vary in power on 
a time-scale of days, which tentatively suggests the delayed shock formation 
scenario is plausible.

In contrast with the two $\gamma$-ray novae discussed by 
\cite{2017MNRAS.469.4341M}, \nova{} shows a single-peaked optical lightcurve. 
In the `two peaks/delayed shock formation' scenario, this means that the shocks
in \nova{} formed quickly, and the fireball and shock-powered optical
lightcurve peaks merge together (or at least are indistinguishable given 
the limited photometric coverage between $t_0$ and $t_0 + 5$\,d, Fig.~\ref{fig:latoptlc}).

Following \cite{2015MNRAS.450.2739M}, \cite{2017NatAs...1..697L}, \cite{2020NatAs...4..776A}, and \cite{2020ApJ...905..114L}, 
we compute the ratio of the $\gamma$-ray flux in the {\em Fermi}/LAT band
(0.1--300\,GeV; \S~\ref{sec:latobs}) to the bolometric optical flux. 
The typical intrinsic colour of a nova near peak brightness is $(B-V)_0 = +0.23$ \citep{1987A&AS...70..125V}. 
For a blackbody with temperature $T < 10000$\,K (corresponding to spectral types later than A0) 
the temperature can be estimated from the $(B-V)_0$ colour as 
$T = \frac{7090}{(B-V)_0 + 0.71} \approx 7500$\,K
(relation derived from the simple comparison of 4400\,\AA\, and 5500\,\AA\, flux
densities predicted by the Rayleigh-Jeans law).
The blackbody bolometric correction \citep[defined in e.g.][]{2009aste.book.....K} for 
$T = 7500$\,K is $-$0.03 according to table~3.1 of \cite{2007iap..book.....B}.
Adopting the observed bolometric magnitude $m_{\rm bol} = 3.67 - A_V$ from the colour-CMOS magnitude of 3.7 
(the best available approximation to the peak $V$ magnitude) 
and following \cite{2015arXiv151006262M}, we obtain a peak bolometric flux of 
$f = 2.518 \times 10^{-5} \times 10^{-0.4 m_{\rm bol}} \,{\rm erg}\,{\rm cm}^{-2}\,{\rm s}^{-1} \approx 8.6 \times 10^{-7}\,{\rm erg}\,{\rm cm}^{-2}\,{\rm s}^{-1}$
corresponding to an optical luminosity of \opticalpeaklum{},
a factor of 6 above the Eddington luminosity of a 1.0\,${\rm M}_\odot$ white dwarf
(see e.g. \S1.2 of \citealt{2002apa..book.....F} and \citealt{1998ApJ...494L.193S}).
The ratio of the peak $\gamma$-ray luminosity (\S~\ref{sec:latobs}) to peak optical luminosity is 
$4.5 \times 10^{-4}$. This value is comparable to what was observed in V339\,Del, 
and an order of magnitude lower than what was found for the other $\gamma$-ray bright novae 
\citep[see supplementary~fig.~14 of][]{2020NatAs...4..776A}.

Using the same technique we estimate the optical bolometric luminosity 
of \nova{} during the {\em NuSTAR} observation to be \opticalnustareplum{} 
based on 27 visual magnitude estimates made during the {\em NuSTAR} observation (mean 5.12\,mag), 
assuming post-peak $(B-V)_0=-0.02$ \citep[\S~\ref{sec:extinction};][]{1987A&AS...70..125V}
corresponding to $T = 10^4$\,K (bolometric correction $-0.28$).
Given the uncertainty of magnitude estimates, nova colour and 
the corresponding bolometric correction, uncertainty of putting the 
visual and unfiltered CMOS photometry on the $V$ magnitude scale as well 
as the uncertainty of $V$ zero-point and distance to \nova{}, 
it is unlikely that the estimated luminosities are accurate to better than 10~per~cent.

The $\gamma$-ray to optical flux ratio places a constraint 
on the particle acceleration efficiency in nova shocks.
If we assume that (i)\,all optical luminosity is powered by shocks; (ii)\,most of the shock energy 
is eventually dissipated as optical radiation; and (iii)\,the accelerated particles 
emit all their energy within the {\em Fermi}/LAT band, the ratio of the {\em Fermi}/LAT to optical fluxes will yield 
the particle acceleration efficiency. Clearly, a large fraction of the optical luminosity comes from the expanded photosphere 
heated directly by the nuclear burning white dwarf, so the GeV to optical flux ratio
sets a lower limit on the acceleration efficiency.

To facilitate comparison with the following paragraphs where we use monochromatic X-ray and $\gamma$-ray fluxes,
we compute the peak monochromatic optical flux at 2.25\,eV (5500\,\AA): $\nu F_\nu = 7.2 \times 10^{-7}$\,erg\,cm$^{-2}$\,s$^{-1}$.
The monochromatic optical flux at the time of the {\em NuSTAR} observation 
is $\nu F_\nu = 1.9 \times 10^{-7}$\,erg\,cm$^{-2}$\,s$^{-1}$.
For the magnitude to flux density conversion we use the absolute fluxes (corresponding to zero magnitude)
from \cite{1998A&A...333..231B}. We note that this conversion is approximate
as it depends on the source spectrum.
The observed magnitudes were corrected for $A_V$ derived in \S~\ref{sec:extinction}.

\subsection{The luminosity of \nova{} at high energies}
\label{sec:luminosity}
 
Here we consider the X-ray and $\gamma$-ray luminosities of \nova{} and compare it to previously observed novae, 
considering order-of-magnitude estimates only. The following factors limit the accuracy of luminosity measurements. 
\begin{itemize}
\item The distances to previously observed 
novae are often not well constrained.
\item The nova flux is changing over the course of its eruption.
While the GeV and optical bands are often well covered by observations and
one can estimate the peak or average flux, the observed X-ray flux is a
strong function of the observation date -- we know this from {\em Swift}/XRT monitoring, 
while in the harder {\em NuSTAR} band the best-covered lightcurve of V906\,Car has only two epochs. 
\item The derived GeV and X-ray fluxes depend on the choice of 
the spectral model and different models have been used in the literature. 
\end{itemize}
Note that while in \S~\ref{sec:lxlg} we will discuss monochromatic flux ratios,
here we discuss luminosities integrated over the specific energy bands.

Integrating the exponentially cut-off power law that fits the {\em Fermi}/LAT 
spectrum (\S~\ref{sec:latobs}) and relying on the {\em Gaia} distance (\S~\ref{sec:extinction}),
we estimate the average 0.1--300\,GeV luminosity of \nova{} over its
$\gamma$-ray bright period to be $1.2 \times 10^{35}$\,erg\,s$^{-1}$.
Scaling this to the $\gamma$-ray photon flux at peak and at the {\em NuSTAR} epoch 
(assuming the spectrum does not change) we obtain the peak luminosity of 
\latpeaklum{} 
and 
the luminosity during the {\em NuSTAR} observation of \latnustareplum{}. 
As the \emph{Fermi} upper limit to the flux at 0.05--0.1\,GeV is well below the value from the extrapolation of the power law fit (Fig.~\ref{fig:latsed}), 
the $\gamma$-ray spectrum is consistent with a substantial drop toward lower energies, so that the 0.1--300\,GeV luminosity may well be representative of the total $\gamma$-ray luminosity of the nova.
The luminosity estimates at different epochs and bands are summarized in
Table~\ref{tab:luminositiestable}.

The GeV luminosity of \nova{} is about an order of magnitude lower than that
of the brightest known $\gamma$-ray nova, V906\,Car \citep{2020NatAs...4..776A}, and
a factor of 5 lower than that of V5855\,Sgr \citep{2019ApJ...872...86N}. 
Taking the \emph{Fermi}/LAT photon fluxes and distances for $\gamma$-ray-detected novae from \cite{2021ApJ...910..134G} and
applying the same photon to energy conversion factor as we adopted for
\nova{} (assuming the other novae have the same spectrum as \nova{}) we find
a median $>100$\,MeV luminosity of $2 \times 10^{35}$\,erg\,s$^{-1}$, close to
that of \nova{}. The lowest-luminosity detected GeV nova in the \cite{2021ApJ...910..134G}
sample \citep[V1369\,Cen, which is also the most nearby, \S~\ref{sec:emissionmech};][]{2016ApJ...826..142C} has the luminosity an order of magnitude lower than \nova{}.
V549\,Vel may be a few times fainter than V1369\,Cen, however there are
questions about the reliability of its distance \citep[and hence luminosity;][]{2020ApJ...905..114L}.

Integrating the thermal plasma model that fits the {\em NuSTAR} spectrum of \nova{} in 
the 3.5--78\,keV energy range we obtain an intrinsic X-ray luminosity of 
\nustarlum{}.
Extrapolating from the model down to a low-energy limit of 0.3 keV, the resulting luminosity increases by a factor of two. 
It is hard to say how
representative these values are of the total X-ray energy output of the 
nova, as soft X-rays are completely hidden by the intrinsic absorption at the 
time of the {\em NuSTAR} observation. A very bright emission component can, in
principle, be completely hidden from view if it is sufficiently soft to
provide no detectable contribution above 3.5\,keV in the {\em NuSTAR} band
\citep[\S~\ref{sec:lxlg};][]{2020MNRAS.497.2569S}. The SSS emission from the
white dwarf is an obvious example, but there might be other shock-related 
emission components hidden at low energies.

The shock-powered X-ray luminosity derived from the {\em NuSTAR}
observation of \nova{} is comparable to that of GeV-bright novae observed by 
{\em Swift} and analysed by \cite{2021ApJ...910..134G}.
Comparing to {\em NuSTAR}-observed novae, \nova{} is 
an order of magnitude fainter than V906\,Car \citep{2020MNRAS.497.2569S} and 
a factor of 8 fainter than V5855\,Sgr \citep{2019ApJ...872...86N}.

\begin{table}
\begin{center}
\caption{\nova{} luminosity}
\label{tab:luminositiestable}
\begin{tabular}{rc}
\hline\hline
Band       & Luminosity \\
\hline
\multicolumn{2}{l}{$\gamma$-ray/optical peak at \topicalpeak{}:} \\
                0.1--300\,GeV & \latpeaklum{} \\
           bolometric optical & \opticalpeaklum{} \\
\hline
\multicolumn{2}{l}{{\em NuSTAR} epoch at \tnustarep{}:} \\
                0.1--300\,GeV & \latnustareplum{} \\
                3.5--78\,keV  & \nustarlum{} \\
extrapolated    0.3--78\,keV  & \nustarlumextrapolated{} \\
           bolometric optical & \opticalnustareplum{} \\
\hline
\end{tabular}
\begin{flushleft}
\end{flushleft}
\end{center}
\end{table}

\subsection{Comparing X-ray properties of \nova{} to nova-quiescent systems}
\label{sec:comparevyscl}

The pre-eruption X-ray upper limits indicate that the nova has brightened 
at least an order of magnitude by the time of the post-eruption {\em XMM-Newton} observation (\S~\ref{sec:xmm}). 
We compare the X-ray properties of \nova{} 10 to 78 days after the nova eruption 
to the more nearby (and, therefore, brighter) non-nova cataclysmic variables. 
\cite{2014MNRAS.445..869Z} examined X-ray properties of four VY\,Scl variables
whose spectra are described with two thermal plasma components (one with ${\rm k}T \lesssim 1$\,keV and the other
with ${\rm k}T \gg 1$\,keV), sometimes requiring super-solar abundances.
The physical interpretation of the two components is unclear. 
It is likely that the true emission is from a multi-temperature plasma, while
the two-temperature model is just the next simplest thing after 
the single-temperature model and provides an acceptable description of the
data just because of the low photon statistics.

No SSS emission (which would indicate continuous nuclear burning) was found
in VY\,Scl systems observed by \cite{2014MNRAS.445..869Z}.
The nova eruption in \nova{} disfavours the suggestion by
\cite{1998A&A...339L..21G}, \cite{1999A&A...343..183G,2001A&A...376.1031G}, \cite{2000NewAR..44..149G} and \cite{2001PASP..113..473H} 
that continuous nuclear burning is a common feature of VY\,Scl systems 
(in accordance with the results of \citealt{2010AN....331..227G} and
\citealt{2014MNRAS.445..869Z}). The emergence of the post-nova SSS in \nova{} (\S~\ref{sec:linesss}; \citealt{2020ATel14043....1S}) 
supports this conclusion suggesting the pre-nova SSS that could have been indicating continuous nuclear burning 
was likely non-existent, rather than somehow hidden from our view.

Combining the unabsorbed flux estimates of VY\,Scl type systems reported by 
\cite{2014MNRAS.445..869Z}, \cite{2010AN....331..227G} and \cite{2014A&A...570A..37P} with 
the {\em Gaia} distances \citep{2018AJ....156...58B}, we estimate the typical luminosity of 
VY\,Scl systems to be $\sim 10^{32}$\,erg\,s$^{-1}$, about an order of
magnitude lower than the post-nova emission of \nova{} (\S~\ref{sec:luminosity}; Table~\ref{tab:luminositiestable}).
The X-ray luminosity of individual VY\,Scl type systems varies 
with time and, possibly, with their optical (high/low) state.
%
%
%
%
%

\nova{} can also be compared to the old nova high accretion rate system V603\,Aql (Nova Aquilae 1918)
observed in 2001 by {\em Chandra} and {\em RXTE}. \cite{2005ApJ...622..602M}
found that V603\,Aql displays strong irregular variability on time-scales of a few ks, 
the 1--7\,keV luminosity $\sim 10^{32}$\,erg\,s$^{-1}$ and the spectrum
described by the cooling flow model.

As expected, the X-ray spectra of \nova{} during its nova eruption 
(bright single-temperature optically thin thermal emission joined later by the super-soft component)
clearly distinguish it from the quiescent spectra of similar systems that
did not show a nova outburst in recent decades. 
Therefore, the X-ray emission we observe in \nova{} is related to the nova 
event rather than any accretion-related phenomena 
\citep[for a detailed discussion of accretion-powered X-rays see][]{2017PASP..129f2001M,2020AdSpR..66.1097B,2020MNRAS.499.3006S}.

\subsection{The mechanisms of X-ray and $\gamma$-ray emission}
\label{sec:emissionmech}

Models for power law emission (\texttt{powerlaw}), thermal bremsstrahlung (\texttt{bremss}; \citealt{1975ApJ...199..299K}),
and thermal plasma emission  \citep[\texttt{vapec};][]{2005AIPC..774..405B} all fit the observed 
{\em NuSTAR} spectrum well (\S~\ref{sec:nustarspec}, Table~\ref{tab:nustarspecmodels}), 
if we allow for non-solar abundances of the absorber \texttt{vphabs} \citep{1992ApJ...400..699B}. 
The intrinsic X-ray emission spectrum in the {\em NuSTAR} band is essentially
smooth and featureless, with few clear signposts allowing us to differentiate between emission models. 

Comptonization of the radioactive MeV lines \citep{1992ApJ...394..217L,2010ApJ...723L..84S,2014ASPC..490..319H}
should produce a flat or rising spectrum below 100\,keV according to \cite{1998MNRAS.296..913G}.
\cite{2019ApJ...872...86N} argue that the Compton optical depth in a nova is
insufficient to produce a detectable hard X-ray flux via this mechanism.
Therefore, we rule out Comptonization as the mechanism behind the X-ray emission
of \nova{}.

The low-energy extension of the energy distribution of particles responsible
for the $\gamma$-ray emission should give rise to \texttt{powerlaw} emission in the hard X-ray band.
\cite{2018ApJ...852...62V} investigate this possibility and predict 
the spectral energy distribution 
$\nu F_{\nu} \propto \nu^{0.8}$ to $\nu^{1.0}$ ($\Gamma = 1.2$ to $1.0$; \S~\ref{sec:thispaper}) 
at energies $\gtrsim10$\,keV. 
The photon index for the power law fit is soft, $\Gamma = 3.3 \pm 0.7$ (the power law index of $-1.3$ in $\nu F_\nu$ units, \S~\ref{sec:thispaper}; Table~\ref{tab:nustarspecmodels}).
The observed spectral slope in the {\em NuSTAR}
band (Table~\ref{tab:nustarspecmodels}) is inconsistent with this prediction. 
It appears likely that the power law model with its soft photon index and high absorbing column just
mimic the intrinsically curved bremsstrahlung spectrum resulting in a good fit. 

Finally, we should mention the possibility of synchrotron emission reaching
all the way to hard X-rays and manifesting itself as a soft power law.
This seems unlikely as no signs of synchrotron emission in novae were
reported at frequencies above the radio band. Generating such emission would
require a very high shock magnetization. Particles emitting synchrotron
X-rays would also emit $>10$\,GeV $\gamma$-rays in the hadronic scenario, 
in contradiction with the observed cut-off around 2\,GeV (\S~\ref{sec:latobs}).

In summary, we suggest that all the emission observed from \nova{} by {\em NuSTAR} is thermal
based on the following two considerations:
\begin{itemize}
\item The power law fit to the {\em NuSTAR} spectrum results in a soft
photon index, while the theory predicts hard spectra for both Comptonization
of MeV line emission and the low-energy extension of the $\gamma$-ray
spectrum \citep{1998MNRAS.296..913G,2018ApJ...852...62V}.
\item The thermal plasma model was clearly preferred over the power law fit 
for a brighter {\em NuSTAR} nova V906\,Car \citep{2020MNRAS.497.2569S}, and
we expect similar emission mechanisms across novae.
\end{itemize}
It is conceivable that some non-thermal emission is mixed into mostly thermal emission as discussed in \S~\ref{sec:nustarspec} (model with two emission components in Table~\ref{tab:nustarspecmodels}), but we have no observational evidence to support this possibility.

\begin{table*}
\begin{center}
\caption{X-ray to $\gamma$-ray monochromatic flux ratio in $\nu F_{\nu}$ units}
\label{tab:lxlgammaratiotable}
\begin{tabular}{cccc}
\hline\hline
Nova       & $({\rm total}~L_{\rm 20\,keV})/L_{\rm 100\,MeV}$ & $({\rm nonthermal}~L_{\rm 20\,keV})/L_{\rm 100\,MeV}$ & Reference \\
\hline
V339\,Del  & $< 4.0 \times 10^{-3}$ & $< 4.0 \times 10^{-3}$ & \cite{2018ApJ...852...62V} \\
V5668\,Sgr & $< 1.7 \times 10^{-3}$ & $< 1.7 \times 10^{-3}$ & \cite{2018ApJ...852...62V} \\
V5855\,Sgr & 0.017                  & $< 1 \times 10^{-3}$   & \cite{2019ApJ...872...86N} \\
V906\,Car  & 0.020                  & $< 5 \times 10^{-4}$   & \cite{2020MNRAS.497.2569S} \\
\nova{}    & $7.0 \times 10^{-3}$   & $< 4 \times 10^{-3}$   & this work \\
\hline
\end{tabular}
\begin{flushleft}
\end{flushleft}
\end{center}
\end{table*}

\cite{2018ApJ...852...62V} make another important prediction: there should be
a lower limit on the ratio of {\it non-thermal} X-ray to $\gamma$-ray fluxes and
this limit depends on the $\gamma$-ray emission mechanism. The predicted monochromatic
flux ratios in $\nu F_{\nu}$ units are $L_X/L_\gamma > 10^{-3}$ for the leptonic model
and $L_X/L_\gamma > 10^{-4}$ for the hadronic model. As no {\it non-thermal}
X-rays were detected by {\em NuSTAR} while the GeV $\gamma$-rays were
observed by {\em Fermi}/LAT, we can constrain the value of this ratio for
\nova{} and compare it to the previously observed novae (Table~\ref{tab:lxlgammaratiotable}). 
To compute the upper limit on the non-thermal monochromatic flux at 20\,keV we use the parameters of the \texttt{powerlaw} component in the model \texttt{constant*phabs*vphabs(vapec+powerlaw)} (Table~\ref{tab:nustarspecmodels}):
%
%
%
%
\begin{equation}
\label{eq:powerlawsed}
\nu F_\nu = C_{\rm erg/keV} K E_{\rm keV}^{2-\Gamma},
\end{equation}
where $C_{\rm erg/keV} = 1.60218 \times 10^{-9}$ is the conversion factor from keV to erg. 
$K$ is the prefactor in the \texttt{powerlaw} component of the model. 
For the absorbed power law model (\texttt{constant*phabs*vphabs*powerlaw} in Table~\ref{tab:nustarspecmodels}) 
$K = (7.06 \pm 0.01) \times 10^{-3}$\,photons\,keV$^{-1}$\,cm$^{-2}$\,s$^{-1}$~at~1\,keV,
while for the absorbed faint power law on top of the bright thermal emission model 
(\texttt{constant*phabs*vphabs(vapec+powerlaw)} in Table~\ref{tab:nustarspecmodels}) 
$K = 8 \times 10^{-6}$\,photons\,keV$^{-1}$\,cm$^{-2}$\,s$^{-1}$~at~1\,keV.
$\Gamma$ is the photon index listed in Table~\ref{tab:nustarspecmodels} and  
$E_{\rm keV} = 20$\,keV, cf.~Eq.~(\ref{eq:latpowerlawexpsed}).
The absorption does not affect the $F_\nu$ calculation (except when fitting
the model) as we are interested in the intrinsic value of $F_\nu$.
The corresponding monochromatic flux at 100\,MeV is computed using Eq.~(\ref{eq:latpowerlawexpsed}) in \S~\ref{sec:latobs}.
The derived upper limit on $({\rm nonthermal}~L_{\rm 20\,keV})/L_{\rm 100\,MeV}$ for \nova{} is consistent 
with both leptonic and hadronic models, 
while the observations of V5855\,Sgr and V906\,Car are consistent only with the hadronic scenario.

\subsection{$L_X/L_\gamma$ and the missing thermal X-ray flux}
\label{sec:lxlg}

While \cite{2018ApJ...852...62V} discuss {\it non-thermal} hard X-ray emission associated with the $\gamma$-ray emitting particle population, 
\cite{2015MNRAS.450.2739M} consider {\it thermal} X-ray emission of the shock responsible for
accelerating these particles. \cite{2015MNRAS.450.2739M} predict bright thermal X-rays that accompany the $\gamma$-rays.
For a radiative shock that accelerates particles with an expected efficiency of $\lesssim$few percent,  thermal X-ray emission should be  $\gtrsim$1--2 orders of magnitude brighter (in $\nu F_{\nu}$ units) than the GeV emission. 
The observed X-ray luminosity is instead 0.007 $L_\gamma$, as measured in the simultaneous 
{\em Fermi}/LAT and {\em NuSTAR} observations (Table~\ref{tab:lxlgammaratiotable}).

For the high densities present early in a nova eruption, a large fraction of the X-ray radiation is absorbed and then re-emitted at longer wavelengths. 
From the shape of the {\em NuSTAR} spectrum, we estimate how much radiation was absorbed and use the unabsorbed (intrinsic) X-ray luminosity to calculate the $L_X/L_\gamma$ ratio. 
The uncertainty in $N_\mathrm{H}$ resulting from uncertain elemental abundances of the neutral absorber
(Table~\ref{tab:nustarspecmodels}; \S~\ref{sec:nustarspec}; \S~\ref{sec:ejectaabund}) 
has less than 1~per~cent effect on $L_X$ as we estimate it at 20\,keV where the absorption is small.

Studies which simulate particle acceleration at shocks find that at most 
20~per~cent of the shock power goes into non-thermal particles \citep{2014ApJ...783...91C}, which effectively sets a lower limit of 5 on the $L_X/L_\gamma$ ratio. 
The modelling of \cite{2018MNRAS.479..687S} suggests that the corrugated geometry of the shock front may suppress X-ray emission by an order of magnitude for the same particle acceleration efficiency.
The X-ray emission may also be Compton scattered away from the line of sight, if the nova ejecta are highly non-spherical \citep{2019ApJ...872...86N}, further lowering the ratio by maybe an order of magnitude.
However, even acting together these effects cannot explain the X-ray emission
in the {\em NuSTAR} band being two to three orders of magnitude {\it fainter}
than the GeV emission. 
The observed $L_X/L_\gamma$ ratio measured from the simultaneous 
{\em Fermi}/LAT and {\em NuSTAR} observations of \nova{} and other novae is
presented in Table~\ref{tab:lxlgammaratiotable}.

Absorption, corrugated shock front geometry and Compton scattering in
an asymmetric shell cannot account for the observed $L_X/L_\gamma$ ratio. 
We are forced to assume that either the shock spends most of its energy on
something other than X-ray radiation, 
such as adiabatic losses or unexpectedly efficient particle acceleration. 
Alternatively, the shock responsible for the X-rays observed by {\em NuSTAR} is not the same
shock that accelerates the $\gamma$-ray emitting particles.
We discuss these possibilities further in the following paragraphs. 

\cite{2018MNRAS.479..687S} and \cite{2020MNRAS.491.4232S} point out that there are two distinct channels for
{\bf adiabatic losses}.
The first is the usual conversion of thermal energy into kinetic energy of the expanding gas. 
The second channel appears as the corrugated shock front has two phases of gas, a cold dense phase and a hot dilute phase. 
In the turbulence behind the shock front, the hot gas can transfer some of its thermal energy to the cold phase in what are technically also adiabatic losses.
This energy is then radiated by the cold phase at long wavelengths (optical).
This is a different mechanism to emit optical radiation than reprocessing X-ray emission from the hot phase 
as the energy transfer from the hot to cold phase is not done via X-ray emission/absorption. 
The shock energy transferred through this channel will not contribute to the unabsorbed X-ray luminosity $L_X$
that we derive from the X-ray spectrum analysis, but it is questionable if
{\it most} of the shock energy can be transferred this way. 

In principle, one can imagine that shocks in novae are somehow especially
efficient at accelerating particles compared to shocks in supernova remnants. 
One possibility is that a shock-accelerated particle interacting with the surrounding matter in dense environment 
of a nova shock may produce secondary particles. The secondary particles may
have sufficient energies to be picked up by the acceleration process.
Such avalanche effect may provide an `infinite' supply of seed particles injected
into diffusive shock acceleration. The idea is similar to runaway electron production
mechanisms \citep{1992PhLA..165..463G,2012JGRA..117.2308D} thought to be responsible 
for terrestrial $\gamma$-ray flashes \citep[e.g.][]{2016JGRA..12111346M} and 
$\gamma$-ray glows \citep[e.g.][]{2021GeoRL..4891910W}. However, this `shock
spending most of its energy accelerating particles' scenario does not
account for the observations of correlated $\gamma$-ray/optical variability
in V5856\,Sgr and V906\,Car that together with the $\gamma$-ray/optical flux
ratio suggested that most of the shock energy is eventually radiated in the
optical band \citep{2017NatAs...1..697L,2020NatAs...4..776A}.

A {\bf second shock}, different from the one responsible for 
the {\em NuSTAR}-detected emission, traveling at a velocity of a few hundred km\,s$^{-1}$ may accelerate
particles that produce {\em Fermi}/LAT-detected $\gamma$-rays. 
The low temperature of this second shock would put the associated X-ray emission below the {\em NuSTAR}
band according to Eq.~(\ref{eq:ktshock}). 
\cite{2016MNRAS.463..394V} mention the possibility that different shocks may
be responsible for emission observed in different bands, or even in the same
band at different times.
Multiple optically thin thermal emission components (that may correspond to
multiple shocks) are observed in some classical novae \citep{2021MNRAS.500.2798N}, but not in others \citep{2020MNRAS.497.2569S}
(multi-temperature emission is commonly observed in novae with an evolved donor; 
\citealt{2008ApJ...673.1067N,2012ApJ...748...43N,2013MNRAS.429.1342O,2015MNRAS.448L..35O,2021MNRAS.tmp.1420O}).
{\em Swift}/XRT 0.3--10\,keV observations of \nova{} were fit with a
single-temperature thermal plasma emission until the emergence of the SSS \citep{2020ATel14043....1S}.
However, the more novae that are observed by {\em NuSTAR}, the harder it becomes to support this somewhat contrived scenario of multiple shocks as the explanation for the low $L_X/L_\gamma$ ratio.

\subsection{Location of the shocked region}
\label{sec:location}

Determining the location of the shocked region(s) within the nova ejecta is
important to draw an accurate physical picture of the eruption and,
specifically, to estimate the influence of $\gamma$-ray opacity 
on the observed GeV spectrum \citep{2016MNRAS.457.1786M}.
The $\gamma$-rays may be absorbed via Bethe-Heitler photo-nuclear pair production 
(the same process used by {\em Fermi}/LAT to detect $\gamma$-rays) and
Breit-Wheeler $\gamma\gamma$ pair production.

The X-ray flux approximately doubled over the 120\,ks duration of the {\em NuSTAR}
observation (Fig.~\ref{fig:nustarlc}; a weighted linear fit to the lightcurve
results in a count rate ratio at time 120\,ks to time 0 of $2.0 \pm 0.2$). 
We can take this as an estimate of the variability time-scale associated with the shock, if we attribute
the X-ray emission to the shock-heated plasma (\S~\ref{sec:emissionmech}).
The post-shock temperature ($T_{\rm shock}$) can be related to the shock velocity ($v_{\rm shock}$)
for a strong shock propagating in monoatomic gas (with polytropic exponent 5/3):
\begin{equation}
{\rm k} T_{\rm shock} = \frac{3}{16} \mu m_p v_{\rm shock}^2
\label{eq:ktshock}
\end{equation}
(equation~[6.58] of \citealt{1997pism.book.....D}),
where $m_p$ is the proton mass, $k$ is the Boltzmann constant, and $\mu$ is the mean molecular weight. 
This relation is derived from the Rankine--Hugoniot jump conditions that follow from conservation of mass, momentum, and energy.
Here we neglect the shock energy losses on particle acceleration \citep{2007ApJ...663L.101T}.
For a fully ionized gas with solar abundances \citep{2009ARA&A..47..481A}
$\mu=0.60$, while the composition derived for V906\,Car by
\cite{2020MNRAS.497.2569S} implies $\mu=0.74$. Assuming the V906\,Car abundances and temperature derived from the {\em NuSTAR} observation (Table~\ref{tab:nustarspecmodels}),
we find $v_{\rm shock} \simeq 2000$\,km\,s$^{-1}$.
Multiplying $v_{\rm shock}$ by the variability time-scale, we constrain 
the shocked region size at \tnustarep{} (the date of the {\em NuSTAR} observation) to be
less than 1.6\,au. The upper limit on the shocked region size allows it to be larger than the binary separation and 
the optical photosphere (\S~\ref{sec:gammaopt}).

At $t_0 + 82$\,d {\em NICER} observed irregular variations on a time-scale of
kiloseconds in soft X-rays \citep{2020ATel14067....1P}, 
corresponding to the size of $>0.01$\,au for the velocities of 1000\,km/s. 
This variability is in the super-soft emission that is directly related to the white
dwarf (and attributed to changes either in emission or absorption in the
vicinity of the white dwarf).

Finally, we mention the possibility that rather than 
having one shock (or a pair of forward and reverse shocks) at 
the interface between the fast and slow components of the nova outflow 
\citep{2014Natur.514..339C,2020ApJ...905...62A}, multiple shocks associated with
individual dense clumps within the nova ejecta may be responsible for the high-energy emission.
This is the mechanism thought to produce X-rays in early-type stars \citep[e.g. sec.~4 of][]{2009A&ARv..17..309G} 
and similar clumps should form in nova ejecta \citep{2001MNRAS.326..126S,2001ApJ...549.1093S}.
If many clumps emit simultaneously, the fast variability associated
with individual clumps may average out.
The ratio of X-ray to bolometric optical luminosity of O-type stars is $\sim 10^{-6}$ 
\citep[fig.~6 of][]{1991ApJ...368..241C}, 
comparable to what we find in \nova{} (\S~\ref{sec:gammaopt}, \ref{sec:luminosity}), 
but single early-type stars are not known to emit $\gamma$-rays. 
In the multiple-clumps/multiple-shocks scenario one may expect a wide range of temperatures
associated with the individual emitting regions.

\subsection{Single-temperature fit to {\em NuSTAR} spectrum}
\label{sec:singletempnustar}

It is somewhat surprising that the hard X-ray spectra of \nova{} (\S~\ref{sec:nustarspec})  
and other novae observed by {\em NuSTAR} 
(V745\,Sco with a giant donor, \citealt{2015MNRAS.448L..35O}; 
V5855\,Sgr, \citealt{2019ApJ...872...86N}; 
and the brightest of them, V906\,Car, \citealt{2020MNRAS.497.2569S}) 
resembles that of a single-temperature plasma, 
even if the emission is produced by one single shock. 
The plasma heated by the shock to a temperature determined by Eq.~(\ref{eq:ktshock}) should 
cool by radiation \citep[e.g.][]{2017MNRAS.469.1314D} producing a
temperature gradient in the post-shock region. The direction-dependent
density profile of the external medium may make the shock propagate with
different velocities in different directions, resulting in a range of shock
temperatures \citep[most relevant in novae with a giant donor][]{2017MNRAS.464.5003O}.
Apparently, in practice there is some characteristic temperature associated 
with the highest emission measure and possibly modified by absorption 
that preferentially affects low-temperature emission. 
The single-temperature approximation
is typically sufficient to describe a {\em NuSTAR} spectrum of a nova.

The exponential cut-off of the bremsstrahlung spectrum is well within 
the {\em NuSTAR} energy range constraining the (highest) temperature of the
emitting plasma. However, the presence of an uncertain amount of intrinsic absorption 
makes it hard to distinguish between the single-temperature and multi-temperature models.
The difference between the models is mainly in the observed slope of the bremsstrahlung
continuum, which may be altered by `adding' more absorbing material to the
model. A high resolution and signal-to-noise spectrum may 
distinguish between the effects of absorption and temperature distribution
by resolving contribution of individual absorption edges and thermal emission lines.
Even then the narrowness of the effective bandpass (between the low-energy cut-off due to absorption 
and the exponential cut-off at ${\rm k} T_{\rm shock}$) may prevent one from reaching 
an unambiguous conclusion.
In summary, the {\em NuSTAR} spectrum of \nova{} allows us to characterize 
the shock temperature, the temperature distribution of the post-shock plasma
cannot be reliably constrained.

\subsection{Line-dominated SSS emission}
\label{sec:linesss}

According to {\em NuSTAR} (\S~\ref{sec:nustarobs}), {\em Swift}/XRT
\citep{2020ATel14043....1S} and {\em NICER} \citep{2020ATel14067....1P}
observations, the early X-ray emission of \nova{} was hard, dominated by 
shock-heated plasma. Around $t_0 + 59$\,d the nova ejecta cleared
sufficiently to reveal super-soft X-rays originating in the vicinity of 
the hydrogen-burning white dwarf. The {\em XMM-Newton} observed \nova{}
on $t_0+77.6$\,d, around the peak of its SSS phase, revealing 
a rare emission-line dominated SSS spectrum.

At low spectral resolution, the SSS component in novae is often 
approximated as a blackbody \citep[e.g.][]{2011ApJS..197...31S}. 
However, grating spectra have revealed two types of SSS: the ones
dominated by a blackbody-like continuum, and in most cases modified by absorption 
lines, and the ones dominated by emission lines on top of a weak blackbody-like continuum \citep{2013A&A...559A..50N}. 
The emission-line-dominated SSS \citep[e.g. U\,Sco;][]{2012ApJ...745...43N}, 
can be interpreted as being the result of obscuration of central continuum emission while emission lines are formed further outside. 
The central blackbody-like continuum emission is thus suppressed increasing the contrast to emission lines that are always present. 
Such obscuration may appear in high-inclination systems (viewed edge-on), where the white dwarf is obscured by the accretion disc (that
survives or gets quickly re-formed following the nova eruption; \citealt{2018A&A...613A...8F}).
The {\em XMM-Newton}/RGS spectrum 
presented in Fig.~\ref{fig:rgsspeccomp} firmly places \nova{} in the latter category, implying it may be a high-inclination system.

The extremely soft emission-line-dominated spectrum of \nova{} was also observed with {\em Chandra} by \cite{2020ATel14214....1D} on $t_0 + 115$\,d. 
The line-dominated nature of \nova{}'s SSS spectrum is apparent only with X-ray grating spectroscopy. 
The low resolution EPIC spectrum could have been easily mistaken for a blackbody.
Since the emission-line spectrum is likely created by scattering of 
the primary (blackbody-like) emission by a medium that has a temperature
similar to that of the stellar surface (as shown by the mirror image absorption/emission
lines), the result is a spectrum that looks blackbody-like at low spectral resolution.

\subsection{Jets in \nova{}?}
\label{sec:jets}

Flows of plasma collimated to an opening angle of $\lesssim10^{\circ}$ 
often producing non-thermal emission are known as jets. 
Jets power astrophysical phenomena emitting in a very wide range of the electromagnetic spectrum, from $\gamma$-rays to radio: 
blazars, microquasars and $\gamma$-ray bursts \citep{2017SSRv..207....5R,2015PhR...561....1K}. 
The exceptions (non-jetted phenomena observed across the electromagnetic spectrum) 
are colliding wind binaries \citep{2016MNRAS.457L..99P,2017A&A...608A..59D}, 
the Galaxy \citep{2015ApJ...815L..25G}, 
the Sun \citep{2011ApJ...734..116A,2021ApJS..252...13A} and Earth
\citep{2012SSRv..173..133D,2020JGRA..12528151M}.
One may ask if jets play a role in novae?

%
%
There were several reports of jets in novae based on optical spectral line profile
studies \citep{2003ApJ...587L..39K,2003A&A...404..997I,2017ApJ...847...35D}
as well as radio \citep{1988IAUS..129..277D,2008ApJ...685L.137S,2008ApJ...688..559R,2020A&A...638A.130G} 
and X-ray \citep{2020MNRAS.495.4372T} imaging. Some earlier claims of 
observations of nova jets were later disputed \citep{1998ApJ...498L..59O,2016A&A...595A..64H}. 
It is actively debated if non-nova cataclysmic variables have jets \citep{2020NewAR..8901540C}. 

\cite{2021MNRAS.503..704M} argue that the H~$\alpha$ line profile in \nova{}
can be explained as a sum of emission from the approaching and receding jets and 
the accretion disc. The authors assume that 
the contribution from the non- or weakly-collimated ejecta to the total line flux is small.
\cite{2021arXiv210605578M} offer the similar line profile interpretation for other novae.

The observations discussed in the present paper do not allow us to deduce
the ejecta geometry, however the following considerations seem to
disfavour the jet scenario. First, the line-dominated SSS (\S~\ref{sec:linesss}) indicates 
that \nova{} is a high-inclination system.
The jets need to be fast in order to produce 
the high-velocity emission line components while being aligned nearly 
perpendicular to the line of sight. Second, \cite{2021MNRAS.503..704M} suggest 
the X-ray emitting shocks are produced by collision of individual blobs 
of material traveling down the jet with various speeds. 
Our observations do not require multi-temperature emission, which would support the jet model.
In the colliding blobs scenario, one could also expect
variability on a time-scale of the blob size over the blob collision velocity. 
The blob size should be of the order of the jet width.  
Instead, the observed variability time-scale and the shock velocity (derived
from the shock temperature) suggest a large emitting region (\S~\ref{sec:location}).

A large nearly-single-temperature shocked region seems to fit more naturally into
the scenario of a slow equatorial outflow (possibly ejected via the common
envelope interaction during the nova eruption) with the fast wind
(accelerated by the white dwarf radiation) -- the scenario favoured by
\cite{2014Natur.514..339C}, \cite{2020arXiv201108751C} and \cite{2020ApJ...905...62A}.
In this scenario, the shock is formed at the interface between the fast and
slow flows while multiple ejections with different velocities and the
complex shape of the ejecta formed via their interaction are responsible for
the complex optical line profiles. In a sense, the question of the existence
of jets is about the degree of collimation that can be achieved by the
fast flow: an opening angle of a few degrees for a jet or a few tens of
degrees for a bipolar outflow. 
We speculate that the presence of particle-accelerating shocks, rather than 
the presence of these shocks specifically in the highly-collimated jets, may be 
the physical mechanism unifying the high-energy to radio emitting phenomena listed above.

\subsection{Ejecta abundances and the white dwarf composition}
\label{sec:ejectaabund}

Optical, infrared and X-ray spectra indicate that nova ejecta are typically enriched
in heavy elements that must be eroded from the white dwarf 
\citep[e.g.][]{1998PASP..110....3G,2012ApJ...755...37H,2020MNRAS.497.2569S}.
Thermonuclear burning in the nova proceeds through the hot carbon-nitrogen-oxygen
(CNO) cycle \citep{2010ARNPS..60..381W}, and may change the relative abundances of C, N and O, 
but will not increase the total abundance of CNO relative to other elements \citep{1972ApJ...176..169S,1986ApJ...308..721T}. 

The {\em XMM-Newton}/RGS spectra (\S~\ref{sec:xmmspec}; Fig.~\ref{fig:rgsspeccomp}) of \nova{} show no signs of
Ne and Mg emission lines that would normally fall into the RGS band.
Such lines are visible in the X-ray grating spectra of 
V382\,Vel \citep{2005MNRAS.364.1015N},
U\,Sco \citep{2012ApJ...745...43N}, 
V959\,Mon \citep{2021MNRAS.500.2798N} and
V3890\,Sgr \citep{2020ApJ...895...80O}.
Instead, the RGS spectra are dominated by emission lines of C and N,
suggesting that the white dwarf in the \nova{} system may be of CO composition rather than ONeMg. 
The ONeMg composition for \nova{} was suggested by \cite{2020ATel14048....1I}
who detected optical lines 
[\ion{Ne}{III}]~3342\,\AA{} and [\ion{Ne}{V}]~3426\,\AA{}.
It is possible that the Ne detected in the optical spectrum is associated with
the material accreted from the companion star (or ablated from the companion
star during eruption?) rather than with the white dwarf material.
A CO white dwarf may have non-zero Ne content on its own \citep{2016ApJ...823...46F}. 
It cannot be excluded that Ne and Mg lines are not visible in the X-ray
spectrum due to the low temperature of the ionizing radiation from the white
dwarf, as Ne and Mg have higher ionization potential than C and N.
An ONeMg white dwarf may have a CO envelope, so the presence of C emission
does not exclude the ONeMg scenario. If the white dwarf in the \nova{} system is of CO composition 
as it seems to be the case, then it qualifies as a Supernova Ia progenitor candidate.

{\em NuSTAR} spectra rule out solar abundances, but are consistent with 
Fe-deficient and/or NO-overabundant plasma (\S~\ref{sec:nustarspec}). 
Accretion of low-metallicity material from the secondary is a possibility given 
the high elevation above the Galactic disc (\S~\ref{sec:extinction}), 
suggesting the system belongs to an old stellar population. 
%
%
But the composition of the accreted matter is not the only possible source of iron deficiency.
Heavy elements, including Fe, are expected to sink below the surface of a
white dwarf and they sink faster the hotter the white dwarf is
\citep{2009A&A...498..517K,2016Sci...352...67K}.
If a large portion of the ejecta originates on the white dwarf,
it is natural to expect it may be both Fe-deficient and CNO-overabundant.
This is exactly what was found in the {\em XMM-Newton} spectroscopy of V906\,Car 
\citep{2020MNRAS.497.2569S}.

\subsection{Ejecta mass}
\label{sec:ejectamass}

One can use the column density, expressed in $N_\mathrm{H}$ and derived from the X-ray spectral fitting 
(\S~\ref{sec:nustarspec}), to estimate the nova ejecta mass under a set of assumptions. 
We assume that the source of hard X-rays is embedded deep within the ejecta 
(shining through most of it). 
The ejecta ahead of the shock (absorbing the X-ray emission) is neutral or
weakly ionized, as atoms stripped of all their electrons will not contribute
to photoelectric absorption.
A spherical absorbing shell is ejected at $t_0$ and expands with velocities ranging from $v_{\rm min}$ to $v_{\rm max}$. 
The ejecta are distributed with a density profile  $\propto r^{-2}$ \citep[e.g.][]{2008clno.book.....B}.
This is the `Hubble flow' model often used to describe thermal radio
emission of novae \citep[e.g.]{2016MNRAS.457..887W,2016MNRAS.460.2687W,2018ApJ...852..108F}.
Following \cite{2014ApJ...788..130C}, we assume $v_{\rm min} = 0.2 v_{\rm max}$.

Our assumptions about the ejecta abundances (\S~\ref{sec:ejectaabund}) have 
a dramatic effect on the derived column density (\S~\ref{sec:nustarspec}),
with the true value likely lying somewhere between the two extremes
listed in Table~\ref{tab:nustarspecmodels}.
Aydi el al. (in prep.) identify two flows on the basis of optical
spectroscopy of \nova{}: the fast flow with velocity of 2700\,km\,s$^{-1}$ and an
intermediate flow with velocity of 1200\,km\,s$^{-1}$.
We do not know which of the two flows carries the most mass.
Combining the assumptions about column density (abundances) and  maximum ejecta velocity for the flow that carries the most mass, we end
up with estimates of the ejected mass of hydrogen in the range 
$2 \times 10^{-6}\, {\rm M}_\odot$ 
($v_{\rm max} = 1200$\,km\,s$^{-1}$, $N_\mathrm{H} = 7.3 \times 10^{22}$\,cm$^{-2}$)
to
$2 \times 10^{-4}\, {\rm M}_\odot$ 
($v_{\rm max} = 2700$\,km\,s$^{-1}$, $N_\mathrm{H} = 131.3 \times 10^{22}$\,cm$^{-2}$).
To obtain the total ejecta mass, the hydrogen mass should be multiplied by a factor of 1.90 for the abundances
of nova V906\,Car \citep{2020MNRAS.497.2569S}, or a factor of 1.36 for the solar abundances of \cite{2009ARA&A..47..481A}.

Our final estimate of the ejecta mass in \nova{} is $\sim 4 \times 10^{-5}\,{\rm M}_\odot$, 
with an order of magnitude uncertainty, as described above. 
The ejecta mass derived for V906\,Car by \cite{2020MNRAS.497.2569S} using
the same technique falls in the middle of the range allowed for \nova{}, so
we speculate that the ejected mass in the two novae may be comparable. 

Comparing the X-ray absorption-based ejecta mass estimate to theoretical 
expectations and ejecta mass estimates made via other methods, we can check
where the X-ray emitting shock is located relative to the bulk of the
ejecta. If the column ahead of the shocks is high (ejecta mass $\sim10^{-4}\, {\rm M}_\odot$),
than the shocks are likely embedded behind the bulk of the ejecta 
(since nova ejecta masses are strained to go above $\sim10^{-4}\, {\rm M}_\odot$; \citealt{2005ApJ...623..398Y}). 
On the other hand, if the column implies the ejecta mass $\lesssim10^{-7}\, {\rm M}_\odot$, 
than the observed emission may be dominated by X-rays escaping from a select few directions 
with a particularly low column. The ejecta mass range estimated above seems
to support the former picture.

\section{Conclusions}
\label{sec:conclusions}

We conducted a joint analysis of {\em Fermi}/LAT, {\em NuSTAR} and {\em XMM-Newton} observations of 
a bright Galactic nova \nova{}. 
The luminosity of \nova{} (Table~\ref{tab:luminositiestable}) 
is well constrained thanks to a {\em Gaia} parallax measurement of the bright 
progenitor: a VY\,Scl type novalike variable with hot accretion disc. 
The GeV, X-ray and optical luminosity of \nova{} is similar to other GeV-bright novae (\S~\ref{sec:luminosity}). 
 
The nova X-ray emission observed by {\em NuSTAR} at \tnustarep{} is 
consistent with being single-temperature thermal (\S~\ref{sec:emissionmech}). 
The low $({\rm thermal~}L_{\rm 20\,keV})/L_{\rm 100\,MeV}$ ratio is at odds with 
the theoretical predictions \citep[\S~\ref{sec:lxlg}, Table~\ref{tab:lxlgammaratiotable}, ][]{2015MNRAS.450.2739M}.
The absence of non-thermal X-rays is consistent with both the leptonic and hadronic scenarios for 
the production of $\gamma$-rays detected by {\em Fermi}/LAT (\S~\ref{sec:emissionmech}).

From the variability time-scale and shock velocity arguments, we constrain the shocked region
size to be less than 1.6\,au on \tnustarep{} (\S~\ref{sec:location}). 
No periodicities were identified in the arrival times of X-ray photons
recorded by {\em NuSTAR} (\S~\ref{sec:nustarvar}) and {\em XMM-Newton} (\S~\ref{sec:xmmtiming}).
The shock-heated region must be associated with the expanding nova shell,
not a structure within the binary system (such as the white dwarf magnetosphere, 
accretion disc, bow shock of the donor star).
The emission line dominated SSS spectrum observed with {\em XMM-Newton} at
$t_0+77.6$\,d suggests \nova{} is a high-inclination system (\S~\ref{sec:linesss})
with a CO white dwarf. 

We use the intrinsic absorption affecting the {\em NuSTAR} spectrum to
estimate an ejecta mass of $4 \times 10^{-5}\, {\rm M}_\odot$ 
(with an order of magnitude uncertainty) 
in the framework of the Hubble flow model (\S~\ref{sec:ejectamass}). 
The estimate is model-dependent and highly sensitive to the assumptions on the range of 
ejecta velocities, abundances, ejection time and location of the X-ray emitting region. 
This results in at least an order of magnitude uncertainty 
in the ejected mass estimate. Also the photoelectric-absorption 
based ejecta mass estimate does not account for any fully ionized material.

\section*{Data availability}

The processed data underlying this work are available at the request to the first author. 
The raw data are publicly available at {\em NuSTAR}, {\em XMM-Newton} and {\em Fermi} science archives.

\section*{Acknowledgements}
%
We thank Dr.~Marina~Orio for sharing the {\em Chandra} spectrum of \nova{}, 
which greatly helped interpreting the results of {\em XMM-Newton} spectroscopy. 
We acknowledge with thanks the variable star observations from the AAVSO International Database contributed by observers worldwide and used in this research.
KVS thanks 
Dr.~Nikolai~N.~Samus for the help in gathering information about the
circumstances surrounding the discovery of \nova{}. 
We acknowledge ESA {\em Gaia}, DPAC and the Photometric Science Alerts Team
(\url{http://gsaweb.ast.cam.ac.uk/alerts}).
This material is based upon work supported by the National Science Foundation under Grant~No.~AST-1751874. 
We acknowledge support for this work from NASA grants 
NASA/NuSTAR 80NSSC21K0277, NASA/Fermi 80NSSC20K1535, and NASA/Swift 80NSSC21K0173 
and from a Cottrell Scholarship from the Research Corporation.
We acknowledge support from the Packard Foundation. 
%
RLO acknowledges financial support from the Brazilian institutions {\it Conselho Nacional de Desenvolvimento Cient\'ifico e Tecnol\'ogico} (CNPq - PQ Grant; 312705/2020-4) and 
{\it Funda\c{c}\~ao de Amparo \`a Pesquisa do Estado de S\~ao Paulo} (FAPESP - Visiting Researcher Grant; 2020/00457-4). 
BDM acknowledges support from NASA (grant 80NSSC20K1557).
KLP acknowledges support from the UK Space Agency. JS acknowledges support from the Packard Foundation. 
KLL is supported by the Ministry of Science and Technology of the Republic of China (Taiwan) through grants 108-2112-M-007-025-MY3 and 109-2636-M-006-017, and he is a Yushan (Young) Scholar of the Ministry of Education of the Republic of China (Taiwan).
IV acknowledges support by the ETAg grant PRG1006 and by EU through the ERDF CoE grant TK133.
%




\bibliographystyle{mnras}
\bibliography{yzret} 

\begin{thebibliography}{}
\makeatletter
\relax
\def\mn@urlcharsother{\let\do\@makeother \do\$\do\&\do\#\do\^\do\_\do\%\do\~}
\def\mn@doi{\begingroup\mn@urlcharsother \@ifnextchar [ {\mn@doi@}
  {\mn@doi@[]}}
\def\mn@doi@[#1]#2{\def\@tempa{#1}\ifx\@tempa\@empty \href
  {http://dx.doi.org/#2} {doi:#2}\else \href {http://dx.doi.org/#2} {#1}\fi
  \endgroup}
\def\mn@eprint#1#2{\mn@eprint@#1:#2::\@nil}
\def\mn@eprint@arXiv#1{\href {http://arxiv.org/abs/#1} {{\tt arXiv:#1}}}
\def\mn@eprint@dblp#1{\href {http://dblp.uni-trier.de/rec/bibtex/#1.xml}
  {dblp:#1}}
\def\mn@eprint@#1:#2:#3:#4\@nil{\def\@tempa {#1}\def\@tempb {#2}\def\@tempc
  {#3}\ifx \@tempc \@empty \let \@tempc \@tempb \let \@tempb \@tempa \fi \ifx
  \@tempb \@empty \def\@tempb {arXiv}\fi \@ifundefined
  {mn@eprint@\@tempb}{\@tempb:\@tempc}{\expandafter \expandafter \csname
  mn@eprint@\@tempb\endcsname \expandafter{\@tempc}}}

\bibitem[\protect\citeauthoryear{{Abdo} et~al.,}{{Abdo}
  et~al.}{2009}]{2009APh....32..193A}
{Abdo} A.~A.,  et~al., 2009, \mn@doi [Astroparticle Physics]
  {10.1016/j.astropartphys.2009.08.002}, \href
  {https://ui.adsabs.harvard.edu/abs/2009APh....32..193A} {32, 193}

\bibitem[\protect\citeauthoryear{{Abdo} et~al.,}{{Abdo}
  et~al.}{2010}]{2010Sci...329..817A}
{Abdo} A.~A.,  et~al., 2010, \mn@doi [Science] {10.1126/science.1192537}, \href
  {http://adsabs.harvard.edu/abs/2010Sci...329..817A} {329, 817}

\bibitem[\protect\citeauthoryear{{Abdo} et~al.,}{{Abdo}
  et~al.}{2011}]{2011ApJ...734..116A}
{Abdo} A.~A.,  et~al., 2011, \mn@doi [\apj] {10.1088/0004-637X/734/2/116},
  \href {https://ui.adsabs.harvard.edu/abs/2011ApJ...734..116A} {734, 116}

\bibitem[\protect\citeauthoryear{{Abdollahi} et~al.,}{{Abdollahi}
  et~al.}{2020}]{2020ApJS..247...33A}
{Abdollahi} S.,  et~al., 2020, \mn@doi [\apjs] {10.3847/1538-4365/ab6bcb},
  \href {https://ui.adsabs.harvard.edu/abs/2020ApJS..247...33A} {247, 33}

\bibitem[\protect\citeauthoryear{{Ackermann} et~al.,}{{Ackermann}
  et~al.}{2012}]{2012ApJS..203....4A}
{Ackermann} M.,  et~al., 2012, \mn@doi [\apjs] {10.1088/0067-0049/203/1/4},
  \href {https://ui.adsabs.harvard.edu/abs/2012ApJS..203....4A} {203, 4}

\bibitem[\protect\citeauthoryear{{Ackermann} et~al.,}{{Ackermann}
  et~al.}{2014}]{2014Sci...345..554A}
{Ackermann} M.,  et~al., 2014, \mn@doi [Science] {10.1126/science.1253947},
  \href {http://adsabs.harvard.edu/abs/2014Sci...345..554A} {345, 554}

\bibitem[\protect\citeauthoryear{{Adams}, {Kochanek}, {Beacom}, {Vagins}  \&
  {Stanek}}{{Adams} et~al.}{2013}]{2013ApJ...778..164A}
{Adams} S.~M.,  {Kochanek} C.~S.,  {Beacom} J.~F.,  {Vagins} M.~R.,   {Stanek}
  K.~Z.,  2013, \mn@doi [\apj] {10.1088/0004-637X/778/2/164}, \href
  {https://ui.adsabs.harvard.edu/abs/2013ApJ...778..164A} {778, 164}

\bibitem[\protect\citeauthoryear{{Ajello} et~al.,}{{Ajello}
  et~al.}{2021}]{2021ApJS..252...13A}
{Ajello} M.,  et~al., 2021, \mn@doi [\apjs] {10.3847/1538-4365/abd32e}, \href
  {https://ui.adsabs.harvard.edu/abs/2021ApJS..252...13A} {252, 13}

\bibitem[\protect\citeauthoryear{{Arnaud}}{{Arnaud}}{1996}]{1996ASPC..101...17A}
{Arnaud} K.~A.,  1996, in {Jacoby} G.~H.,  {Barnes} J.,  eds,  Astronomical
  Society of the Pacific Conference Series Vol. 101, Astronomical Data Analysis
  Software and Systems V. p.~17

\bibitem[\protect\citeauthoryear{{Arnaud}, {Smith}  \&
  {Siemiginowska}}{{Arnaud} et~al.}{2011}]{2011hxra.book.....A}
{Arnaud} K.,  {Smith} R.,   {Siemiginowska} A.,  2011, {Handbook of X-ray
  Astronomy}

\bibitem[\protect\citeauthoryear{{Asplund}, {Grevesse}, {Sauval}  \&
  {Scott}}{{Asplund} et~al.}{2009}]{2009ARA&A..47..481A}
{Asplund} M.,  {Grevesse} N.,  {Sauval} A.~J.,   {Scott} P.,  2009, \mn@doi
  [\araa] {10.1146/annurev.astro.46.060407.145222}, \href
  {http://adsabs.harvard.edu/abs/2009ARA%26A..47..481A} {47, 481}

\bibitem[\protect\citeauthoryear{{Atwood} et~al.,}{{Atwood}
  et~al.}{2009}]{2009ApJ...697.1071A}
{Atwood} W.~B.,  et~al., 2009, \mn@doi [\apj] {10.1088/0004-637X/697/2/1071},
  \href {https://ui.adsabs.harvard.edu/abs/2009ApJ...697.1071A} {697, 1071}

\bibitem[\protect\citeauthoryear{{Aydi} et~al.,}{{Aydi}
  et~al.}{2020a}]{2020NatAs...4..776A}
{Aydi} E.,  et~al., 2020a, \mn@doi [Nature Astronomy]
  {10.1038/s41550-020-1070-y}, \href
  {https://ui.adsabs.harvard.edu/abs/2020NatAs...4..776A} {4, 776}

\bibitem[\protect\citeauthoryear{{Aydi} et~al.,}{{Aydi}
  et~al.}{2020b}]{2020ApJ...905...62A}
{Aydi} E.,  et~al., 2020b, \mn@doi [\apj] {10.3847/1538-4357/abc3bb}, \href
  {https://ui.adsabs.harvard.edu/abs/2020ApJ...905...62A} {905, 62}

\bibitem[\protect\citeauthoryear{{Aydi} et~al.,}{{Aydi}
  et~al.}{2020c}]{2020ATel13867....1A}
{Aydi} E.,  et~al., 2020c, The Astronomer's Telegram, \href
  {https://ui.adsabs.harvard.edu/abs/2020ATel13867....1A} {13867, 1}

\bibitem[\protect\citeauthoryear{{Aydi} et~al.,}{{Aydi}
  et~al.}{2021}]{2021arXiv210807868A}
{Aydi} E.,  et~al., 2021, arXiv e-prints, \href
  {https://ui.adsabs.harvard.edu/abs/2021arXiv210807868A} {p. arXiv:2108.07868}

\bibitem[\protect\citeauthoryear{{Bailer-Jones}, {Rybizki}, {Fouesneau},
  {Mantelet}  \& {Andrae}}{{Bailer-Jones} et~al.}{2018}]{2018AJ....156...58B}
{Bailer-Jones} C.~A.~L.,  {Rybizki} J.,  {Fouesneau} M.,  {Mantelet} G.,
  {Andrae} R.,  2018, \mn@doi [\aj] {10.3847/1538-3881/aacb21}, \href
  {https://ui.adsabs.harvard.edu/abs/2018AJ....156...58B} {156, 58}

\bibitem[\protect\citeauthoryear{{Bajaja}, {Arnal}, {Larrarte}, {Morras},
  {P{\"o}ppel}  \& {Kalberla}}{{Bajaja} et~al.}{2005}]{2005A&A...440..767B}
{Bajaja} E.,  {Arnal} E.~M.,  {Larrarte} J.~J.,  {Morras} R.,  {P{\"o}ppel}
  W.~G.~L.,   {Kalberla} P.~M.~W.,  2005, \mn@doi [\aap]
  {10.1051/0004-6361:20041863}, \href
  {https://ui.adsabs.harvard.edu/abs/2005A&A...440..767B} {440, 767}

\bibitem[\protect\citeauthoryear{{Balman}}{{Balman}}{2020}]{2020AdSpR..66.1097B}
{Balman} {\c{S}}.,  2020, \mn@doi [Advances in Space Research]
  {10.1016/j.asr.2020.05.031}, \href
  {https://ui.adsabs.harvard.edu/abs/2020AdSpR..66.1097B} {66, 1097}

\bibitem[\protect\citeauthoryear{{Balucinska-Church} \&
  {McCammon}}{{Balucinska-Church} \& {McCammon}}{1992}]{1992ApJ...400..699B}
{Balucinska-Church} M.,  {McCammon} D.,  1992, \mn@doi [\apj] {10.1086/172032},
  \href {http://adsabs.harvard.edu/abs/1992ApJ...400..699B} {400, 699}

\bibitem[\protect\citeauthoryear{{Bambi}}{{Bambi}}{2020}]{2020tgxg.book.....B}
{Bambi} C.,  2020, {Tutorial Guide to X-ray and Gamma-ray Astronomy; Data
  Reduction and Analysis}, \mn@doi{10.1007/978-981-15-6337-9.
}

\bibitem[\protect\citeauthoryear{{Bessell}, {Castelli}  \& {Plez}}{{Bessell}
  et~al.}{1998}]{1998A&A...333..231B}
{Bessell} M.~S.,  {Castelli} F.,   {Plez} B.,  1998, \aap, \href
  {https://ui.adsabs.harvard.edu/abs/1998A&A...333..231B} {333, 231}

\bibitem[\protect\citeauthoryear{{Blandford} \& {Ostriker}}{{Blandford} \&
  {Ostriker}}{1978}]{1978ApJ...221L..29B}
{Blandford} R.~D.,  {Ostriker} J.~P.,  1978, \mn@doi [\apjl] {10.1086/182658},
  \href {https://ui.adsabs.harvard.edu/abs/1978ApJ...221L..29B} {221, L29}

\bibitem[\protect\citeauthoryear{{Bode} \& {Evans}}{{Bode} \&
  {Evans}}{2008}]{2008clno.book.....B}
{Bode} M.~F.,  {Evans} A.,  2008, {Classical Novae}

\bibitem[\protect\citeauthoryear{{Boller}, {Freyberg}, {Tr{\"u}mper}, {Haberl},
  {Voges}  \& {Nandra}}{{Boller} et~al.}{2016}]{2016A&A...588A.103B}
{Boller} T.,  {Freyberg} M.~J.,  {Tr{\"u}mper} J.,  {Haberl} F.,  {Voges} W.,
  {Nandra} K.,  2016, \mn@doi [\aap] {10.1051/0004-6361/201525648}, \href
  {https://ui.adsabs.harvard.edu/abs/2016A&A...588A.103B} {588, A103}

\bibitem[\protect\citeauthoryear{{Bond}}{{Bond}}{2020}]{2020ATel13825....1B}
{Bond} H.~E.,  2020, The Astronomer's Telegram, \href
  {https://ui.adsabs.harvard.edu/abs/2020ATel13825....1B} {13825, 1}

\bibitem[\protect\citeauthoryear{{Bond} \& {Miszalski}}{{Bond} \&
  {Miszalski}}{2018}]{2018PASP..130i4201B}
{Bond} H.~E.,  {Miszalski} B.,  2018, \mn@doi [\pasp]
  {10.1088/1538-3873/aace3e}, \href
  {https://ui.adsabs.harvard.edu/abs/2018PASP..130i4201B} {130, 094201}

\bibitem[\protect\citeauthoryear{{B{\"o}ttcher}, {Reimer}, {Sweeney}  \&
  {Prakash}}{{B{\"o}ttcher} et~al.}{2013}]{2013ApJ...768...54B}
{B{\"o}ttcher} M.,  {Reimer} A.,  {Sweeney} K.,   {Prakash} A.,  2013, \mn@doi
  [\apj] {10.1088/0004-637X/768/1/54}, \href
  {https://ui.adsabs.harvard.edu/abs/2013ApJ...768...54B} {768, 54}

\bibitem[\protect\citeauthoryear{{Brickhouse}, {Desai}, {Hoogerwerf}, {Liedahl}
   \& {Smith}}{{Brickhouse} et~al.}{2005}]{2005AIPC..774..405B}
{Brickhouse} N.~S.,  {Desai} P.,  {Hoogerwerf} R.,  {Liedahl} D.~A.,   {Smith}
  R.~K.,  2005, in {Smith} R.,  ed.,  American Institute of Physics Conference
  Series Vol. 774, X-ray Diagnostics of Astrophysical Plasmas: Theory,
  Experiment, and Observation. pp 405--407, \mn@doi{10.1063/1.1960961}

\bibitem[\protect\citeauthoryear{{Brinkman} et~al.,}{{Brinkman}
  et~al.}{2000}]{2000ApJ...530L.111B}
{Brinkman} A.~C.,  et~al., 2000, \mn@doi [\apjl] {10.1086/312504}, \href
  {https://ui.adsabs.harvard.edu/abs/2000ApJ...530L.111B} {530, L111}

\bibitem[\protect\citeauthoryear{{Bruel}, {Burnett}, {Digel}, {Johannesson},
  {Omodei}  \& {Wood}}{{Bruel} et~al.}{2018}]{2018arXiv181011394B}
{Bruel} P.,  {Burnett} T.~H.,  {Digel} S.~W.,  {Johannesson} G.,  {Omodei} N.,
   {Wood} M.,  2018, arXiv e-prints, \href
  {https://ui.adsabs.harvard.edu/abs/2018arXiv181011394B} {p. arXiv:1810.11394}

\bibitem[\protect\citeauthoryear{{Budding} \& {Demircan}}{{Budding} \&
  {Demircan}}{2007}]{2007iap..book.....B}
{Budding} E.,  {Demircan} O.,  2007, {Introduction to Astronomical Photometry}.
 Cambridge observing handbooks for research astronomers Vol. 6

\bibitem[\protect\citeauthoryear{{Buson}, {Jean}  \& {Cheung}}{{Buson}
  et~al.}{2019}]{2019ATel13114....1B}
{Buson} S.,  {Jean} P.,   {Cheung} C.~C.,  2019, The Astronomer's Telegram,
  \href {https://ui.adsabs.harvard.edu/abs/2019ATel13114....1B} {13114, 1}

\bibitem[\protect\citeauthoryear{{Buson}, {Cheung}  \& {Jean}}{{Buson}
  et~al.}{2021}]{2021ATel14658....1B}
{Buson} S.,  {Cheung} C.~C.,   {Jean} P.,  2021, The Astronomer's Telegram,
  \href {https://ui.adsabs.harvard.edu/abs/2021ATel14658....1B} {14658, 1}

\bibitem[\protect\citeauthoryear{{Byckling}, {Mukai}, {Thorstensen}  \&
  {Osborne}}{{Byckling} et~al.}{2010}]{2010MNRAS.408.2298B}
{Byckling} K.,  {Mukai} K.,  {Thorstensen} J.~R.,   {Osborne} J.~P.,  2010,
  \mn@doi [\mnras] {10.1111/j.1365-2966.2010.17276.x}, \href
  {https://ui.adsabs.harvard.edu/abs/2010MNRAS.408.2298B} {408, 2298}

\bibitem[\protect\citeauthoryear{{Caprioli} \& {Spitkovsky}}{{Caprioli} \&
  {Spitkovsky}}{2014}]{2014ApJ...783...91C}
{Caprioli} D.,  {Spitkovsky} A.,  2014, \mn@doi [\apj]
  {10.1088/0004-637X/783/2/91}, \href
  {https://ui.adsabs.harvard.edu/abs/2014ApJ...783...91C} {783, 91}

\bibitem[\protect\citeauthoryear{{Carr}, {Said}, {Davis}, {Lidman}  \&
  {Tucker}}{{Carr} et~al.}{2020}]{2020ATel13874....1C}
{Carr} A.,  {Said} K.,  {Davis} T.~M.,  {Lidman} C.,   {Tucker} B.~E.,  2020,
  The Astronomer's Telegram, \href
  {https://ui.adsabs.harvard.edu/abs/2020ATel13874....1C} {13874, 1}

\bibitem[\protect\citeauthoryear{{Cerruti}}{{Cerruti}}{2020}]{2020Galax...8...72C}
{Cerruti} M.,  2020, \mn@doi [Galaxies] {10.3390/galaxies8040072}, \href
  {https://ui.adsabs.harvard.edu/abs/2020Galax...8...72C} {8, 72}

\bibitem[\protect\citeauthoryear{{Cheung}, {Shore}, {De Gennaro Aquino},
  {Charbonnel}, {Edlin}, {Hays}, {Corbet}  \& {Wood}}{{Cheung}
  et~al.}{2012}]{2012ATel.4310....1C}
{Cheung} C.~C.,  {Shore} S.~N.,  {De Gennaro Aquino} I.,  {Charbonnel} S.,
  {Edlin} J.,  {Hays} E.,  {Corbet} R.~H.~D.,   {Wood} D.~L.,  2012, The
  Astronomer's Telegram, \href
  {https://ui.adsabs.harvard.edu/abs/2012ATel.4310....1C} {4310, 1}

\bibitem[\protect\citeauthoryear{{Cheung} et~al.,}{{Cheung}
  et~al.}{2016}]{2016ApJ...826..142C}
{Cheung} C.~C.,  et~al., 2016, \mn@doi [\apj] {10.3847/0004-637X/826/2/142},
  \href {https://ui.adsabs.harvard.edu/abs/2016ApJ...826..142C} {826, 142}

\bibitem[\protect\citeauthoryear{{Chlebowski} \& {Garmany}}{{Chlebowski} \&
  {Garmany}}{1991}]{1991ApJ...368..241C}
{Chlebowski} T.,  {Garmany} C.~D.,  1991, \mn@doi [\apj] {10.1086/169687},
  \href {https://ui.adsabs.harvard.edu/abs/1991ApJ...368..241C} {368, 241}

\bibitem[\protect\citeauthoryear{{Chochol}, {Shugarov}, {Hamb{\'a}lek},
  {Skopal}, {Parimucha}  \& {Dubovsk{\'y}}}{{Chochol}
  et~al.}{2021}]{2020arXiv200713337C}
{Chochol} D.,  {Shugarov} S.,  {Hamb{\'a}lek} {\v{L}}.,  {Skopal} A.,
  {Parimucha} {\v{S}}.,   {Dubovsk{\'y}} P.,  2021, in The Golden Age of
  Cataclysmic Variables and Related Objects V. 2-7 September 2019. Palermo.
  p.~29 (\mn@eprint {arXiv} {2007.13337})

\bibitem[\protect\citeauthoryear{{Chomiuk} et~al.,}{{Chomiuk}
  et~al.}{2012}]{2012ApJ...761..173C}
{Chomiuk} L.,  et~al., 2012, \mn@doi [\apj] {10.1088/0004-637X/761/2/173},
  \href {https://ui.adsabs.harvard.edu/abs/2012ApJ...761..173C} {761, 173}

\bibitem[\protect\citeauthoryear{{Chomiuk} et~al.,}{{Chomiuk}
  et~al.}{2014a}]{2014Natur.514..339C}
{Chomiuk} L.,  et~al., 2014a, \mn@doi [\nat] {10.1038/nature13773}, \href
  {https://ui.adsabs.harvard.edu/abs/2014Natur.514..339C} {514, 339}

\bibitem[\protect\citeauthoryear{{Chomiuk} et~al.,}{{Chomiuk}
  et~al.}{2014b}]{2014ApJ...788..130C}
{Chomiuk} L.,  et~al., 2014b, \mn@doi [\apj] {10.1088/0004-637X/788/2/130},
  \href {https://ui.adsabs.harvard.edu/abs/2014ApJ...788..130C} {788, 130}

\bibitem[\protect\citeauthoryear{{Chomiuk}, {Metzger}  \& {Shen}}{{Chomiuk}
  et~al.}{2021}]{2020arXiv201108751C}
{Chomiuk} L.,  {Metzger} B.~D.,   {Shen} K.~J.,  2021, \mn@doi [\araa]
  {10.1146/annurev-astro-112420-114502}, \href
  {https://ui.adsabs.harvard.edu/abs/2021ARA&A..59..391C} {59}

\bibitem[\protect\citeauthoryear{{Collazzi}, {Schaefer}, {Xiao}, {Pagnotta},
  {Kroll}, {L{\"o}chel}  \& {Henden}}{{Collazzi}
  et~al.}{2009}]{2009AJ....138.1846C}
{Collazzi} A.~C.,  {Schaefer} B.~E.,  {Xiao} L.,  {Pagnotta} A.,  {Kroll} P.,
  {L{\"o}chel} K.,   {Henden} A.~A.,  2009, \mn@doi [\aj]
  {10.1088/0004-6256/138/6/1846}, \href
  {https://ui.adsabs.harvard.edu/abs/2009AJ....138.1846C} {138, 1846}

\bibitem[\protect\citeauthoryear{{Coppejans} \& {Knigge}}{{Coppejans} \&
  {Knigge}}{2020}]{2020NewAR..8901540C}
{Coppejans} D.~L.,  {Knigge} C.,  2020, \mn@doi [\nar]
  {10.1016/j.newar.2020.101540}, \href
  {https://ui.adsabs.harvard.edu/abs/2020NewAR..8901540C} {89, 101540}

\bibitem[\protect\citeauthoryear{{Coppejans} et~al.,}{{Coppejans}
  et~al.}{2016}]{2016MNRAS.463.2229C}
{Coppejans} D.~L.,  et~al., 2016, \mn@doi [\mnras] {10.1093/mnras/stw2133},
  \href {https://ui.adsabs.harvard.edu/abs/2016MNRAS.463.2229C} {463, 2229}

\bibitem[\protect\citeauthoryear{{Cowley} \& {MacConnell}}{{Cowley} \&
  {MacConnell}}{1972}]{1972ApJ...176L..27C}
{Cowley} A.~P.,  {MacConnell} D.~J.,  1972, \mn@doi [\apjl] {10.1086/181013},
  \href {https://ui.adsabs.harvard.edu/abs/1972ApJ...176L..27C} {176, L27}

\bibitem[\protect\citeauthoryear{{Craig} et~al.,}{{Craig}
  et~al.}{2011}]{2011SPIE.8147E..0HC}
{Craig} W.~W.,  et~al., 2011, in {O'Dell} S.~L.,  {Pareschi} G.,  eds,  Society
  of Photo-Optical Instrumentation Engineers (SPIE) Conference Series Vol.
  8147, Society of Photo-Optical Instrumentation Engineers (SPIE) Conference
  Series. p. 81470H, \mn@doi{10.1117/12.895278}

\bibitem[\protect\citeauthoryear{{Darnley} \& {Starrfield}}{{Darnley} \&
  {Starrfield}}{2018}]{2018RNAAS...2...24D}
{Darnley} M.~J.,  {Starrfield} S.,  2018, \mn@doi [Research Notes of the
  American Astronomical Society] {10.3847/2515-5172/aac26c}, \href
  {https://ui.adsabs.harvard.edu/abs/2018RNAAS...2...24D} {2, 24}

\bibitem[\protect\citeauthoryear{{Darnley} et~al.,}{{Darnley}
  et~al.}{2017}]{2017ApJ...847...35D}
{Darnley} M.~J.,  et~al., 2017, \mn@doi [\apj] {10.3847/1538-4357/aa8867},
  \href {https://ui.adsabs.harvard.edu/abs/2017ApJ...847...35D} {847, 35}

\bibitem[\protect\citeauthoryear{{Darnley}, {Page}, {Beardmore}, {Henze}  \&
  {Starrfield}}{{Darnley} et~al.}{2018}]{2018ATel11905....1D}
{Darnley} M.~J.,  {Page} K.~L.,  {Beardmore} A.~P.,  {Henze} M.,   {Starrfield}
  S.,  2018, The Astronomer's Telegram, \href
  {https://ui.adsabs.harvard.edu/abs/2018ATel11905....1D} {11905, 1}

\bibitem[\protect\citeauthoryear{{Davis}, {Taylor}, {Bode}  \&
  {Porcas}}{{Davis} et~al.}{1988}]{1988IAUS..129..277D}
{Davis} R.~J.,  {Taylor} A.~R.,  {Bode} M.~F.,   {Porcas} R.~W.,  1988, in
  {Reid} M.~J.,  {Moran} J.~M.,  eds,  Proceedings of the 129th IAU Symposium,
  Cambridge, MA, May 10-15, 1987 Vol. 129, The Impact of VLBI on Astrophysics
  and Geophysics. p.~277

\bibitem[\protect\citeauthoryear{{De} et~al.,}{{De}
  et~al.}{2021}]{2021arXiv210104045D}
{De} K.,  et~al., 2021, \mn@doi [\apj] {10.3847/1538-4357/abeb75}, \href
  {https://ui.adsabs.harvard.edu/abs/2021ApJ...912...19D} {912, 19}

\bibitem[\protect\citeauthoryear{{Deeming}}{{Deeming}}{1975}]{1975Ap&SS..36..137D}
{Deeming} T.~J.,  1975, \mn@doi [\apss] {10.1007/BF00681947}, \href
  {https://ui.adsabs.harvard.edu/abs/1975Ap&SS..36..137D} {36, 137}

\bibitem[\protect\citeauthoryear{{Della Valle} \& {Izzo}}{{Della Valle} \&
  {Izzo}}{2020}]{2020A&ARv..28....3D}
{Della Valle} M.,  {Izzo} L.,  2020, \mn@doi [\aapr]
  {10.1007/s00159-020-0124-6}, \href
  {https://ui.adsabs.harvard.edu/abs/2020A&ARv..28....3D} {28, 3}

\bibitem[\protect\citeauthoryear{{Denisenko}}{{Denisenko}}{2020}]{2020ATel13829....1D}
{Denisenko} D.,  2020, The Astronomer's Telegram, \href
  {https://ui.adsabs.harvard.edu/abs/2020ATel13829....1D} {13829, 1}

\bibitem[\protect\citeauthoryear{{Derdzinski}, {Metzger}  \&
  {Lazzati}}{{Derdzinski} et~al.}{2017}]{2017MNRAS.469.1314D}
{Derdzinski} A.~M.,  {Metzger} B.~D.,   {Lazzati} D.,  2017, \mn@doi [\mnras]
  {10.1093/mnras/stx829}, \href
  {https://ui.adsabs.harvard.edu/abs/2017MNRAS.469.1314D} {469, 1314}

\bibitem[\protect\citeauthoryear{{Dhillon}}{{Dhillon}}{1996}]{1996ASSL..208....3D}
{Dhillon} V.~S.,  1996, {The Nova-like variables}.
p.~3, \mn@doi{10.1007/978-94-009-0325-8_1}

\bibitem[\protect\citeauthoryear{{Done}, {Gierli{\'n}ski}  \& {Kubota}}{{Done}
  et~al.}{2007}]{2007A&ARv..15....1D}
{Done} C.,  {Gierli{\'n}ski} M.,   {Kubota} A.,  2007, \mn@doi [\aapr]
  {10.1007/s00159-007-0006-1}, \href
  {https://ui.adsabs.harvard.edu/abs/2007A&ARv..15....1D} {15, 1}

\bibitem[\protect\citeauthoryear{{Drake} et~al.,}{{Drake}
  et~al.}{2009}]{2009ApJ...696..870D}
{Drake} A.~J.,  et~al., 2009, \mn@doi [\apj] {10.1088/0004-637X/696/1/870},
  \href {https://ui.adsabs.harvard.edu/abs/2009ApJ...696..870D} {696, 870}

\bibitem[\protect\citeauthoryear{{Drake}, {Orio}, {Bearmore}, {Behar}, {Luna}
  \& {Ness}}{{Drake} et~al.}{2020}]{2020ATel14214....1D}
{Drake} J.~J.,  {Orio} M.,  {Bearmore} A.,  {Behar} E.,  {Luna} G. J.~M.,
  {Ness} J.-U.,  2020, The Astronomer's Telegram, \href
  {https://ui.adsabs.harvard.edu/abs/2020ATel14214....1D} {14214, 1}

\bibitem[\protect\citeauthoryear{{Dubus}, {Guillard}, {Petrucci}  \&
  {Martin}}{{Dubus} et~al.}{2017}]{2017A&A...608A..59D}
{Dubus} G.,  {Guillard} N.,  {Petrucci} P.-O.,   {Martin} P.,  2017, \mn@doi
  [\aap] {10.1051/0004-6361/201731084}, \href
  {https://ui.adsabs.harvard.edu/abs/2017A&A...608A..59D} {608, A59}

\bibitem[\protect\citeauthoryear{{Dwyer}}{{Dwyer}}{2012}]{2012JGRA..117.2308D}
{Dwyer} J.~R.,  2012, \mn@doi [Journal of Geophysical Research (Space Physics)]
  {10.1029/2011JA017160}, \href
  {https://ui.adsabs.harvard.edu/abs/2012JGRA..117.2308D} {117, A02308}

\bibitem[\protect\citeauthoryear{{Dwyer}, {Smith}  \& {Cummer}}{{Dwyer}
  et~al.}{2012}]{2012SSRv..173..133D}
{Dwyer} J.~R.,  {Smith} D.~M.,   {Cummer} S.~A.,  2012, \mn@doi [\ssr]
  {10.1007/s11214-012-9894-0}, \href
  {https://ui.adsabs.harvard.edu/abs/2012SSRv..173..133D} {173, 133}

\bibitem[\protect\citeauthoryear{{Dyson} \& {Williams}}{{Dyson} \&
  {Williams}}{1997}]{1997pism.book.....D}
{Dyson} J.~E.,  {Williams} D.~A.,  1997, {The physics of the interstellar
  medium}, \mn@doi{10.1201/9780585368115.
}

\bibitem[\protect\citeauthoryear{{Evans}, {Beardmore}, {Osborne}  \&
  {Wynn}}{{Evans} et~al.}{2009}]{2009MNRAS.399.1167E}
{Evans} P.~A.,  {Beardmore} A.~P.,  {Osborne} J.~P.,   {Wynn} G.~A.,  2009,
  \mn@doi [\mnras] {10.1111/j.1365-2966.2009.15376.x}, \href
  {https://ui.adsabs.harvard.edu/abs/2009MNRAS.399.1167E} {399, 1167}

\bibitem[\protect\citeauthoryear{{Fang}, {Metzger}, {Vurm}, {Aydi}  \&
  {Chomiuk}}{{Fang} et~al.}{2020}]{2020ApJ...904....4F}
{Fang} K.,  {Metzger} B.~D.,  {Vurm} I.,  {Aydi} E.,   {Chomiuk} L.,  2020,
  \mn@doi [\apj] {10.3847/1538-4357/abbc6e}, \href
  {https://ui.adsabs.harvard.edu/abs/2020ApJ...904....4F} {904, 4}

\bibitem[\protect\citeauthoryear{{Ferlet}, {Vidal-Madjar}  \& {Gry}}{{Ferlet}
  et~al.}{1985}]{1985ApJ...298..838F}
{Ferlet} R.,  {Vidal-Madjar} A.,   {Gry} C.,  1985, \mn@doi [\apj]
  {10.1086/163666}, \href
  {https://ui.adsabs.harvard.edu/abs/1985ApJ...298..838F} {298, 838}

\bibitem[\protect\citeauthoryear{{Fields}, {Farmer}, {Petermann}, {Iliadis}  \&
  {Timmes}}{{Fields} et~al.}{2016}]{2016ApJ...823...46F}
{Fields} C.~E.,  {Farmer} R.,  {Petermann} I.,  {Iliadis} C.,   {Timmes} F.~X.,
   2016, \mn@doi [\apj] {10.3847/0004-637X/823/1/46}, \href
  {https://ui.adsabs.harvard.edu/abs/2016ApJ...823...46F} {823, 46}

\bibitem[\protect\citeauthoryear{{Figueira}, {Jos{\'e}}, {Garc{\'\i}a-Berro},
  {Campbell}, {Garc{\'\i}a-Senz}  \& {Mohamed}}{{Figueira}
  et~al.}{2018}]{2018A&A...613A...8F}
{Figueira} J.,  {Jos{\'e}} J.,  {Garc{\'\i}a-Berro} E.,  {Campbell} S.~W.,
  {Garc{\'\i}a-Senz} D.,   {Mohamed} S.,  2018, \mn@doi [\aap]
  {10.1051/0004-6361/201731545}, \href
  {https://ui.adsabs.harvard.edu/abs/2018A&A...613A...8F} {613, A8}

\bibitem[\protect\citeauthoryear{{Figuera Jaimes}, {Arellano Ferro}, {Bramich},
  {Giridhar}  \& {Kuppuswamy}}{{Figuera Jaimes}
  et~al.}{2013}]{2013A&A...556A..20F}
{Figuera Jaimes} R.,  {Arellano Ferro} A.,  {Bramich} D.~M.,  {Giridhar} S.,
  {Kuppuswamy} K.,  2013, \mn@doi [\aap] {10.1051/0004-6361/201220824}, \href
  {https://ui.adsabs.harvard.edu/abs/2013A&A...556A..20F} {556, A20}

\bibitem[\protect\citeauthoryear{{Finzell} et~al.,}{{Finzell}
  et~al.}{2018}]{2018ApJ...852..108F}
{Finzell} T.,  et~al., 2018, \mn@doi [\apj] {10.3847/1538-4357/aaa12a}, \href
  {https://ui.adsabs.harvard.edu/abs/2018ApJ...852..108F} {852, 108}

\bibitem[\protect\citeauthoryear{{Franckowiak}, {Jean}, {Wood}, {Cheung}  \&
  {Buson}}{{Franckowiak} et~al.}{2018}]{2018A&A...609A.120F}
{Franckowiak} A.,  {Jean} P.,  {Wood} M.,  {Cheung} C.~C.,   {Buson} S.,  2018,
  \mn@doi [\aap] {10.1051/0004-6361/201731516}, \href
  {http://adsabs.harvard.edu/abs/2018A%26A...609A.120F} {609, A120}

\bibitem[\protect\citeauthoryear{{Frank}, {King}  \& {Raine}}{{Frank}
  et~al.}{2002}]{2002apa..book.....F}
{Frank} J.,  {King} A.,   {Raine} D.~J.,  2002, {Accretion Power in
  Astrophysics: Third Edition}

\bibitem[\protect\citeauthoryear{{Gaggero}, {Grasso}, {Marinelli}, {Urbano}  \&
  {Valli}}{{Gaggero} et~al.}{2015}]{2015ApJ...815L..25G}
{Gaggero} D.,  {Grasso} D.,  {Marinelli} A.,  {Urbano} A.,   {Valli} M.,  2015,
  \mn@doi [\apjl] {10.1088/2041-8205/815/2/L25}, \href
  {https://ui.adsabs.harvard.edu/abs/2015ApJ...815L..25G} {815, L25}

\bibitem[\protect\citeauthoryear{{Gaia Collaboration} et~al.,}{{Gaia
  Collaboration} et~al.}{2018}]{2018A&A...616A...1G}
{Gaia Collaboration} et~al., 2018, \mn@doi [\aap]
  {10.1051/0004-6361/201833051}, \href
  {https://ui.adsabs.harvard.edu/abs/2018A&A...616A...1G} {616, A1}

\bibitem[\protect\citeauthoryear{{Galan} \& {Mikolajewska}}{{Galan} \&
  {Mikolajewska}}{2020}]{2020ATel14149....1G}
{Galan} C.,  {Mikolajewska} J.,  2020, The Astronomer's Telegram, \href
  {https://ui.adsabs.harvard.edu/abs/2020ATel14149....1G} {14149, 1}

\bibitem[\protect\citeauthoryear{{Geballe} et~al.,}{{Geballe}
  et~al.}{2019}]{2019ApJ...886L..14G}
{Geballe} T.~R.,  et~al., 2019, \mn@doi [\apjl] {10.3847/2041-8213/ab5310},
  \href {https://ui.adsabs.harvard.edu/abs/2019ApJ...886L..14G} {886, L14}

\bibitem[\protect\citeauthoryear{{Gehrels}}{{Gehrels}}{1997}]{1997NCimB.112...11G}
{Gehrels} N.,  1997, Nuovo Cimento B Serie, \href
  {https://ui.adsabs.harvard.edu/abs/1997NCimB.112...11G} {112B, 11}

\bibitem[\protect\citeauthoryear{{Gehrz}, {Truran}, {Williams}  \&
  {Starrfield}}{{Gehrz} et~al.}{1998}]{1998PASP..110....3G}
{Gehrz} R.~D.,  {Truran} J.~W.,  {Williams} R.~E.,   {Starrfield} S.,  1998,
  \mn@doi [\pasp] {10.1086/316107}, \href
  {http://adsabs.harvard.edu/abs/1998PASP..110....3G} {110, 3}

\bibitem[\protect\citeauthoryear{{Ginzburg} \& {Quataert}}{{Ginzburg} \&
  {Quataert}}{2021}]{2021MNRAS.507..475G}
{Ginzburg} S.,  {Quataert} E.,  2021, \mn@doi [\mnras]
  {10.1093/mnras/stab2170}, \href
  {https://ui.adsabs.harvard.edu/abs/2021MNRAS.507..475G} {507, 475}

\bibitem[\protect\citeauthoryear{{Giroletti} et~al.,}{{Giroletti}
  et~al.}{2020}]{2020A&A...638A.130G}
{Giroletti} M.,  et~al., 2020, \mn@doi [\aap] {10.1051/0004-6361/202038142},
  \href {https://ui.adsabs.harvard.edu/abs/2020A&A...638A.130G} {638, A130}

\bibitem[\protect\citeauthoryear{{Gomez-Gomar}, {Hernanz}, {Jose}  \&
  {Isern}}{{Gomez-Gomar} et~al.}{1998}]{1998MNRAS.296..913G}
{Gomez-Gomar} J.,  {Hernanz} M.,  {Jose} J.,   {Isern} J.,  1998, \mn@doi
  [\mnras] {10.1046/j.1365-8711.1998.01421.x}, \href
  {https://ui.adsabs.harvard.edu/abs/1998MNRAS.296..913G} {296, 913}

\bibitem[\protect\citeauthoryear{{Gordon}, {Aydi}, {Page}, {Li}, {Chomiuk},
  {Sokolovsky}, {Mukai}  \& {Seitz}}{{Gordon}
  et~al.}{2021}]{2021ApJ...910..134G}
{Gordon} A.~C.,  {Aydi} E.,  {Page} K.~L.,  {Li} K.-L.,  {Chomiuk} L.,
  {Sokolovsky} K.~V.,  {Mukai} K.,   {Seitz} J.,  2021, \mn@doi [\apj]
  {10.3847/1538-4357/abe547}, \href
  {https://ui.adsabs.harvard.edu/abs/2021ApJ...910..134G} {910, 134}

\bibitem[\protect\citeauthoryear{Gough}{Gough}{2009}]{gough2009gnu}
Gough B.,  2009, GNU scientific library reference manual.
Network Theory Ltd.

\bibitem[\protect\citeauthoryear{{Greiner}}{{Greiner}}{2000}]{2000NewAR..44..149G}
{Greiner} J.,  2000, \mn@doi [\nar] {10.1016/S1387-6473(00)00029-4}, \href
  {https://ui.adsabs.harvard.edu/abs/2000NewAR..44..149G} {44, 149}

\bibitem[\protect\citeauthoryear{{Greiner} \& {Teeseling}}{{Greiner} \&
  {Teeseling}}{1998}]{1998A&A...339L..21G}
{Greiner} J.,  {Teeseling} A.,  1998, \aap, \href
  {https://ui.adsabs.harvard.edu/abs/1998A&A...339L..21G} {339, L21}

\bibitem[\protect\citeauthoryear{{Greiner}, {Tovmassian}, {Di Stefano},
  {Prestwich}, {Gonz{\'a}lez-Riestra}, {Szentasko}  \&
  {Chavarr{\'\i}a}}{{Greiner} et~al.}{1999}]{1999A&A...343..183G}
{Greiner} J.,  {Tovmassian} G.~H.,  {Di Stefano} R.,  {Prestwich} A.,
  {Gonz{\'a}lez-Riestra} R.,  {Szentasko} L.,   {Chavarr{\'\i}a} C.,  1999,
  \aap, \href {https://ui.adsabs.harvard.edu/abs/1999A&A...343..183G} {343,
  183}

\bibitem[\protect\citeauthoryear{{Greiner} et~al.,}{{Greiner}
  et~al.}{2001}]{2001A&A...376.1031G}
{Greiner} J.,  et~al., 2001, \mn@doi [\aap] {10.1051/0004-6361:20011048}, \href
  {https://ui.adsabs.harvard.edu/abs/2001A&A...376.1031G} {376, 1031}

\bibitem[\protect\citeauthoryear{{Greiner}, {Schwarz}, {Tappert}, {Mennickent},
  {Reinsch}  \& {Sala}}{{Greiner} et~al.}{2010}]{2010AN....331..227G}
{Greiner} J.,  {Schwarz} R.,  {Tappert} C.,  {Mennickent} R.~E.,  {Reinsch} K.,
    {Sala} G.,  2010, \mn@doi [Astronomische Nachrichten]
  {10.1002/asna.200911332}, \href
  {https://ui.adsabs.harvard.edu/abs/2010AN....331..227G} {331, 227}

\bibitem[\protect\citeauthoryear{{Guainazzi} \& {Bianchi}}{{Guainazzi} \&
  {Bianchi}}{2007}]{2007MNRAS.374.1290G}
{Guainazzi} M.,  {Bianchi} S.,  2007, \mn@doi [\mnras]
  {10.1111/j.1365-2966.2006.11229.x}, \href
  {https://ui.adsabs.harvard.edu/abs/2007MNRAS.374.1290G} {374, 1290}

\bibitem[\protect\citeauthoryear{{G{\"u}del} \& {Naz{\'e}}}{{G{\"u}del} \&
  {Naz{\'e}}}{2009}]{2009A&ARv..17..309G}
{G{\"u}del} M.,  {Naz{\'e}} Y.,  2009, \mn@doi [\aapr]
  {10.1007/s00159-009-0022-4}, \href
  {https://ui.adsabs.harvard.edu/abs/2009A&ARv..17..309G} {17, 309}

\bibitem[\protect\citeauthoryear{{Gulati}, {Murphy}, {Wang}, {Leung},
  {Pritchard}, {Lenc}  \& {Kaplan}}{{Gulati}
  et~al.}{2022}]{2022ATel15264....1G}
{Gulati} A.,  {Murphy} T.,  {Wang} Y.,  {Leung} J.,  {Pritchard} J.,  {Lenc}
  E.,   {Kaplan} D.,  2022, The Astronomer's Telegram, \href
  {https://ui.adsabs.harvard.edu/abs/2022ATel15264....1G} {15264, 1}

\bibitem[\protect\citeauthoryear{{Gurevich}, {Milikh}  \&
  {Roussel-Dupre}}{{Gurevich} et~al.}{1992}]{1992PhLA..165..463G}
{Gurevich} A.~V.,  {Milikh} G.~M.,   {Roussel-Dupre} R.,  1992, \mn@doi
  [Physics Letters A] {10.1016/0375-9601(92)90348-P}, \href
  {https://ui.adsabs.harvard.edu/abs/1992PhLA..165..463G} {165, 463}

\bibitem[\protect\citeauthoryear{{G{\"u}ver} \& {{\"O}zel}}{{G{\"u}ver} \&
  {{\"O}zel}}{2009}]{2009MNRAS.400.2050G}
{G{\"u}ver} T.,  {{\"O}zel} F.,  2009, \mn@doi [\mnras]
  {10.1111/j.1365-2966.2009.15598.x}, \href
  {https://ui.adsabs.harvard.edu/abs/2009MNRAS.400.2050G} {400, 2050}

\bibitem[\protect\citeauthoryear{{Hameury}}{{Hameury}}{2020}]{2020AdSpR..66.1004H}
{Hameury} J.~M.,  2020, \mn@doi [Advances in Space Research]
  {10.1016/j.asr.2019.10.022}, \href
  {https://ui.adsabs.harvard.edu/abs/2020AdSpR..66.1004H} {66, 1004}

\bibitem[\protect\citeauthoryear{{Hameury} \& {Lasota}}{{Hameury} \&
  {Lasota}}{2002}]{2002A&A...394..231H}
{Hameury} J.~M.,  {Lasota} J.~P.,  2002, \mn@doi [\aap]
  {10.1051/0004-6361:20021136}, \href
  {https://ui.adsabs.harvard.edu/abs/2002A&A...394..231H} {394, 231}

\bibitem[\protect\citeauthoryear{{Harrison} et~al.,}{{Harrison}
  et~al.}{2013}]{2013ApJ...770..103H}
{Harrison} F.~A.,  et~al., 2013, \mn@doi [\apj] {10.1088/0004-637X/770/2/103},
  \href {https://ui.adsabs.harvard.edu/abs/2013ApJ...770..103H} {770, 103}

\bibitem[\protect\citeauthoryear{{Harvey}, {Redman}, {Boumis}  \&
  {Akras}}{{Harvey} et~al.}{2016}]{2016A&A...595A..64H}
{Harvey} E.,  {Redman} M.~P.,  {Boumis} P.,   {Akras} S.,  2016, \mn@doi [\aap]
  {10.1051/0004-6361/201628132}, \href
  {https://ui.adsabs.harvard.edu/abs/2016A&A...595A..64H} {595, A64}

\bibitem[\protect\citeauthoryear{{Hasinger}}{{Hasinger}}{1994}]{1994RvMA....7..129H}
{Hasinger} G.,  1994, Reviews in Modern Astronomy, \href
  {https://ui.adsabs.harvard.edu/abs/1994RvMA....7..129H} {7, 129}

\bibitem[\protect\citeauthoryear{{Heirtzler}}{{Heirtzler}}{2002}]{2002JASTP..64.1701H}
{Heirtzler} J.,  2002, \mn@doi [Journal of Atmospheric and Solar-Terrestrial
  Physics] {10.1016/S1364-6826(02)00120-7}, \href
  {https://ui.adsabs.harvard.edu/abs/2002JASTP..64.1701H} {64, 1701}

\bibitem[\protect\citeauthoryear{{Hellier}}{{Hellier}}{2001}]{2001cvs..book.....H}
{Hellier} C.,  2001, {Cataclysmic Variable Stars}

\bibitem[\protect\citeauthoryear{{Hellier} \& {Naylor}}{{Hellier} \&
  {Naylor}}{1998}]{1998MNRAS.295L..50H}
{Hellier} C.,  {Naylor} T.,  1998, \mn@doi [\mnras]
  {10.1046/j.1365-8711.1998.01556.x}, \href
  {https://ui.adsabs.harvard.edu/abs/1998MNRAS.295L..50H} {295, L50}

\bibitem[\protect\citeauthoryear{{Helton} et~al.,}{{Helton}
  et~al.}{2012}]{2012ApJ...755...37H}
{Helton} L.~A.,  et~al., 2012, \mn@doi [\apj] {10.1088/0004-637X/755/1/37},
  \href {https://ui.adsabs.harvard.edu/abs/2012ApJ...755...37H} {755, 37}

\bibitem[\protect\citeauthoryear{{Hernanz}}{{Hernanz}}{2014}]{2014ASPC..490..319H}
{Hernanz} M.,  2014, in {Woudt} P.~A.,  {Ribeiro} V.~A.~R.~M.,  eds,
  Astronomical Society of the Pacific Conference Series Vol. 490, Stellar
  Novae: Past and Future Decades. p.~319 (\mn@eprint {arXiv} {1305.0769})

\bibitem[\protect\citeauthoryear{{Hernanz} et~al.,}{{Hernanz}
  et~al.}{2002}]{2002AIPC..637..435H}
{Hernanz} M.,  et~al., 2002, in {Hernanz} M.,  {Jos{\'e}} J.,  eds,  American
  Institute of Physics Conference Series Vol. 637, Classical Nova Explosions.
  pp 435--439, \mn@doi{10.1063/1.1518243}

\bibitem[\protect\citeauthoryear{{Hillman}, {Shara}, {Prialnik}  \&
  {Kovetz}}{{Hillman} et~al.}{2020}]{2020NatAs...4..886H}
{Hillman} Y.,  {Shara} M.~M.,  {Prialnik} D.,   {Kovetz} A.,  2020, \mn@doi
  [Nature Astronomy] {10.1038/s41550-020-1062-y}, \href
  {https://ui.adsabs.harvard.edu/abs/2020NatAs...4..886H} {4, 886}

\bibitem[\protect\citeauthoryear{{Hoffmeister}, {Richter}  \&
  {Wenzel}}{{Hoffmeister} et~al.}{1984}]{1984vest.book.....H}
{Hoffmeister} C.,  {Richter} G.,   {Wenzel} W.,  1984, {Veraenderliche Sterne}

\bibitem[\protect\citeauthoryear{{Honeycutt}}{{Honeycutt}}{2001}]{2001PASP..113..473H}
{Honeycutt} R.~K.,  2001, \mn@doi [\pasp] {10.1086/319543}, \href
  {https://ui.adsabs.harvard.edu/abs/2001PASP..113..473H} {113, 473}

\bibitem[\protect\citeauthoryear{{Honeycutt} \& {Kafka}}{{Honeycutt} \&
  {Kafka}}{2004}]{2004AJ....128.1279H}
{Honeycutt} R.~K.,  {Kafka} S.,  2004, \mn@doi [\aj] {10.1086/422737}, \href
  {https://ui.adsabs.harvard.edu/abs/2004AJ....128.1279H} {128, 1279}

\bibitem[\protect\citeauthoryear{{Honeycutt}, {Robertson}  \&
  {Turner}}{{Honeycutt} et~al.}{1998}]{1998AJ....115.2527H}
{Honeycutt} R.~K.,  {Robertson} J.~W.,   {Turner} G.~W.,  1998, \mn@doi [\aj]
  {10.1086/300381}, \href
  {https://ui.adsabs.harvard.edu/abs/1998AJ....115.2527H} {115, 2527}

\bibitem[\protect\citeauthoryear{{Honeycutt}, {Robertson}  \&
  {Kafka}}{{Honeycutt} et~al.}{2011}]{2011AJ....141..121H}
{Honeycutt} R.~K.,  {Robertson} J.~W.,   {Kafka} S.,  2011, \mn@doi [\aj]
  {10.1088/0004-6256/141/4/121}, \href
  {https://ui.adsabs.harvard.edu/abs/2011AJ....141..121H} {141, 121}

\bibitem[\protect\citeauthoryear{{Howell}, {Nelson}  \& {Rappaport}}{{Howell}
  et~al.}{2001}]{2001ApJ...550..897H}
{Howell} S.~B.,  {Nelson} L.~A.,   {Rappaport} S.,  2001, \mn@doi [\apj]
  {10.1086/319776}, \href
  {https://ui.adsabs.harvard.edu/abs/2001ApJ...550..897H} {550, 897}

\bibitem[\protect\citeauthoryear{{Iijima} \& {Esenoglu}}{{Iijima} \&
  {Esenoglu}}{2003}]{2003A&A...404..997I}
{Iijima} T.,  {Esenoglu} H.~H.,  2003, \mn@doi [\aap]
  {10.1051/0004-6361:20030528}, \href
  {https://ui.adsabs.harvard.edu/abs/2003A&A...404..997I} {404, 997}

\bibitem[\protect\citeauthoryear{{Izzo} et~al.,}{{Izzo}
  et~al.}{2020}]{2020ATel14048....1I}
{Izzo} L.,  et~al., 2020, The Astronomer's Telegram, \href
  {https://ui.adsabs.harvard.edu/abs/2020ATel14048....1I} {14048, 1}

\bibitem[\protect\citeauthoryear{{Jose}}{{Jose}}{2016}]{2016stex.book.....J}
{Jose} J.,  2016, {Stellar Explosions: Hydrodynamics and Nucleosynthesis},
  \mn@doi{10.1201/b19165.
}

\bibitem[\protect\citeauthoryear{{Joye} \& {Mandel}}{{Joye} \&
  {Mandel}}{2003}]{2003ASPC..295..489J}
{Joye} W.~A.,  {Mandel} E.,  2003, {New Features of SAOImage DS9}.
p.~489

\bibitem[\protect\citeauthoryear{{Kafka}}{{Kafka}}{2021}]{AAVSODATA}
{Kafka} S.,  2021, {Observations from the AAVSO International Database,
  \url{https://www.aavso.org}}

\bibitem[\protect\citeauthoryear{{Kahabka} \& {van den Heuvel}}{{Kahabka} \&
  {van den Heuvel}}{1997}]{1997ARA&A..35...69K}
{Kahabka} P.,  {van den Heuvel} E.~P.~J.,  1997, \mn@doi [\araa]
  {10.1146/annurev.astro.35.1.69}, \href
  {https://ui.adsabs.harvard.edu/abs/1997ARA&A..35...69K} {35, 69}

\bibitem[\protect\citeauthoryear{{Kalberla}, {Burton}, {Hartmann}, {Arnal},
  {Bajaja}, {Morras}  \& {P{\"o}ppel}}{{Kalberla}
  et~al.}{2005}]{2005A&A...440..775K}
{Kalberla} P.~M.~W.,  {Burton} W.~B.,  {Hartmann} D.,  {Arnal} E.~M.,  {Bajaja}
  E.,  {Morras} R.,   {P{\"o}ppel} W.~G.~L.,  2005, \mn@doi [\aap]
  {10.1051/0004-6361:20041864}, \href
  {https://ui.adsabs.harvard.edu/abs/2005A&A...440..775K} {440, 775}

\bibitem[\protect\citeauthoryear{{Kato}}{{Kato}}{2010}]{2010AN....331..140K}
{Kato} M.,  2010, \mn@doi [Astronomische Nachrichten] {10.1002/asna.200911315},
  \href {https://ui.adsabs.harvard.edu/abs/2010AN....331..140K} {331, 140}

\bibitem[\protect\citeauthoryear{{Kato} \& {Hachisu}}{{Kato} \&
  {Hachisu}}{2003}]{2003ApJ...587L..39K}
{Kato} M.,  {Hachisu} I.,  2003, \mn@doi [\apjl] {10.1086/375043}, \href
  {https://ui.adsabs.harvard.edu/abs/2003ApJ...587L..39K} {587, L39}

\bibitem[\protect\citeauthoryear{{Kato} et~al.,}{{Kato}
  et~al.}{2016}]{2016ApJ...830...40K}
{Kato} M.,  et~al., 2016, \mn@doi [\apj] {10.3847/0004-637X/830/1/40}, \href
  {https://ui.adsabs.harvard.edu/abs/2016ApJ...830...40K} {830, 40}

\bibitem[\protect\citeauthoryear{{Kaufman} et~al.,}{{Kaufman}
  et~al.}{2020}]{2020CBET.4812....1K}
{Kaufman} R.,  et~al., 2020, Central Bureau Electronic Telegrams, 4812, 1

\bibitem[\protect\citeauthoryear{{Kazarovets}, {O'Meara}, {McNaught}, {Pearce},
  {de S. Aguiar}  \& {Souza}}{{Kazarovets} et~al.}{2020}]{2020CBET.4826....1K}
{Kazarovets} E.,  {O'Meara} S.~J.,  {McNaught} R.~H.,  {Pearce} A.,  {de S.
  Aguiar} J.~G.,   {Souza} W.,  2020, Central Bureau Electronic Telegrams,
  4826, 1

\bibitem[\protect\citeauthoryear{{Kellogg}, {Baldwin}  \& {Koch}}{{Kellogg}
  et~al.}{1975}]{1975ApJ...199..299K}
{Kellogg} E.,  {Baldwin} J.~R.,   {Koch} D.,  1975, \mn@doi [\apj]
  {10.1086/153692}, \href
  {https://ui.adsabs.harvard.edu/#abs/1975ApJ...199..299K} {199, 299}

\bibitem[\protect\citeauthoryear{{Kepler}, {Koester}  \& {Ourique}}{{Kepler}
  et~al.}{2016}]{2016Sci...352...67K}
{Kepler} S.~O.,  {Koester} D.,   {Ourique} G.,  2016, \mn@doi [Science]
  {10.1126/science.aad6705}, \href
  {https://ui.adsabs.harvard.edu/abs/2016Sci...352...67K} {352, 67}

\bibitem[\protect\citeauthoryear{{Kerr}}{{Kerr}}{2011}]{2011ApJ...732...38K}
{Kerr} M.,  2011, \mn@doi [\apj] {10.1088/0004-637X/732/1/38}, \href
  {https://ui.adsabs.harvard.edu/abs/2011ApJ...732...38K} {732, 38}

\bibitem[\protect\citeauthoryear{{Kilkenny}, {O'Donoghue}, {Worters}, {Koen},
  {Hambly}  \& {MacGillivray}}{{Kilkenny} et~al.}{2015}]{2015MNRAS.453.1879K}
{Kilkenny} D.,  {O'Donoghue} D.,  {Worters} H.~L.,  {Koen} C.,  {Hambly} N.,
  {MacGillivray} H.,  2015, \mn@doi [\mnras] {10.1093/mnras/stv1771}, \href
  {https://ui.adsabs.harvard.edu/abs/2015MNRAS.453.1879K} {453, 1879}

\bibitem[\protect\citeauthoryear{{Kitchin}}{{Kitchin}}{2009}]{2009aste.book.....K}
{Kitchin} C.~R.,  2009, {Astrophysical Techniques, Fifth Edition}

\bibitem[\protect\citeauthoryear{{Knigge}, {Baraffe}  \& {Patterson}}{{Knigge}
  et~al.}{2011}]{2011ApJS..194...28K}
{Knigge} C.,  {Baraffe} I.,   {Patterson} J.,  2011, \mn@doi [\apjs]
  {10.1088/0067-0049/194/2/28}, \href
  {https://ui.adsabs.harvard.edu/abs/2011ApJS..194...28K} {194, 28}

\bibitem[\protect\citeauthoryear{{Kochanek} et~al.,}{{Kochanek}
  et~al.}{2017}]{2017PASP..129j4502K}
{Kochanek} C.~S.,  et~al., 2017, \mn@doi [\pasp] {10.1088/1538-3873/aa80d9},
  \href {https://ui.adsabs.harvard.edu/abs/2017PASP..129j4502K} {129, 104502}

\bibitem[\protect\citeauthoryear{{Koester}}{{Koester}}{2009}]{2009A&A...498..517K}
{Koester} D.,  2009, \mn@doi [\aap] {10.1051/0004-6361/200811468}, \href
  {https://ui.adsabs.harvard.edu/abs/2009A&A...498..517K} {498, 517}

\bibitem[\protect\citeauthoryear{K{\"o}nig et~al.,}{K{\"o}nig
  et~al.}{2022}]{Konig2022}
K{\"o}nig O.,  et~al., 2022, \mn@doi [Nature] {10.1038/s41586-022-04635-y},
  605, 248

\bibitem[\protect\citeauthoryear{{Kraft}, {Burrows}  \& {Nousek}}{{Kraft}
  et~al.}{1991}]{1991ApJ...374..344K}
{Kraft} R.~P.,  {Burrows} D.~N.,   {Nousek} J.~A.,  1991, \mn@doi [\apj]
  {10.1086/170124}, \href
  {https://ui.adsabs.harvard.edu/abs/1991ApJ...374..344K} {374, 344}

\bibitem[\protect\citeauthoryear{{Krautter}}{{Krautter}}{2002}]{2002AIPC..637..345K}
{Krautter} J.,  2002, in {Hernanz} M.,  {Jos{\'e}} J.,  eds,  American
  Institute of Physics Conference Series Vol. 637, Classical Nova Explosions.
  pp 345--354, \mn@doi{10.1063/1.1518228}

\bibitem[\protect\citeauthoryear{{Krautter}}{{Krautter}}{2008}]{2008ASPC..401..139K}
{Krautter} J.,  2008, in {Evans} A.,  {Bode} M.~F.,  {O'Brien} T.~J.,
  {Darnley} M.~J.,  eds,  Astronomical Society of the Pacific Conference Series
  Vol. 401, RS Ophiuchi (2006) and the Recurrent Nova Phenomenon. p.~139

\bibitem[\protect\citeauthoryear{{Kumar} \& {Zhang}}{{Kumar} \&
  {Zhang}}{2015}]{2015PhR...561....1K}
{Kumar} P.,  {Zhang} B.,  2015, \mn@doi [\physrep]
  {10.1016/j.physrep.2014.09.008}, \href
  {https://ui.adsabs.harvard.edu/abs/2015PhR...561....1K} {561, 1}

\bibitem[\protect\citeauthoryear{{Leach}, {Hessman}, {King}, {Stehle}  \&
  {Mattei}}{{Leach} et~al.}{1999}]{1999MNRAS.305..225L}
{Leach} R.,  {Hessman} F.~V.,  {King} A.~R.,  {Stehle} R.,   {Mattei} J.,
  1999, \mn@doi [\mnras] {10.1046/j.1365-8711.1999.02450.x}, \href
  {https://ui.adsabs.harvard.edu/abs/1999MNRAS.305..225L} {305, 225}

\bibitem[\protect\citeauthoryear{{Li}}{{Li}}{2021}]{2021ATel14705....1L}
{Li} K.-L.,  2021, The Astronomer's Telegram, \href
  {https://ui.adsabs.harvard.edu/abs/2021ATel14705....1L} {14705, 1}

\bibitem[\protect\citeauthoryear{{Li} et~al.,}{{Li}
  et~al.}{2012}]{2012ApJ...761...99L}
{Li} K.~L.,  et~al., 2012, \mn@doi [\apj] {10.1088/0004-637X/761/2/99}, \href
  {https://ui.adsabs.harvard.edu/abs/2012ApJ...761...99L} {761, 99}

\bibitem[\protect\citeauthoryear{{Li} et~al.,}{{Li}
  et~al.}{2017}]{2017NatAs...1..697L}
{Li} K.-L.,  et~al., 2017, \mn@doi [Nature Astronomy]
  {10.1038/s41550-017-0222-1}, \href
  {http://adsabs.harvard.edu/abs/2017NatAs...1..697L} {1, 697}

\bibitem[\protect\citeauthoryear{{Li}, {Chomiuk}  \& {Strader}}{{Li}
  et~al.}{2018}]{2018ATel11590....1L}
{Li} K.-L.,  {Chomiuk} L.,   {Strader} J.,  2018, The Astronomer's Telegram,
  \href {https://ui.adsabs.harvard.edu/abs/2018ATel11590....1L} {11590, 1}

\bibitem[\protect\citeauthoryear{{Li} et~al.,}{{Li}
  et~al.}{2019}]{2019ATel13116....1L}
{Li} K.-L.,  et~al., 2019, The Astronomer's Telegram, \href
  {https://ui.adsabs.harvard.edu/abs/2019ATel13116....1L} {13116, 1}

\bibitem[\protect\citeauthoryear{{Li}, {Hambsch}, {Munari}, {Metzger},
  {Chomiuk}, {Frigo}  \& {Strader}}{{Li} et~al.}{2020a}]{2020ApJ...905..114L}
{Li} K.-L.,  {Hambsch} F.-J.,  {Munari} U.,  {Metzger} B.~D.,  {Chomiuk} L.,
  {Frigo} A.,   {Strader} J.,  2020a, \mn@doi [\apj]
  {10.3847/1538-4357/abc3be}, \href
  {https://ui.adsabs.harvard.edu/abs/2020ApJ...905..114L} {905, 114}

\bibitem[\protect\citeauthoryear{{Li}, {Kong}, {Aydi}, {Sokolovsky}, {Chomiuk},
  {Kawash}  \& {Strader}}{{Li} et~al.}{2020b}]{2020ATel13868....1L}
{Li} K.-L.,  {Kong} A.,  {Aydi} E.,  {Sokolovsky} K.,  {Chomiuk} L.,  {Kawash}
  A.,   {Strader} J.,  2020b, The Astronomer's Telegram, \href
  {https://ui.adsabs.harvard.edu/abs/2020ATel13868....1L} {13868, 1}

\bibitem[\protect\citeauthoryear{{Linford} et~al.,}{{Linford}
  et~al.}{2018}]{2018ATel11647....1L}
{Linford} J.~D.,  et~al., 2018, The Astronomer's Telegram, \href
  {https://ui.adsabs.harvard.edu/abs/2018ATel11647....1L} {11647, 1}

\bibitem[\protect\citeauthoryear{{Liu} \& {Hu}}{{Liu} \&
  {Hu}}{2000}]{2000ApJS..128..387L}
{Liu} W.,  {Hu} J.~Y.,  2000, \mn@doi [\apjs] {10.1086/313380}, \href
  {https://ui.adsabs.harvard.edu/abs/2000ApJS..128..387L} {128, 387}

\bibitem[\protect\citeauthoryear{{Livio} \& {Truran}}{{Livio} \&
  {Truran}}{1994}]{1994ApJ...425..797L}
{Livio} M.,  {Truran} J.~W.,  1994, \mn@doi [\apj] {10.1086/174024}, \href
  {http://adsabs.harvard.edu/abs/1994ApJ...425..797L} {425, 797}

\bibitem[\protect\citeauthoryear{{Livio}, {Mastichiadis}, {Oegelman}  \&
  {Truran}}{{Livio} et~al.}{1992}]{1992ApJ...394..217L}
{Livio} M.,  {Mastichiadis} A.,  {Oegelman} H.,   {Truran} J.~W.,  1992,
  \mn@doi [\apj] {10.1086/171573}, \href
  {https://ui.adsabs.harvard.edu/abs/1992ApJ...394..217L} {394, 217}

\bibitem[\protect\citeauthoryear{{Madlee}, {Mitthumsiri}, {Ruffolo}, {Digel}
  \& {Nuntiyakul}}{{Madlee} et~al.}{2020}]{2020JGRA..12528151M}
{Madlee} S.,  {Mitthumsiri} W.,  {Ruffolo} D.,  {Digel} S.,   {Nuntiyakul} W.,
  2020, \mn@doi [Journal of Geophysical Research (Space Physics)]
  {10.1029/2020JA028151}, \href
  {https://ui.adsabs.harvard.edu/abs/2020JGRA..12528151M} {125, e28151}

\bibitem[\protect\citeauthoryear{{Madsen} et~al.,}{{Madsen}
  et~al.}{2015}]{2015ApJS..220....8M}
{Madsen} K.~K.,  et~al., 2015, \mn@doi [\apjs] {10.1088/0067-0049/220/1/8},
  \href {https://ui.adsabs.harvard.edu/abs/2015ApJS..220....8M} {220, 8}

\bibitem[\protect\citeauthoryear{{Madsen}, {Grefenstette}, {Pike}, {Miyasaka},
  {Brightman}, {Forster}  \& {Harrison}}{{Madsen}
  et~al.}{2020}]{2020arXiv200500569M}
{Madsen} K.~K.,  {Grefenstette} B.~W.,  {Pike} S.,  {Miyasaka} H.,  {Brightman}
  M.,  {Forster} K.,   {Harrison} F.~A.,  2020, arXiv e-prints, \href
  {https://ui.adsabs.harvard.edu/abs/2020arXiv200500569M} {p. arXiv:2005.00569}

\bibitem[\protect\citeauthoryear{{Mailyan} et~al.,}{{Mailyan}
  et~al.}{2016}]{2016JGRA..12111346M}
{Mailyan} B.~G.,  et~al., 2016, \mn@doi [Journal of Geophysical Research (Space
  Physics)] {10.1002/2016JA022702}, \href
  {https://ui.adsabs.harvard.edu/abs/2016JGRA..12111346M} {121, 11,346}

\bibitem[\protect\citeauthoryear{{Mamajek} et~al.,}{{Mamajek}
  et~al.}{2015}]{2015arXiv151006262M}
{Mamajek} E.~E.,  et~al., 2015, arXiv e-prints, \href
  {https://ui.adsabs.harvard.edu/abs/2015arXiv151006262M} {p. arXiv:1510.06262}

\bibitem[\protect\citeauthoryear{{Martin}, {Dubus}, {Jean}, {Tatischeff}  \&
  {Dosne}}{{Martin} et~al.}{2018}]{2018A&A...612A..38M}
{Martin} P.,  {Dubus} G.,  {Jean} P.,  {Tatischeff} V.,   {Dosne} C.,  2018,
  \mn@doi [\aap] {10.1051/0004-6361/201731692}, \href
  {http://adsabs.harvard.edu/abs/2018A%26A...612A..38M} {612, A38}

\bibitem[\protect\citeauthoryear{{Mason} et~al.,}{{Mason}
  et~al.}{2001}]{2001A&A...365L..36M}
{Mason} K.~O.,  et~al., 2001, \mn@doi [\aap] {10.1051/0004-6361:20000044},
  \href {https://ui.adsabs.harvard.edu/abs/2001A&A...365L..36M} {365, L36}

\bibitem[\protect\citeauthoryear{{Mattox} et~al.,}{{Mattox}
  et~al.}{1996}]{1996ApJ...461..396M}
{Mattox} J.~R.,  et~al., 1996, \mn@doi [\apj] {10.1086/177068}, \href
  {https://ui.adsabs.harvard.edu/abs/1996ApJ...461..396M} {461, 396}

\bibitem[\protect\citeauthoryear{{Max-Moerbeck}, {Richards}, {Hovatta},
  {Pavlidou}, {Pearson}  \& {Readhead}}{{Max-Moerbeck}
  et~al.}{2014}]{2014MNRAS.445..437M}
{Max-Moerbeck} W.,  {Richards} J.~L.,  {Hovatta} T.,  {Pavlidou} V.,  {Pearson}
  T.~J.,   {Readhead} A.~C.~S.,  2014, \mn@doi [\mnras]
  {10.1093/mnras/stu1707}, \href
  {https://ui.adsabs.harvard.edu/abs/2014MNRAS.445..437M} {445, 437}

\bibitem[\protect\citeauthoryear{{McLoughlin}, {Blundell}, {Lee}  \&
  {McCowage}}{{McLoughlin} et~al.}{2021a}]{2021MNRAS.503..704M}
{McLoughlin} D.,  {Blundell} K.~M.,  {Lee} S.,   {McCowage} C.,  2021a, \mn@doi
  [\mnras] {10.1093/mnras/stab581}, \href
  {https://ui.adsabs.harvard.edu/abs/2021MNRAS.503..704M} {503, 704}

\bibitem[\protect\citeauthoryear{{McLoughlin}, {Blundell}, {Lee}  \&
  {McCowage}}{{McLoughlin} et~al.}{2021b}]{2021arXiv210605578M}
{McLoughlin} D.,  {Blundell} K.~M.,  {Lee} S.,   {McCowage} C.,  2021b, \mn@doi
  [\mnras] {10.1093/mnras/stab1364}, \href
  {https://ui.adsabs.harvard.edu/abs/2021MNRAS.505.2518M} {505, 2518}

\bibitem[\protect\citeauthoryear{{McNaught}}{{McNaught}}{2020}]{2020CBET.4811....1M}
{McNaught} R.~H.,  2020, Central Bureau Electronic Telegrams, 4811, 1

\bibitem[\protect\citeauthoryear{{Metzger}, {Hasco{\"e}t}, {Vurm},
  {Beloborodov}, {Chomiuk}, {Sokoloski}  \& {Nelson}}{{Metzger}
  et~al.}{2014}]{2014MNRAS.442..713M}
{Metzger} B.~D.,  {Hasco{\"e}t} R.,  {Vurm} I.,  {Beloborodov} A.~M.,
  {Chomiuk} L.,  {Sokoloski} J.~L.,   {Nelson} T.,  2014, \mn@doi [\mnras]
  {10.1093/mnras/stu844}, \href
  {http://adsabs.harvard.edu/abs/2014MNRAS.442..713M} {442, 713}

\bibitem[\protect\citeauthoryear{{Metzger}, {Finzell}, {Vurm}, {Hasco{\"e}t},
  {Beloborodov}  \& {Chomiuk}}{{Metzger} et~al.}{2015}]{2015MNRAS.450.2739M}
{Metzger} B.~D.,  {Finzell} T.,  {Vurm} I.,  {Hasco{\"e}t} R.,  {Beloborodov}
  A.~M.,   {Chomiuk} L.,  2015, \mn@doi [\mnras] {10.1093/mnras/stv742}, \href
  {http://adsabs.harvard.edu/abs/2015MNRAS.450.2739M} {450, 2739}

\bibitem[\protect\citeauthoryear{{Metzger}, {Caprioli}, {Vurm}, {Beloborodov},
  {Bartos}  \& {Vlasov}}{{Metzger} et~al.}{2016}]{2016MNRAS.457.1786M}
{Metzger} B.~D.,  {Caprioli} D.,  {Vurm} I.,  {Beloborodov} A.~M.,  {Bartos}
  I.,   {Vlasov} A.,  2016, \mn@doi [\mnras] {10.1093/mnras/stw123}, \href
  {http://adsabs.harvard.edu/abs/2016MNRAS.457.1786M} {457, 1786}

\bibitem[\protect\citeauthoryear{{Miszalski} et~al.,}{{Miszalski}
  et~al.}{2016}]{2016MNRAS.456..633M}
{Miszalski} B.,  et~al., 2016, \mn@doi [\mnras] {10.1093/mnras/stv2689}, \href
  {https://ui.adsabs.harvard.edu/abs/2016MNRAS.456..633M} {456, 633}

\bibitem[\protect\citeauthoryear{{Morii} et~al.,}{{Morii}
  et~al.}{2013}]{2013ApJ...779..118M}
{Morii} M.,  et~al., 2013, \mn@doi [\apj] {10.1088/0004-637X/779/2/118}, \href
  {https://ui.adsabs.harvard.edu/abs/2013ApJ...779..118M} {779, 118}

\bibitem[\protect\citeauthoryear{{Morii}, {Yamaoka}, {Mihara}, {Matsuoka}  \&
  {Kawai}}{{Morii} et~al.}{2016}]{2016PASJ...68S..11M}
{Morii} M.,  {Yamaoka} H.,  {Mihara} T.,  {Matsuoka} M.,   {Kawai} N.,  2016,
  \mn@doi [\pasj] {10.1093/pasj/psw007}, \href
  {https://ui.adsabs.harvard.edu/abs/2016PASJ...68S..11M} {68, S11}

\bibitem[\protect\citeauthoryear{{Mr{\'o}z} et~al.,}{{Mr{\'o}z}
  et~al.}{2015}]{2015AcA....65..313M}
{Mr{\'o}z} P.,  et~al., 2015, \actaa, \href
  {https://ui.adsabs.harvard.edu/abs/2015AcA....65..313M} {65, 313}

\bibitem[\protect\citeauthoryear{{Mr{\'o}z} et~al.,}{{Mr{\'o}z}
  et~al.}{2016}]{2016Natur.537..649M}
{Mr{\'o}z} P.,  et~al., 2016, \mn@doi [\nat] {10.1038/nature19066}, \href
  {https://ui.adsabs.harvard.edu/abs/2016Natur.537..649M} {537, 649}

\bibitem[\protect\citeauthoryear{{Mukai}}{{Mukai}}{2017}]{2017PASP..129f2001M}
{Mukai} K.,  2017, \mn@doi [\pasp] {10.1088/1538-3873/aa6736}, \href
  {http://adsabs.harvard.edu/abs/2017PASP..129f2001M} {129, 062001}

\bibitem[\protect\citeauthoryear{{Mukai} \& {Orio}}{{Mukai} \&
  {Orio}}{2005}]{2005ApJ...622..602M}
{Mukai} K.,  {Orio} M.,  2005, \mn@doi [\apj] {10.1086/427915}, \href
  {https://ui.adsabs.harvard.edu/abs/2005ApJ...622..602M} {622, 602}

\bibitem[\protect\citeauthoryear{{Mukai} et~al.,}{{Mukai}
  et~al.}{2014}]{2014ASPC..490..327M}
{Mukai} K.,  et~al., 2014, in {Woudt} P.~A.,  {Ribeiro} V.~A.~R.~M.,  eds,
  Astronomical Society of the Pacific Conference Series Vol. 490, Stellar
  Novae: Past and Future Decades. p.~327

\bibitem[\protect\citeauthoryear{{Munari} et~al.,}{{Munari}
  et~al.}{2011}]{2011MNRAS.410L..52M}
{Munari} U.,  et~al., 2011, \mn@doi [\mnras]
  {10.1111/j.1745-3933.2010.00979.x}, \href
  {https://ui.adsabs.harvard.edu/abs/2011MNRAS.410L..52M} {410, L52}

\bibitem[\protect\citeauthoryear{{Munari}, {Hambsch}  \& {Frigo}}{{Munari}
  et~al.}{2017}]{2017MNRAS.469.4341M}
{Munari} U.,  {Hambsch} F.~J.,   {Frigo} A.,  2017, \mn@doi [\mnras]
  {10.1093/mnras/stx1116}, \href
  {https://ui.adsabs.harvard.edu/abs/2017MNRAS.469.4341M} {469, 4341}

\bibitem[\protect\citeauthoryear{{Nasa High Energy Astrophysics Science Archive
  Research Center (Heasarc)}}{{Nasa High Energy Astrophysics Science Archive
  Research Center (Heasarc)}}{2014}]{2014ascl.soft08004N}
{Nasa High Energy Astrophysics Science Archive Research Center (Heasarc)} 2014,
  {HEAsoft: Unified Release of FTOOLS and XANADU} (\mn@eprint {ascl}
  {1408.004})

\bibitem[\protect\citeauthoryear{{Nelemans}, {Siess}, {Repetto}, {Toonen}  \&
  {Phinney}}{{Nelemans} et~al.}{2016}]{2016ApJ...817...69N}
{Nelemans} G.,  {Siess} L.,  {Repetto} S.,  {Toonen} S.,   {Phinney} E.~S.,
  2016, \mn@doi [\apj] {10.3847/0004-637X/817/1/69}, \href
  {https://ui.adsabs.harvard.edu/abs/2016ApJ...817...69N} {817, 69}

\bibitem[\protect\citeauthoryear{{Nelson}, {Orio}, {Cassinelli}, {Still},
  {Leibowitz}  \& {Mucciarelli}}{{Nelson} et~al.}{2008}]{2008ApJ...673.1067N}
{Nelson} T.,  {Orio} M.,  {Cassinelli} J.~P.,  {Still} M.,  {Leibowitz} E.,
  {Mucciarelli} P.,  2008, \mn@doi [\apj] {10.1086/524054}, \href
  {https://ui.adsabs.harvard.edu/abs/2008ApJ...673.1067N} {673, 1067}

\bibitem[\protect\citeauthoryear{{Nelson}, {Donato}, {Mukai}, {Sokoloski}  \&
  {Chomiuk}}{{Nelson} et~al.}{2012}]{2012ApJ...748...43N}
{Nelson} T.,  {Donato} D.,  {Mukai} K.,  {Sokoloski} J.,   {Chomiuk} L.,  2012,
  \mn@doi [\apj] {10.1088/0004-637X/748/1/43}, \href
  {https://ui.adsabs.harvard.edu/abs/2012ApJ...748...43N} {748, 43}

\bibitem[\protect\citeauthoryear{{Nelson} et~al.,}{{Nelson}
  et~al.}{2019}]{2019ApJ...872...86N}
{Nelson} T.,  et~al., 2019, \mn@doi [\apj] {10.3847/1538-4357/aafb6d}, \href
  {https://ui.adsabs.harvard.edu/abs/2019ApJ...872...86N} {872, 86}

\bibitem[\protect\citeauthoryear{{Nelson} et~al.,}{{Nelson}
  et~al.}{2021}]{2021MNRAS.500.2798N}
{Nelson} T.,  et~al., 2021, \mn@doi [\mnras] {10.1093/mnras/staa3367}, \href
  {https://ui.adsabs.harvard.edu/abs/2021MNRAS.500.2798N} {500, 2798}

\bibitem[\protect\citeauthoryear{{Ness}}{{Ness}}{2012}]{2012BASI...40..353N}
{Ness} J.~U.,  2012, Bulletin of the Astronomical Society of India, \href
  {https://ui.adsabs.harvard.edu/abs/2012BASI...40..353N} {40, 353}

\bibitem[\protect\citeauthoryear{{Ness} et~al.,}{{Ness}
  et~al.}{2003}]{2003ApJ...594L.127N}
{Ness} J.~U.,  et~al., 2003, \mn@doi [\apjl] {10.1086/378664}, \href
  {https://ui.adsabs.harvard.edu/abs/2003ApJ...594L.127N} {594, L127}

\bibitem[\protect\citeauthoryear{{Ness}, {Starrfield}, {Jordan}, {Krautter}  \&
  {Schmitt}}{{Ness} et~al.}{2005}]{2005MNRAS.364.1015N}
{Ness} J.-U.,  {Starrfield} S.,  {Jordan} C.,  {Krautter} J.,   {Schmitt}
  J.~H.~M.~M.,  2005, \mn@doi [\mnras] {10.1111/j.1365-2966.2005.09664.x},
  \href {http://adsabs.harvard.edu/abs/2005MNRAS.364.1015N} {364, 1015}

\bibitem[\protect\citeauthoryear{{Ness}, {Schwarz}, {Retter}, {Starrfield},
  {Schmitt}, {Gehrels}, {Burrows}  \& {Osborne}}{{Ness}
  et~al.}{2007a}]{2007ApJ...663..505N}
{Ness} J.-U.,  {Schwarz} G.~J.,  {Retter} A.,  {Starrfield} S.,  {Schmitt}
  J.~H.~M.~M.,  {Gehrels} N.,  {Burrows} D.,   {Osborne} J.~P.,  2007a, \mn@doi
  [\apj] {10.1086/518084}, \href
  {http://adsabs.harvard.edu/abs/2007ApJ...663..505N} {663, 505}

\bibitem[\protect\citeauthoryear{{Ness} et~al.,}{{Ness}
  et~al.}{2007b}]{2007ApJ...665.1334N}
{Ness} J.~U.,  et~al., 2007b, \mn@doi [\apj] {10.1086/519676}, \href
  {https://ui.adsabs.harvard.edu/abs/2007ApJ...665.1334N} {665, 1334}

\bibitem[\protect\citeauthoryear{{Ness} et~al.,}{{Ness}
  et~al.}{2009}]{2009AJ....137.3414N}
{Ness} J.~U.,  et~al., 2009, \mn@doi [\aj] {10.1088/0004-6256/137/2/3414},
  \href {https://ui.adsabs.harvard.edu/abs/2009AJ....137.3414N} {137, 3414}

\bibitem[\protect\citeauthoryear{{Ness} et~al.,}{{Ness}
  et~al.}{2011}]{2011ApJ...733...70N}
{Ness} J.~U.,  et~al., 2011, \mn@doi [\apj] {10.1088/0004-637X/733/1/70}, \href
  {https://ui.adsabs.harvard.edu/abs/2011ApJ...733...70N} {733, 70}

\bibitem[\protect\citeauthoryear{{Ness} et~al.,}{{Ness}
  et~al.}{2012}]{2012ApJ...745...43N}
{Ness} J.~U.,  et~al., 2012, \mn@doi [\apj] {10.1088/0004-637X/745/1/43}, \href
  {https://ui.adsabs.harvard.edu/abs/2012ApJ...745...43N} {745, 43}

\bibitem[\protect\citeauthoryear{{Ness} et~al.,}{{Ness}
  et~al.}{2013}]{2013A&A...559A..50N}
{Ness} J.~U.,  et~al., 2013, \mn@doi [\aap] {10.1051/0004-6361/201322415},
  \href {https://ui.adsabs.harvard.edu/abs/2013A&A...559A..50N} {559, A50}

\bibitem[\protect\citeauthoryear{{Ness} et~al.,}{{Ness}
  et~al.}{2015}]{2015A&A...578A..39N}
{Ness} J.~U.,  et~al., 2015, \mn@doi [\aap] {10.1051/0004-6361/201425178},
  \href {https://ui.adsabs.harvard.edu/abs/2015A&A...578A..39N} {578, A39}

\bibitem[\protect\citeauthoryear{{Ness} et~al.,}{{Ness}
  et~al.}{2022}]{2022A&A...658A.169N}
{Ness} J.~U.,  et~al., 2022, \mn@doi [\aap] {10.1051/0004-6361/202142037},
  \href {https://ui.adsabs.harvard.edu/abs/2022A&A...658A.169N} {658, A169}

\bibitem[\protect\citeauthoryear{{O'Brien} \& {Cohen}}{{O'Brien} \&
  {Cohen}}{1998}]{1998ApJ...498L..59O}
{O'Brien} T.~J.,  {Cohen} J.~G.,  1998, \mn@doi [\apjl] {10.1086/311300}, \href
  {https://ui.adsabs.harvard.edu/abs/1998ApJ...498L..59O} {498, L59}

\bibitem[\protect\citeauthoryear{{Orio}}{{Orio}}{2020}]{2020AdSpR..66.1193O}
{Orio} M.,  2020, \mn@doi [Advances in Space Research]
  {10.1016/j.asr.2020.02.017}, \href
  {https://ui.adsabs.harvard.edu/abs/2020AdSpR..66.1193O} {66, 1193}

\bibitem[\protect\citeauthoryear{{Orio} et~al.,}{{Orio}
  et~al.}{2013}]{2013MNRAS.429.1342O}
{Orio} M.,  et~al., 2013, \mn@doi [\mnras] {10.1093/mnras/sts421}, \href
  {https://ui.adsabs.harvard.edu/abs/2013MNRAS.429.1342O} {429, 1342}

\bibitem[\protect\citeauthoryear{{Orio}, {Rana}, {Page}, {Sokoloski}  \&
  {Harrison}}{{Orio} et~al.}{2015}]{2015MNRAS.448L..35O}
{Orio} M.,  {Rana} V.,  {Page} K.~L.,  {Sokoloski} J.,   {Harrison} F.,  2015,
  \mn@doi [\mnras] {10.1093/mnrasl/slu195}, \href
  {https://ui.adsabs.harvard.edu/abs/2015MNRAS.448L..35O} {448, L35}

\bibitem[\protect\citeauthoryear{{Orio} et~al.,}{{Orio}
  et~al.}{2018}]{2018ApJ...862..164O}
{Orio} M.,  et~al., 2018, \mn@doi [\apj] {10.3847/1538-4357/aacf06}, \href
  {https://ui.adsabs.harvard.edu/abs/2018ApJ...862..164O} {862, 164}

\bibitem[\protect\citeauthoryear{{Orio} et~al.,}{{Orio}
  et~al.}{2020}]{2020ApJ...895...80O}
{Orio} M.,  et~al., 2020, \mn@doi [\apj] {10.3847/1538-4357/ab8c4d}, \href
  {https://ui.adsabs.harvard.edu/abs/2020ApJ...895...80O} {895, 80}

\bibitem[\protect\citeauthoryear{{Orio} et~al.,}{{Orio}
  et~al.}{2021}]{2021MNRAS.tmp.1420O}
{Orio} M.,  et~al., 2021, \mn@doi [\mnras] {10.1093/mnras/stab1391}, \href
  {https://ui.adsabs.harvard.edu/abs/2021MNRAS.505.3113O} {505, 3113}

\bibitem[\protect\citeauthoryear{{Orlando}, {Drake}  \& {Laming}}{{Orlando}
  et~al.}{2009}]{2009A&A...493.1049O}
{Orlando} S.,  {Drake} J.~J.,   {Laming} J.~M.,  2009, \mn@doi [\aap]
  {10.1051/0004-6361:200810109}, \href
  {https://ui.adsabs.harvard.edu/abs/2009A&A...493.1049O} {493, 1049}

\bibitem[\protect\citeauthoryear{{Orlando}, {Drake}  \& {Miceli}}{{Orlando}
  et~al.}{2017}]{2017MNRAS.464.5003O}
{Orlando} S.,  {Drake} J.~J.,   {Miceli} M.,  2017, \mn@doi [\mnras]
  {10.1093/mnras/stw2718}, \href
  {https://ui.adsabs.harvard.edu/abs/2017MNRAS.464.5003O} {464, 5003}

\bibitem[\protect\citeauthoryear{{Osaki}}{{Osaki}}{2005}]{2005PJAB...81..291O}
{Osaki} Y.,  2005, \mn@doi [Proceeding of the Japan Academy, Series B]
  {10.2183/pjab.81.291}, \href
  {https://ui.adsabs.harvard.edu/abs/2005PJAB...81..291O} {81, 291}

\bibitem[\protect\citeauthoryear{{Page}, {Osborne}, {Beardmore}, {Evans},
  {Rosen}  \& {Watson}}{{Page} et~al.}{2014}]{2014A&A...570A..37P}
{Page} K.~L.,  {Osborne} J.~P.,  {Beardmore} A.~P.,  {Evans} P.~A.,  {Rosen}
  S.~R.,   {Watson} M.~G.,  2014, \mn@doi [\aap] {10.1051/0004-6361/201424777},
  \href {https://ui.adsabs.harvard.edu/abs/2014A&A...570A..37P} {570, A37}

\bibitem[\protect\citeauthoryear{{Page} et~al.,}{{Page}
  et~al.}{2020}]{2020MNRAS.499.4814P}
{Page} K.~L.,  et~al., 2020, \mn@doi [\mnras] {10.1093/mnras/staa3083}, \href
  {https://ui.adsabs.harvard.edu/abs/2020MNRAS.499.4814P} {499, 4814}

\bibitem[\protect\citeauthoryear{{Patterson}}{{Patterson}}{2011}]{2011MNRAS.411.2695P}
{Patterson} J.,  2011, \mn@doi [\mnras] {10.1111/j.1365-2966.2010.17881.x},
  \href {https://ui.adsabs.harvard.edu/abs/2011MNRAS.411.2695P} {411, 2695}

\bibitem[\protect\citeauthoryear{{Patterson} et~al.,}{{Patterson}
  et~al.}{2013}]{2013MNRAS.434.1902P}
{Patterson} J.,  et~al., 2013, \mn@doi [\mnras] {10.1093/mnras/stt1085}, \href
  {https://ui.adsabs.harvard.edu/abs/2013MNRAS.434.1902P} {434, 1902}

\bibitem[\protect\citeauthoryear{{Pei} et~al.,}{{Pei}
  et~al.}{2020}]{2020ATel14067....1P}
{Pei} S.,  et~al., 2020, The Astronomer's Telegram, \href
  {https://ui.adsabs.harvard.edu/abs/2020ATel14067....1P} {14067, 1}

\bibitem[\protect\citeauthoryear{{Peretz}, {Orio}, {Behar}, {Bianchini},
  {Gallagher}, {Rauch}, {Tofflemire}  \& {Zemko}}{{Peretz}
  et~al.}{2016}]{2016ApJ...829....2P}
{Peretz} U.,  {Orio} M.,  {Behar} E.,  {Bianchini} A.,  {Gallagher} J.,
  {Rauch} T.,  {Tofflemire} B.,   {Zemko} P.,  2016, \mn@doi [\apj]
  {10.3847/0004-637X/829/1/2}, \href
  {https://ui.adsabs.harvard.edu/abs/2016ApJ...829....2P} {829, 2}

\bibitem[\protect\citeauthoryear{{Poggiani}}{{Poggiani}}{2018}]{2018arXiv180311529P}
{Poggiani} R.,  2018, arXiv e-prints, \href
  {https://ui.adsabs.harvard.edu/abs/2018arXiv180311529P} {p. arXiv:1803.11529}

\bibitem[\protect\citeauthoryear{{Pshirkov}}{{Pshirkov}}{2016}]{2016MNRAS.457L..99P}
{Pshirkov} M.~S.,  2016, \mn@doi [\mnras] {10.1093/mnrasl/slv205}, \href
  {https://ui.adsabs.harvard.edu/abs/2016MNRAS.457L..99P} {457, L99}

\bibitem[\protect\citeauthoryear{{Read} et~al.,}{{Read}
  et~al.}{2008}]{2008A&A...482L...1R}
{Read} A.~M.,  et~al., 2008, \mn@doi [\aap] {10.1051/0004-6361:200809456},
  \href {https://ui.adsabs.harvard.edu/abs/2008A&A...482L...1R} {482, L1}

\bibitem[\protect\citeauthoryear{{Romero}, {Boettcher}, {Markoff}  \&
  {Tavecchio}}{{Romero} et~al.}{2017}]{2017SSRv..207....5R}
{Romero} G.~E.,  {Boettcher} M.,  {Markoff} S.,   {Tavecchio} F.,  2017,
  \mn@doi [\ssr] {10.1007/s11214-016-0328-2}, \href
  {https://ui.adsabs.harvard.edu/abs/2017SSRv..207....5R} {207, 5}

\bibitem[\protect\citeauthoryear{{Rudy}, {Russell}  \& {Sitko}}{{Rudy}
  et~al.}{2021}]{2021RNAAS...5...48R}
{Rudy} R.~J.,  {Russell} R.~W.,   {Sitko} M.~L.,  2021, \mn@doi [Research Notes
  of the American Astronomical Society] {10.3847/2515-5172/abeefc}, \href
  {https://ui.adsabs.harvard.edu/abs/2021RNAAS...5...48R} {5, 48}

\bibitem[\protect\citeauthoryear{{Rupen}, {Mioduszewski}  \&
  {Sokoloski}}{{Rupen} et~al.}{2008}]{2008ApJ...688..559R}
{Rupen} M.~P.,  {Mioduszewski} A.~J.,   {Sokoloski} J.~L.,  2008, \mn@doi
  [\apj] {10.1086/525555}, \href
  {https://ui.adsabs.harvard.edu/abs/2008ApJ...688..559R} {688, 559}

\bibitem[\protect\citeauthoryear{{Salazar}, {LeBleu}, {Schaefer}, {Land olt}
  \& {Dvorak}}{{Salazar} et~al.}{2017}]{2017MNRAS.469.4116V}
{Salazar} I.~V.,  {LeBleu} A.,  {Schaefer} B.~E.,  {Land olt} A.~U.,   {Dvorak}
  S.,  2017, \mn@doi [\mnras] {10.1093/mnras/stx1161}, \href
  {https://ui.adsabs.harvard.edu/abs/2017MNRAS.469.4116V} {469, 4116}

\bibitem[\protect\citeauthoryear{{Samus'}, {Kazarovets}, {Durlevich}, {Kireeva}
   \& {Pastukhova}}{{Samus'} et~al.}{2017}]{2017ARep...61...80S}
{Samus'} N.~N.,  {Kazarovets} E.~V.,  {Durlevich} O.~V.,  {Kireeva} N.~N.,
  {Pastukhova} E.~N.,  2017, \mn@doi [Astronomy Reports]
  {10.1134/S1063772917010085}, \href
  {https://ui.adsabs.harvard.edu/abs/2017ARep...61...80S} {61, 80}

\bibitem[\protect\citeauthoryear{{Saxton} \& {Gimeno}}{{Saxton} \&
  {Gimeno}}{2011}]{2011ASPC..442..567S}
{Saxton} R.,  {Gimeno} C.~D.~T.,  2011, in {Evans} I.~N.,  {Accomazzi} A.,
  {Mink} D.~J.,   {Rots} A.~H.,  eds,  Astronomical Society of the Pacific
  Conference Series Vol. 442, Astronomical Data Analysis Software and Systems
  XX. p.~567

\bibitem[\protect\citeauthoryear{{Saxton}, {Read}, {Esquej}, {Freyberg},
  {Altieri}  \& {Bermejo}}{{Saxton} et~al.}{2008}]{2008A&A...480..611S}
{Saxton} R.~D.,  {Read} A.~M.,  {Esquej} P.,  {Freyberg} M.~J.,  {Altieri} B.,
   {Bermejo} D.,  2008, \mn@doi [\aap] {10.1051/0004-6361:20079193}, \href
  {https://ui.adsabs.harvard.edu/abs/2008A&A...480..611S} {480, 611}

\bibitem[\protect\citeauthoryear{{Schaefer}}{{Schaefer}}{2010}]{2010ApJS..187..275S}
{Schaefer} B.~E.,  2010, \mn@doi [\apjs] {10.1088/0067-0049/187/2/275}, \href
  {https://ui.adsabs.harvard.edu/abs/2010ApJS..187..275S} {187, 275}

\bibitem[\protect\citeauthoryear{{Schaefer}}{{Schaefer}}{2018}]{2018MNRAS.481.3033S}
{Schaefer} B.~E.,  2018, \mn@doi [\mnras] {10.1093/mnras/sty2388}, \href
  {https://ui.adsabs.harvard.edu/abs/2018MNRAS.481.3033S} {481, 3033}

\bibitem[\protect\citeauthoryear{{Schaefer}}{{Schaefer}}{2021}]{2021RNAAS...5..150S}
{Schaefer} B.~E.,  2021, \mn@doi [Research Notes of the American Astronomical
  Society] {10.3847/2515-5172/ac0d5b}, \href
  {https://ui.adsabs.harvard.edu/abs/2021RNAAS...5..150S} {5, 150}

\bibitem[\protect\citeauthoryear{{Schenker}, {Kolb}  \& {Ritter}}{{Schenker}
  et~al.}{1998}]{1998MNRAS.297..633S}
{Schenker} K.,  {Kolb} U.,   {Ritter} H.,  1998, \mn@doi [\mnras]
  {10.1046/j.1365-8711.1998.01529.x}, \href
  {https://ui.adsabs.harvard.edu/abs/1998MNRAS.297..633S} {297, 633}

\bibitem[\protect\citeauthoryear{{Schmidtobreick}, {Shara}, {Tappert}, {Bayo}
  \& {Ederoclite}}{{Schmidtobreick} et~al.}{2015}]{2015MNRAS.449.2215S}
{Schmidtobreick} L.,  {Shara} M.,  {Tappert} C.,  {Bayo} A.,   {Ederoclite} A.,
   2015, \mn@doi [\mnras] {10.1093/mnras/stv250}, \href
  {https://ui.adsabs.harvard.edu/abs/2015MNRAS.449.2215S} {449, 2215}

\bibitem[\protect\citeauthoryear{{Schure}, {Bell}, {O'C Drury}  \&
  {Bykov}}{{Schure} et~al.}{2012}]{2012SSRv..173..491S}
{Schure} K.~M.,  {Bell} A.~R.,  {O'C Drury} L.,   {Bykov} A.~M.,  2012, \mn@doi
  [\ssr] {10.1007/s11214-012-9871-7}, \href
  {https://ui.adsabs.harvard.edu/abs/2012SSRv..173..491S} {173, 491}

\bibitem[\protect\citeauthoryear{{Schwarz}, {Shore}, {Starrfield},
  {Hauschildt}, {Della Valle}  \& {Baron}}{{Schwarz}
  et~al.}{2001}]{2001MNRAS.320..103S}
{Schwarz} G.~J.,  {Shore} S.~N.,  {Starrfield} S.,  {Hauschildt} P.~H.,  {Della
  Valle} M.,   {Baron} E.,  2001, \mn@doi [\mnras]
  {10.1046/j.1365-8711.2001.03960.x}, \href
  {http://adsabs.harvard.edu/abs/2001MNRAS.320..103S} {320, 103}

\bibitem[\protect\citeauthoryear{{Schwarz} et~al.,}{{Schwarz}
  et~al.}{2011}]{2011ApJS..197...31S}
{Schwarz} G.~J.,  et~al., 2011, \mn@doi [\apjs] {10.1088/0067-0049/197/2/31},
  \href {http://adsabs.harvard.edu/abs/2011ApJS..197...31S} {197, 31}

\bibitem[\protect\citeauthoryear{{Shafter}}{{Shafter}}{2017}]{2017ApJ...834..196S}
{Shafter} A.~W.,  2017, \mn@doi [\apj] {10.3847/1538-4357/834/2/196}, \href
  {https://ui.adsabs.harvard.edu/abs/2017ApJ...834..196S} {834, 196}

\bibitem[\protect\citeauthoryear{{Shappee} et~al.,}{{Shappee}
  et~al.}{2014}]{2014ApJ...788...48S}
{Shappee} B.~J.,  et~al., 2014, \mn@doi [\apj] {10.1088/0004-637X/788/1/48},
  \href {https://ui.adsabs.harvard.edu/abs/2014ApJ...788...48S} {788, 48}

\bibitem[\protect\citeauthoryear{{Shara}}{{Shara}}{1989}]{1989PASP..101....5S}
{Shara} M.~M.,  1989, \mn@doi [\pasp] {10.1086/132400}, \href
  {https://ui.adsabs.harvard.edu/abs/1989PASP..101....5S} {101, 5}

\bibitem[\protect\citeauthoryear{{Shara} et~al.,}{{Shara}
  et~al.}{2007}]{2007Natur.446..159S}
{Shara} M.~M.,  et~al., 2007, \mn@doi [\nat] {10.1038/nature05576}, \href
  {https://ui.adsabs.harvard.edu/abs/2007Natur.446..159S} {446, 159}

\bibitem[\protect\citeauthoryear{{Shara}, {Mizusawa}, {Wehinger}, {Zurek},
  {Martin}, {Neill}, {Forster}  \& {Seibert}}{{Shara}
  et~al.}{2012}]{2012ApJ...758..121S}
{Shara} M.~M.,  {Mizusawa} T.,  {Wehinger} P.,  {Zurek} D.,  {Martin} C.~D.,
  {Neill} J.~D.,  {Forster} K.,   {Seibert} M.,  2012, \mn@doi [\apj]
  {10.1088/0004-637X/758/2/121}, \href
  {https://ui.adsabs.harvard.edu/abs/2012ApJ...758..121S} {758, 121}

\bibitem[\protect\citeauthoryear{{Shara} et~al.,}{{Shara}
  et~al.}{2017}]{2017Natur.548..558S}
{Shara} M.~M.,  et~al., 2017, \mn@doi [\nat] {10.1038/nature23644}, \href
  {https://ui.adsabs.harvard.edu/abs/2017Natur.548..558S} {548, 558}

\bibitem[\protect\citeauthoryear{{Shaviv}}{{Shaviv}}{1998}]{1998ApJ...494L.193S}
{Shaviv} N.~J.,  1998, \mn@doi [\apjl] {10.1086/311182}, \href
  {https://ui.adsabs.harvard.edu/abs/1998ApJ...494L.193S} {494, L193}

\bibitem[\protect\citeauthoryear{{Shaviv}}{{Shaviv}}{2001a}]{2001MNRAS.326..126S}
{Shaviv} N.~J.,  2001a, \mn@doi [\mnras] {10.1046/j.1365-8711.2001.04574.x},
  \href {https://ui.adsabs.harvard.edu/abs/2001MNRAS.326..126S} {326, 126}

\bibitem[\protect\citeauthoryear{{Shaviv}}{{Shaviv}}{2001b}]{2001ApJ...549.1093S}
{Shaviv} N.~J.,  2001b, \mn@doi [\apj] {10.1086/319428}, \href
  {https://ui.adsabs.harvard.edu/abs/2001ApJ...549.1093S} {549, 1093}

\bibitem[\protect\citeauthoryear{{Siegert} et~al.,}{{Siegert}
  et~al.}{2018}]{2018A&A...615A.107S}
{Siegert} T.,  et~al., 2018, \mn@doi [\aap] {10.1051/0004-6361/201732514},
  \href {https://ui.adsabs.harvard.edu/abs/2018A&A...615A.107S} {615, A107}

\bibitem[\protect\citeauthoryear{{Singh}, {Girish}, {Pavana}, {Ness}, {Anupama}
   \& {Orio}}{{Singh} et~al.}{2021}]{2021MNRAS.501...36S}
{Singh} K.~P.,  {Girish} V.,  {Pavana} M.,  {Ness} J.-U.,  {Anupama} G.~C.,
  {Orio} M.,  2021, \mn@doi [\mnras] {10.1093/mnras/staa3303}, \href
  {https://ui.adsabs.harvard.edu/abs/2021MNRAS.501...36S} {501, 36}

\bibitem[\protect\citeauthoryear{{Sitko}, {Rudy}  \& {Russell}}{{Sitko}
  et~al.}{2020}]{2020ATel14205....1S}
{Sitko} M.~L.,  {Rudy} R.~J.,   {Russell} R.~W.,  2020, The Astronomer's
  Telegram, \href {https://ui.adsabs.harvard.edu/abs/2020ATel14205....1S}
  {14205, 1}

\bibitem[\protect\citeauthoryear{{Sokoloski} et~al.,}{{Sokoloski}
  et~al.}{2006}]{2006ApJ...636.1002S}
{Sokoloski} J.~L.,  et~al., 2006, \mn@doi [\apj] {10.1086/498206}, \href
  {https://ui.adsabs.harvard.edu/abs/2006ApJ...636.1002S} {636, 1002}

\bibitem[\protect\citeauthoryear{{Sokoloski}, {Rupen}  \&
  {Mioduszewski}}{{Sokoloski} et~al.}{2008}]{2008ApJ...685L.137S}
{Sokoloski} J.~L.,  {Rupen} M.~P.,   {Mioduszewski} A.~J.,  2008, \mn@doi
  [\apjl] {10.1086/592602}, \href
  {https://ui.adsabs.harvard.edu/abs/2008ApJ...685L.137S} {685, L137}

\bibitem[\protect\citeauthoryear{{Sokolovsky} et~al.,}{{Sokolovsky}
  et~al.}{2017}]{2017MNRAS.464..274S}
{Sokolovsky} K.~V.,  et~al., 2017, \mn@doi [\mnras] {10.1093/mnras/stw2262},
  \href {https://ui.adsabs.harvard.edu/abs/2017MNRAS.464..274S} {464, 274}

\bibitem[\protect\citeauthoryear{{Sokolovsky} et~al.,}{{Sokolovsky}
  et~al.}{2020a}]{2020MNRAS.497.2569S}
{Sokolovsky} K.~V.,  et~al., 2020a, \mn@doi [\mnras] {10.1093/mnras/staa2104},
  \href {https://ui.adsabs.harvard.edu/abs/2020MNRAS.497.2569S} {497, 2569}

\bibitem[\protect\citeauthoryear{{Sokolovsky} et~al.,}{{Sokolovsky}
  et~al.}{2020b}]{2020ATel13900....1S}
{Sokolovsky} K.~V.,  et~al., 2020b, The Astronomer's Telegram, \href
  {https://ui.adsabs.harvard.edu/abs/2020ATel13900....1S} {13900, 1}

\bibitem[\protect\citeauthoryear{{Sokolovsky} et~al.,}{{Sokolovsky}
  et~al.}{2020c}]{2020ATel14043....1S}
{Sokolovsky} K.,  et~al., 2020c, The Astronomer's Telegram, \href
  {https://ui.adsabs.harvard.edu/abs/2020ATel14043....1S} {14043, 1}

\bibitem[\protect\citeauthoryear{{Starrfield}, {Truran}, {Sparks}  \&
  {Kutter}}{{Starrfield} et~al.}{1972}]{1972ApJ...176..169S}
{Starrfield} S.,  {Truran} J.~W.,  {Sparks} W.~M.,   {Kutter} G.~S.,  1972,
  \mn@doi [\apj] {10.1086/151619}, \href
  {https://ui.adsabs.harvard.edu/abs/1972ApJ...176..169S} {176, 169}

\bibitem[\protect\citeauthoryear{{Starrfield}, {Iliadis}  \&
  {Hix}}{{Starrfield} et~al.}{2016}]{2016PASP..128e1001S}
{Starrfield} S.,  {Iliadis} C.,   {Hix} W.~R.,  2016, \mn@doi [\pasp]
  {10.1088/1538-3873/128/963/051001}, \href
  {http://adsabs.harvard.edu/abs/2016PASP..128e1001S} {128, 051001}

\bibitem[\protect\citeauthoryear{{Steinberg} \& {Metzger}}{{Steinberg} \&
  {Metzger}}{2018}]{2018MNRAS.479..687S}
{Steinberg} E.,  {Metzger} B.~D.,  2018, \mn@doi [\mnras]
  {10.1093/mnras/sty1641}, \href
  {https://ui.adsabs.harvard.edu/abs/2018MNRAS.479..687S} {479, 687}

\bibitem[\protect\citeauthoryear{{Steinberg} \& {Metzger}}{{Steinberg} \&
  {Metzger}}{2020}]{2020MNRAS.491.4232S}
{Steinberg} E.,  {Metzger} B.~D.,  2020, \mn@doi [\mnras]
  {10.1093/mnras/stz3300}, \href
  {https://ui.adsabs.harvard.edu/abs/2020MNRAS.491.4232S} {491, 4232}

\bibitem[\protect\citeauthoryear{{Str{\"u}der} et~al.,}{{Str{\"u}der}
  et~al.}{2001}]{2001A&A...365L..18S}
{Str{\"u}der} L.,  et~al., 2001, \mn@doi [\aap] {10.1051/0004-6361:20000066},
  \href {https://ui.adsabs.harvard.edu/abs/2001A&A...365L..18S} {365, L18}

\bibitem[\protect\citeauthoryear{{Sun}, {Orio}, {Dobrotka}, {Luna}, {Shugarov}
  \& {Zemko}}{{Sun} et~al.}{2020}]{2020MNRAS.499.3006S}
{Sun} B.,  {Orio} M.,  {Dobrotka} A.,  {Luna} G. J.~M.,  {Shugarov} S.,
  {Zemko} P.,  2020, \mn@doi [\mnras] {10.1093/mnras/staa3012}, \href
  {https://ui.adsabs.harvard.edu/abs/2020MNRAS.499.3006S} {499, 3006}

\bibitem[\protect\citeauthoryear{{Suzuki} \& {Shigeyama}}{{Suzuki} \&
  {Shigeyama}}{2010}]{2010ApJ...723L..84S}
{Suzuki} A.,  {Shigeyama} T.,  2010, \mn@doi [\apjl]
  {10.1088/2041-8205/723/1/L84}, \href
  {http://adsabs.harvard.edu/abs/2010ApJ...723L..84S} {723, L84}

\bibitem[\protect\citeauthoryear{{Taguchi}, {Maehara}, {Isogai}, {Tampo},
  {Kojiguchi}, {Kato}  \& {Nogami}}{{Taguchi}
  et~al.}{2021}]{2021ATel14472....1T}
{Taguchi} K.,  {Maehara} H.,  {Isogai} K.,  {Tampo} Y.,  {Kojiguchi} N.,
  {Kato} T.,   {Nogami} D.,  2021, The Astronomer's Telegram, \href
  {https://ui.adsabs.harvard.edu/abs/2021ATel14472....1T} {14472, 1}

\bibitem[\protect\citeauthoryear{{Tamuz}, {Mazeh}  \& {North}}{{Tamuz}
  et~al.}{2006}]{2006MNRAS.367.1521T}
{Tamuz} O.,  {Mazeh} T.,   {North} P.,  2006, \mn@doi [\mnras]
  {10.1111/j.1365-2966.2006.10049.x}, \href
  {https://ui.adsabs.harvard.edu/abs/2006MNRAS.367.1521T} {367, 1521}

\bibitem[\protect\citeauthoryear{{Tappert}, {Barria}, {Fuentes Morales},
  {Vogt}, {Ederoclite}  \& {Schmidtobreick}}{{Tappert}
  et~al.}{2016}]{2016MNRAS.462.1371T}
{Tappert} C.,  {Barria} D.,  {Fuentes Morales} I.,  {Vogt} N.,  {Ederoclite}
  A.,   {Schmidtobreick} L.,  2016, \mn@doi [\mnras] {10.1093/mnras/stw1748},
  \href {https://ui.adsabs.harvard.edu/abs/2016MNRAS.462.1371T} {462, 1371}

\bibitem[\protect\citeauthoryear{{Tatischeff} \& {Hernanz}}{{Tatischeff} \&
  {Hernanz}}{2007}]{2007ApJ...663L.101T}
{Tatischeff} V.,  {Hernanz} M.,  2007, \mn@doi [\apjl] {10.1086/520049}, \href
  {https://ui.adsabs.harvard.edu/abs/2007ApJ...663L.101T} {663, L101}

\bibitem[\protect\citeauthoryear{{Toal{\'a}}, {Guerrero}, {Santamar{\'\i}a},
  {Ramos-Larios}  \& {Sabin}}{{Toal{\'a}} et~al.}{2020}]{2020MNRAS.495.4372T}
{Toal{\'a}} J.~A.,  {Guerrero} M.~A.,  {Santamar{\'\i}a} E.,  {Ramos-Larios}
  G.,   {Sabin} L.,  2020, \mn@doi [\mnras] {10.1093/mnras/staa1502}, \href
  {https://ui.adsabs.harvard.edu/abs/2020MNRAS.495.4372T} {495, 4372}

\bibitem[\protect\citeauthoryear{{Truran} \& {Livio}}{{Truran} \&
  {Livio}}{1986}]{1986ApJ...308..721T}
{Truran} J.~W.,  {Livio} M.,  1986, \mn@doi [\apj] {10.1086/164544}, \href
  {https://ui.adsabs.harvard.edu/abs/1986ApJ...308..721T} {308, 721}

\bibitem[\protect\citeauthoryear{{Turner} et~al.,}{{Turner}
  et~al.}{2001}]{2001A&A...365L..27T}
{Turner} M.~J.~L.,  et~al., 2001, \mn@doi [\aap] {10.1051/0004-6361:20000087},
  \href {https://ui.adsabs.harvard.edu/abs/2001A&A...365L..27T} {365, L27}

\bibitem[\protect\citeauthoryear{{Vasilopoulos}, {Koliopanos}, {Woods},
  {Haberl}, {Soraisam}  \& {Udalski}}{{Vasilopoulos}
  et~al.}{2020}]{2020MNRAS.499.2007V}
{Vasilopoulos} G.,  {Koliopanos} F.,  {Woods} T.~E.,  {Haberl} F.,  {Soraisam}
  M.~D.,   {Udalski} A.,  2020, \mn@doi [\mnras] {10.1093/mnras/staa2922},
  \href {https://ui.adsabs.harvard.edu/abs/2020MNRAS.499.2007V} {499, 2007}

\bibitem[\protect\citeauthoryear{{Verbunt}}{{Verbunt}}{1984}]{1984MNRAS.209..227V}
{Verbunt} F.,  1984, \mn@doi [\mnras] {10.1093/mnras/209.2.227}, \href
  {https://ui.adsabs.harvard.edu/abs/1984MNRAS.209..227V} {209, 227}

\bibitem[\protect\citeauthoryear{{Vlasov}, {Vurm}  \& {Metzger}}{{Vlasov}
  et~al.}{2016}]{2016MNRAS.463..394V}
{Vlasov} A.,  {Vurm} I.,   {Metzger} B.~D.,  2016, \mn@doi [\mnras]
  {10.1093/mnras/stw1949}, \href
  {http://adsabs.harvard.edu/abs/2016MNRAS.463..394V} {463, 394}

\bibitem[\protect\citeauthoryear{{Vogt}, {Tappert}, {Puebla},
  {Fuentes-Morales}, {Ederoclite}  \& {Schmidtobreick}}{{Vogt}
  et~al.}{2018}]{2018MNRAS.478.5427V}
{Vogt} N.,  {Tappert} C.,  {Puebla} E.~C.,  {Fuentes-Morales} I.,  {Ederoclite}
  A.,   {Schmidtobreick} L.,  2018, \mn@doi [\mnras] {10.1093/mnras/sty1445},
  \href {https://ui.adsabs.harvard.edu/abs/2018MNRAS.478.5427V} {478, 5427}

\bibitem[\protect\citeauthoryear{{Vurm} \& {Metzger}}{{Vurm} \&
  {Metzger}}{2018}]{2018ApJ...852...62V}
{Vurm} I.,  {Metzger} B.~D.,  2018, \mn@doi [\apj] {10.3847/1538-4357/aa9c4a},
  \href {https://ui.adsabs.harvard.edu/abs/2018ApJ...852...62V} {852, 62}

\bibitem[\protect\citeauthoryear{{Wada} et~al.,}{{Wada}
  et~al.}{2021}]{2021GeoRL..4891910W}
{Wada} Y.,  et~al., 2021, \mn@doi [\grl] {10.1029/2020GL091910}, \href
  {https://ui.adsabs.harvard.edu/abs/2021GeoRL..4891910W} {48, e91910}

\bibitem[\protect\citeauthoryear{{Warner}}{{Warner}}{2003}]{2003cvs..book.....W}
{Warner} B.,  2003, {Cataclysmic Variable Stars},
  \mn@doi{10.1017/CBO9780511586491.
}

\bibitem[\protect\citeauthoryear{{Weston} et~al.,}{{Weston}
  et~al.}{2016a}]{2016MNRAS.457..887W}
{Weston} J. H.~S.,  et~al., 2016a, \mn@doi [\mnras] {10.1093/mnras/stv3019},
  \href {https://ui.adsabs.harvard.edu/abs/2016MNRAS.457..887W} {457, 887}

\bibitem[\protect\citeauthoryear{{Weston} et~al.,}{{Weston}
  et~al.}{2016b}]{2016MNRAS.460.2687W}
{Weston} J. H.~S.,  et~al., 2016b, \mn@doi [\mnras] {10.1093/mnras/stw1161},
  \href {https://ui.adsabs.harvard.edu/abs/2016MNRAS.460.2687W} {460, 2687}

\bibitem[\protect\citeauthoryear{{Whewell} et~al.,}{{Whewell}
  et~al.}{2015}]{2015A&A...581A..79W}
{Whewell} M.,  et~al., 2015, \mn@doi [\aap] {10.1051/0004-6361/201526742},
  \href {https://ui.adsabs.harvard.edu/abs/2015A&A...581A..79W} {581, A79}

\bibitem[\protect\citeauthoryear{{Wiescher}, {G{\"o}rres}, {Uberseder},
  {Imbriani}  \& {Pignatari}}{{Wiescher} et~al.}{2010}]{2010ARNPS..60..381W}
{Wiescher} M.,  {G{\"o}rres} J.,  {Uberseder} E.,  {Imbriani} G.,   {Pignatari}
  M.,  2010, \mn@doi [Annual Review of Nuclear and Particle Science]
  {10.1146/annurev.nucl.012809.104505}, \href
  {https://ui.adsabs.harvard.edu/abs/2010ARNPS..60..381W} {60, 381}

\bibitem[\protect\citeauthoryear{{Williams}}{{Williams}}{1985}]{1985ESOC...21..225W}
{Williams} R.~E.,  1985, in {Danziger} I.~J.,  {Matteucci} F.,   {Kjar} K.,
  eds,  European Southern Observatory Conference and Workshop Proceedings Vol.
  21, European Southern Observatory Conference and Workshop Proceedings. pp
  225--232

\bibitem[\protect\citeauthoryear{{Williams}}{{Williams}}{1992}]{1992AJ....104..725W}
{Williams} R.~E.,  1992, \mn@doi [\aj] {10.1086/116268}, \href
  {https://ui.adsabs.harvard.edu/abs/1992AJ....104..725W} {104, 725}

\bibitem[\protect\citeauthoryear{{Wolf}, {Bildsten}, {Brooks}  \&
  {Paxton}}{{Wolf} et~al.}{2013}]{2013ApJ...777..136W}
{Wolf} W.~M.,  {Bildsten} L.,  {Brooks} J.,   {Paxton} B.,  2013, \mn@doi
  [\apj] {10.1088/0004-637X/777/2/136}, \href
  {https://ui.adsabs.harvard.edu/abs/2013ApJ...777..136W} {777, 136}

\bibitem[\protect\citeauthoryear{{Wolf}, {Townsend}  \& {Bildsten}}{{Wolf}
  et~al.}{2018}]{2018ApJ...855..127W}
{Wolf} W.~M.,  {Townsend} R. H.~D.,   {Bildsten} L.,  2018, \mn@doi [\apj]
  {10.3847/1538-4357/aaad05}, \href
  {https://ui.adsabs.harvard.edu/abs/2018ApJ...855..127W} {855, 127}

\bibitem[\protect\citeauthoryear{{Yaron}, {Prialnik}, {Shara}  \&
  {Kovetz}}{{Yaron} et~al.}{2005}]{2005ApJ...623..398Y}
{Yaron} O.,  {Prialnik} D.,  {Shara} M.~M.,   {Kovetz} A.,  2005, \mn@doi
  [\apj] {10.1086/428435}, \href
  {https://ui.adsabs.harvard.edu/abs/2005ApJ...623..398Y} {623, 398}

\bibitem[\protect\citeauthoryear{{Zel'dovich} \& {Raizer}}{{Zel'dovich} \&
  {Raizer}}{1967}]{1967pswh.book.....Z}
{Zel'dovich} Y.~B.,  {Raizer} Y.~P.,  1967, {Physics of shock waves and
  high-temperature hydrodynamic phenomena}

\bibitem[\protect\citeauthoryear{{Zemko}, {Orio}, {Mukai}  \&
  {Shugarov}}{{Zemko} et~al.}{2014}]{2014MNRAS.445..869Z}
{Zemko} P.,  {Orio} M.,  {Mukai} K.,   {Shugarov} S.,  2014, \mn@doi [\mnras]
  {10.1093/mnras/stu1783}, \href
  {https://ui.adsabs.harvard.edu/abs/2014MNRAS.445..869Z} {445, 869}

\bibitem[\protect\citeauthoryear{{de Diego}}{{de
  Diego}}{2010}]{2010AJ....139.1269D}
{de Diego} J.~A.,  2010, \mn@doi [\aj] {10.1088/0004-6256/139/3/1269}, \href
  {https://ui.adsabs.harvard.edu/abs/2010AJ....139.1269D} {139, 1269}

\bibitem[\protect\citeauthoryear{{de Jager} \& {B{\"u}sching}}{{de Jager} \&
  {B{\"u}sching}}{2010}]{2010A&A...517L...9D}
{de Jager} O.~C.,  {B{\"u}sching} I.,  2010, \mn@doi [\aap]
  {10.1051/0004-6361/201014362}, \href
  {https://ui.adsabs.harvard.edu/abs/2010A&A...517L...9D} {517, L9}

\bibitem[\protect\citeauthoryear{{de Jager}, {Raubenheimer}  \&
  {Swanepoel}}{{de Jager} et~al.}{1989}]{1989A&A...221..180D}
{de Jager} O.~C.,  {Raubenheimer} B.~C.,   {Swanepoel} J.~W.~H.,  1989, \aap,
  \href {https://ui.adsabs.harvard.edu/abs/1989A&A...221..180D} {221, 180}

\bibitem[\protect\citeauthoryear{{den Herder} et~al.,}{{den Herder}
  et~al.}{2001}]{2001A&A...365L...7D}
{den Herder} J.~W.,  et~al., 2001, \mn@doi [\aap] {10.1051/0004-6361:20000058},
  \href {https://ui.adsabs.harvard.edu/abs/2001A&A...365L...7D} {365, L7}

\bibitem[\protect\citeauthoryear{{{\v{S}}imon}}{{{\v{S}}imon}}{2018}]{2018A&A...614A.141S}
{{\v{S}}imon} V.,  2018, \mn@doi [\aap] {10.1051/0004-6361/201731308}, \href
  {https://ui.adsabs.harvard.edu/abs/2018A&A...614A.141S} {614, A141}

\bibitem[\protect\citeauthoryear{{van den Bergh} \& {Younger}}{{van den Bergh}
  \& {Younger}}{1987}]{1987A&AS...70..125V}
{van den Bergh} S.,  {Younger} P.~F.,  1987, \aaps, \href
  {https://ui.adsabs.harvard.edu/abs/1987A&AS...70..125V} {70, 125}

\makeatother
\end{thebibliography}








\bsp 
\label{lastpage}
\end{document}